\documentclass[12pt]{iopart}

\usepackage{xspace}
\usepackage{epsfig}
\usepackage{cancel}

\usepackage{graphicx}

\newcommand{\nc}{\newcommand}

\nc{\beq}{\begin{equation}}
\nc{\eeq}{\end{equation}}
\nc{\beqa}{\begin{eqnarray}}
\nc{\eeqa}{\end{eqnarray}}
\nc{\bea}{\begin{eqnarray}}
\nc{\eea}{\end{eqnarray}}
\nc{\ra}{\rightarrow}

\nc{\LL}{L}

\def\lozenge{\boxit{\hbox to 1.5pt{\vrule height 1pt width 0pt \hfill}}}

\def\to{\rightarrow}

\newcommand{\ord}[1]{\mathcal{O}{(#1)}}
\newcommand{\vev}[1]{\langle {#1} \rangle}
\newcommand{\mP}{{\bar M}_P}

\newcommand{\dalam}{\raise-1mm\hbox{\large$\Box$}}
\newcommand{\fb}[1]{{#1}~fb$^{-1}$}

\def\bea{\begin{eqnarray}}
\def\eea{\end{eqnarray}}

\def\lsim{\mathrel{\raise.3ex\hbox{$<$\kern-.75em\lower1ex\hbox{$\sim$}}}}
\def\gsim{\mathrel{\raise.3ex\hbox{$>$\kern-.75em\lower1ex\hbox{$\sim$}}}}
\newcommand{ \slashchar }[1]{\setbox0=\hbox{$#1$}   
   \dimen0=\wd0                                     
   \setbox1=\hbox{/} \dimen1=\wd1                   
   \ifdim\dimen0>\dimen1                            
      \rlap{\hbox to \dimen0{\hfil/\hfil}}          
      #1                                            
   \else                                            
      \rlap{\hbox to \dimen1{\hfil$#1$\hfil}}       
      /                                             
   \fi}                                             %

\providecommand{\tabularnewline}{\\}

\def\zpri{{Z^\prime}}

\def\mzpri{M_{Z^\prime}}

\def\wpri{{W^\prime}}

\def\mwpri{M_{W^\prime}}

\def\to{\rightarrow}

\begin{document}

\title{Warped 5-Dimensional Models: \\
Phenomenological Status and Experimental Prospects}

\author{Hooman Davoudiasl}
\ead{hooman@bnl.gov}
\address{Department of Physics,
Brookhaven National Laboratory, Upton, NY 11973-5000, USA}
\author{Shrihari Gopalakrishna}
\ead{shri@quark.phy.bnl.gov}
\address{Department of Physics,
Brookhaven National Laboratory, Upton, NY 11973-5000, USA, and \\
The Institute of Mathematical Sciences (IMSc), C.I.T. Campus,
Taramani, Chennai 600113. India.}
\author{Eduardo Pont\'{o}n}
\ead{eponton@phys.columbia.edu}
\address{Department of Physics, Columbia University, 538 W. 120th St, New York,
NY 10027, USA}
\author{Jos\'{e} Santiago}
\ead{jsantiago@ugr.es}
\address{Institute for Theoretical Physics, ETH, CH-8093, Z\"urich, Switzerland, and \\
AFPE and Departamento de F\'{\i}sica Te\'orica y del Cosmos,
Universidad de Granada, E-18071 Granada, Spain}

\begin{abstract}

Warped 5-dimensional models, based on the original Randall-Sundrum
geometry, have been extended beyond their initial purpose of resolving
the gauge hierarchy problem.  Over the past decade, various
ingredients have been added to their basic structure in order to
provide natural and predictive models of flavor and also to address
existing constraints from precision data.  In this review, we examine
the theoretical and experimental status of realistic models that
accommodate current data, while addressing the hierarchy and flavor
puzzles of the Standard Model.  We also discuss the prospects for
future discovery of the TeV-scale Kaluza-Klein states that are
predicted to emerge in these models, and outline some of the
challenges that the detection of such particles pose for experiments
at the Large Hadron Collider.

\end{abstract}

\maketitle


\section{Introduction}

The Standard Model (SM) of particle physics is remarkably successful
in explaining a wide range of microscopic phenomena and has passed
numerous tests over the past few decades.  The only ingredient of this
model that has yet to be discovered is the Higgs boson.  It is the
vacuum expectation value (vev) of this field that breaks the
electroweak (EW) symmetry $SU(2)_L \times U(1)_Y$ down to $U(1)_{EM}$
of quantum electrodynamics and provides masses for elementary
particles in the SM. However, this picture, though economical, leaves
some intriguing questions unanswered.  One obvious and central
question is related to the patterns of the SM fermion masses (that,
including the neutrinos, span 12 orders of magnitude!)  and mixing;
this is the {\it flavor puzzle} and is based on firm experimental
evidence.

Another question, which is more theoretical in nature, arises when one
considers the effect of quantum corrections on the Higgs vev.  These
corrections are quadratically sensitive to any physical mass scales
that could emerge at energies above the weak scale $M_W \sim 100$~GeV.
However, stringent requirements from precision data strongly suggest
that new physics can only appear at scales much larger than $M_W$.
Besides, very high physical scales are well-motivated in the context
of a Grand Unified Theory (GUT) or theories of quantum gravity.
Therefore, one is faced with the question of what stabilizes the Higgs
potential well below such large scales.  The severest version of this
problem arises when the fundamental cutoff scale is assumed to be near
the 4D Planck mass $M_P\sim 10^{19}$~GeV; this is the {\it hierarchy
problem}.

In this review, we will focus on a framework based on a model
originally proposed by Randall and Sundrum \cite{Randall:1999ee} to
address the hierarchy problem.  The Randall-Sundrum (RS) model is
based on a slice of AdS$_5$ spacetime between two flat 4D boundaries,
often referred to as the UV (Planck) and IR (TeV) branes.  The branes
are assumed to be separated by a distance of order the curvature
radius of the background, which is in turn of order the 5D fundamental
scale $M_5$.  This model provides a natural resolution of the
hierarchy, by exponentially generating the weak scale from scales of
order $M_5 \sim M_P$.  To achieve this, the Higgs field is localized
at the IR-brane and the warping along the fifth dimension redshifts
the Planckian 5D Higgs vev down to the weak scale.  The requisite
brane separation was shown to be easily obtained in simple models
\cite{Goldberger:1999uk}, resulting in a stable geometry.

In the original RS model all SM fields were localized on the IR-brane
and the most distinct signature of this setup was the emergence of a
spin-2 tower of Kaluza-Klein (KK) gravitons \cite{Davoudiasl:1999jd}
at the TeV scale.  However, it was soon realized that resolving the
hierarchy only required the Higgs to be IR-localized
\cite{Goldberger:1999wh} and other fields could propagate in the 5D
bulk.  The SM gauge sector \cite{Davoudiasl:1999tf,Pomarol:1999ad} and
the fermions \cite{Grossman:1999ra} were then promoted to 5D fields.
It was shown that localization of fermion zero modes could be achieved
using 5D mass terms \cite{Grossman:1999ra}, with the heavy fermions
localized toward the Higgs and the light fields localized away from
the Higgs, resulting in a natural and predictive model of flavor.  In
particular, given that 5D location sets the relevant physical mass
scale in the RS background, light-flavor 4-fermion operators are
governed by large cutoff scales in warped flavor models
\cite{Gherghetta:2000qt}, as required by precision data.  This
interesting feature allows one to address both the hierarchy and
flavor puzzles within RS-type models, making them a very attractive
model building framework that can be experimentally tested at the weak
scale.

While warped 5D models explain some of the important unresolved issues
of the SM to a large extent, they are still subject to a degree of
(fine-)tuning, once confronted with precision EW \cite{Carena:2007ua}
and flavor data
\cite{Huber:2000ie}-\cite{Agashe:2006iy}.   
The gist of this
tension is traced back to the resolution of the hierarchy problem that
would prefer the scale of new physics to be near $M_W$.  This gives
rise to an effective 4D cutoff scale in the TeV range, where all
non-renormalizable interactions that are not forbidden by a symmetry
will emerge, whereas precision data generally require the cutoff scale
to be at $\ord{10}$~TeV or more.  To eliminate such ``little
hierarchies," the basic 5D SM structure of warped models have been
augmented by various new ingredients, such as larger gauge symmetry
groups \cite{Agashe:2003zs,Agashe:2006at}, in order to
keep the scale of KK modes in the few-TeV range.  This would remove
the need for inordinate amounts of (fine-)tuning and also make the
models more likely to be testable in various high energy experiments.

Even though the most recent warped models of hierarchy and flavor can
accommodate KK masses as low as 2-3~TeV \cite{Carena:2007ua}, the
discovery of the signature states quite generically poses a
significant challenge.  The basic reason for this originates in the
way various states are localized in the bulk: light fermions are
UV-localized, whereas the KK modes and heavy fermions are
IR-localized.  This suppresses the couplings of the KK modes to light
flavors that are important initial (such as light quarks) and final
(such as $\mu^\pm$) states in collider production and detection,
respectively.  A volume suppression also affects the couplings of the
highly localized KK modes to gauge fields that are spread over the
bulk.  Hence, in models that explain the flavor problem we end up with
suppressed production of KK modes that eventually decay into heavy
fields, such as $t {\bar t}$, requiring complicated event
reconstructions.  Also, the heavy final states are highly boosted due
to the high mass of the KK parent.  This makes the eventual decays of
the heavy final sates into di-jets difficult to distinguish from
mono-jets, hence rendering QCD background suppression challenging.

In the first half of this review, we will discuss the theoretical
techniques that have been used to analyze the phenomenology of RS-type
models.  Although our results are of more general applicability, we
focus on the most recent prototypical models that incorporate a
realistic flavor structure (bulk SM gauge and fermion fields) and
possibly allow for new physics within reach of the Large Hadron
Collider (LHC).  We restrict ourselves to the original RS background
and assume that there is a Higgs degree of freedom in the low-energy
theory.  We survey the phenomenological status of these realistic
scenarios, thus providing a guide as to the plausible values of
parameters in viable models.  Having determined the phenomenologically
relevant range of parameters, we will discuss the key collider
signatures of warped 5D models in the second half of the review.  Our
main focus will be models endowed with bulk symmetries that allow for
the KK masses to be within the reach of the LHC, without the need for
unnatural choices of parameters.  Some of the experimental obstacles
that these models pose, as well as proposals to overcome such
problems, will be outlined.  We will also briefly discuss the expected
enhancement of clean signals in ``truncated" warped models whose UV
cutoff is well below $M_P$, but still explain flavor.  We will
conclude with a summary.

Before closing this introduction, we would like to emphasize that this
review is not meant to be exhaustive or comprehensive.  Such an
undertaking will require a much longer article due to the large amount
of research performed on warped models, over the past several years.
In particular, important cosmological and gravitational effects that have been discussed in a number of interesting works lie outside the scope of our review, but certainly deserve attention in separate or more comprehensive surveys of warped physics.  
As such, we have limited our survey to some of the most generic
features of model building and phenomenology.  
To the extent possible,
we have cited works that are directly or closely related to our
discussions.  However, unfortunately, many worthy papers have been
left out in our review.  
We hope that this article will serve to
present a sufficiently inclusive account of the status of warped
models to motivate the interested reader to pursue further references
and delve more deeply into some of the subjects that we have (not)
considered here.


\title{Part I \,\, General Tools and Electroweak Constraints}
\vspace{2mm}
\noindent
\textbf{\it Contributed by E.~Pont\'{o}n and J.~Santiago}
\vspace{4mm}

In the first part of this review, we describe the tools that have been
used to study the physics of warped extra-dimensional scenarios.  We
have put special emphasis on the explanation and comparison of the
different techniques used.  We focus on models that incorporate a
custodial symmetry, and analyze the constraints from EW precision
measurements.  There are also important constraints from flavour data
that, under the assumption of strict ``anarchy'' (that the 5D flavor
couplings present no structure at all and are all of the same order,
including complex phases), can result in rather severe bounds on the
new physics.  These constraints depend on parameters different from
those relevant for the EW precision constraints and can be evaded
without significantly modifying the latter.  Due to space constraints,
we focus on the EW analysis, except for a brief section on flavour
where we collect some of the recent references.

\section{Methods in models with warped extra dimensions}
\label{sec:methods}

We consider a 5D space-time with a warped metric,
written in conformally flat coordinates as~\footnote{Another commonly
used form for the AdS metric is $ds^2 = e^{-2k y} \eta_{\mu\nu} dx^\mu
dx^\nu - dy^2$, where $k = 1/L_{0}$ is the AdS curvature.  The
formulas in conformal coordinates can be transcribed into the ``proper
distance" coordinates by setting $z/L_{0} = e^{k y}$, $\partial_{y} =
a(z)^{-1} \partial_{z}$ and $dy = a(z) dz$.}
\begin{equation}
ds^2 = a(z)^2 \big( \eta_{\mu\nu} dx^\mu dx^\nu - dz^2)~,
\qquad
a(z)=\frac{L_0}{z}~,
\label{metric}
\end{equation}
where $L_0 \leq z \leq L_1$.  Solving the hierarchy problem requires
$L_0 \approx M_P^{-1}$, and $L_1 \approx (\mbox{TeV})^{-1}$, although
other choices of scales have proved useful in addressing e.g. the flavour
problem~\cite{Davoudiasl:2008hx}.

\subsection{Kaluza-Klein expansions}

We start by collecting the main results for the Kaluza-Klein (KK)
wavefunctions associated with spin-$2$~\cite{Davoudiasl:1999jd},
spin-$1$~\cite{Davoudiasl:1999tf,Pomarol:1999ad} and
spin-$1/2$~\cite{Grossman:1999ra} fields propagating on the background
of Eq.~(\ref{metric}) (for a unified derivation,
see~\cite{Gherghetta:2000qt}).  The KK wavefunctions are especially
useful in studying the collider phenomenology of warped extra
dimensions, to be discussed in the second part of this review.

\subsubsection{Gauge Bosons}

The 5D gauge action is
\beqa
S_{\rm gauge} &=& -\frac{1}{4g^2_{5}} \int \! d^5x \sqrt{g} 
  F^{a}_{MN} F^{a\,MN} 
+ S^{\rm UV}_{\rm gauge} + S^{\rm IR}_{\rm gauge}~,
\label{GaugeAction}
\eeqa
where $M,N = 0,1,2,3,z$ run over the 5D coordinates, $g_{5}$ is the 5D
gauge coupling (with mass dimension $-1/2$), $F^{a}_{MN}$ is the gauge
field strength, $a$ is a gauge index, and $S^{\rm UV}_{\rm gauge}$,
$S^{\rm IR}_{\rm gauge}$ contain possible brane-localized terms for
the gauge fields, to be specified shortly.  If the gauge field does
not satisfy Dirichlet boundary conditions at both branes (see below)
we can go to the unitary gauge, $A_5=0$ \footnote{Other gauge choices
can also be useful~\cite{Randall:2001gb}.  A gauge independent
expansion can be obtained by choosing the KK modes of $A_5$ equal to
$\partial_z f_n/m_n$.}, in which the KK decomposition reads
\beqa
A_{\mu}(x,z) &=& \sum_{n} A^n_{\mu}(x) f_n(z)~.
\label{GaugeKK}
\eeqa
The KK wavefunctions satisfy $O(m_{n},z) f_n(z) = 0$, where $m_{n}$
are the gauge KK masses and the differential operator in the
coordinates of Eq.~(\ref{metric}) is
\begin{equation}
O(p,z) = \partial_z a(z) \partial_z + a(z) p^2~.
\label{GaugeOp}
\end{equation}
The solutions are written in terms of Bessel functions of the first
and second kind, $J_{\alpha}(x)$ and $Y_{\alpha}(x)$, for which we
shall use the shorthand notation
\beqa
J_{\alpha}^{0,1,z} \equiv J_{\alpha}(m_{n} L_0), J_{\alpha}(m_{n} L_1), J_{\alpha}(m_{n} z)~,
\label{Bessel}
\eeqa
and similarly for $Y^{0,1,z}_{\alpha}$. Then
\beqa
f_n(z) &=& A_{n} \, a(z)^{-1} \left[ J_1 ^{z} + B_{n} Y_{1}^z \right]~,
\label{Gaugef}
\eeqa
where the overall constants $A_{n}$ are determined by the
orthonormality condition
\beqa
\int^{L_{1}}_{L_{0}} \! dz \, a(z) f_n(z) f_m(z) &=& L \, \delta_{mn}~,
\label{GaugeNormalization}
\eeqa
and we defined the volume factor
\beqa
L \equiv \int_{L_0}^{L_1} \! dz\, a(z) = L_0 \log\frac{L_1}{L_0}~.
\label{volume}
\eeqa
The boundary conditions (b.c.'s) take the form
\beqa
\big[L_{0} \partial_z - a(z) b_{\rm UV} \big] f_{n}(z) \Big|_{z=L_0} &=& 0~,
\label{bcUVGauge}
\\
\big[L_{0} \partial_z + a(z) b_{\rm IR} \big] f_{n}(z) \Big|_{z=L_1} &=& 0~,
\label{bcIRGauge}
\eeqa
where the $b_{i}$ depend on the brane localized terms.  For instance,
in the presence of localized kinetic terms with coefficients $r_{i}$
and localized mass terms $m_{i}$:
\beqa
b_{\rm UV, IR} \equiv a^{-2}_{0,1} \, 
\hat{r}_{\rm UV, IR} \, \hat{m}^2_{n} - \hat{m}_{\rm UV, IR}~,
\label{GaugeLocalizedTerms}
\eeqa
where $a_0 \equiv a(L_0)$, $a_1 \equiv a(L_1)$, and we defined
dimensionless quantities $\hat{m}_{n} = m_{n} L_{0}$, $\hat{r}_{i} =
r_{i}/L_{0}$ and $\hat{m}_{i} = m_{i} L_{0}$ for $i = {\rm UV, IR}$
(the natural scale of the dimensionful microscopic parameters is
expected to be of order $L_{0}$).  The b.c. on the IR brane determines
$B_{n} = -\tilde{J}^{\rm IR}_{1}/\tilde{Y}^{\rm IR}_{1}$, where
\beqa
\tilde{J}_{\alpha}^{\rm UV, IR} \equiv \hat{m}_{n} J_{\alpha-1}^{0,1} 
\pm a_{0,1} b_{\rm UV, IR} J_{\alpha}^{0,1}~,
\label{Besseltilde}
\eeqa
and an analogous definition for $\tilde{Y}^{\rm UV, IR}_{\alpha}$.
The b.c. on the UV brane then leads to the eigenvalue equation
\beqa
\tilde{J}^{\rm UV}_{\alpha} \tilde{Y}^{\rm IR}_{\alpha} - 
\tilde{Y}^{\rm UV}_{\alpha} \tilde{J}^{\rm IR}_{\alpha} &=& 0~,
\quad \quad \quad
\textrm{with}~\alpha = 1~,
\label{eigenGauge}
\eeqa
which determines the KK masses, $m_{n}$. 

The case with $\hat{m}_{\rm UV} = \hat{m}_{\rm IR} = 0$ leads to a
zero mass eigenvalue, $m_{0} = 0$, corresponding to an unbroken 4D
gauge symmetry.  In the SM electroweak sector, an IR localized Higgs
field leads to $\hat{m}_{\rm IR} \neq 0$ and non-zero $Z$ and $W^\pm$
masses.  The Higgsless limit~\cite{Csaki:2003dt} corresponds to
$\hat{m}_{\rm IR} \rightarrow \infty$.  A case of interest in the
models with custodial symmetry to be introduced in
Section~\ref{CustodialModel} corresponds to Dirichlet boundary
conditions on the UV brane, and can be obtained by taking
$\hat{m}_{\rm UV} \rightarrow \infty$.  In the following, for
simplicity we will assume that the localized kinetic terms are
sufficiently small to be neglected.  However, it should be remarked
that when $\hat{r}_{\rm IR} > 1$ these can have important
phenomenological consequences~\cite{Davoudiasl:2002ua,Carena:2002dz}.
If both b.c.'s are Dirichlet, $\hat{m}_{\mathrm{UV},\mathrm{IR}}\to
\infty$, a zero mode for $A_5$ remains in the
spectrum~\cite{Manton:1979kb}-\cite{Hatanaka:1998yp}.  This can have
interesting consequences for EWSB~\cite{Contino:2003ve,Agashe:2004rs}.

\subsubsection{Fermions}

The fermion action is
\beqa
S_{\rm fermion} &=&
\int\! d^5x \sqrt{g} \left\{ \frac{i}{2}
\overline{\Psi} e^M_{A} \Gamma^{A} D_{M} \Psi - \frac{i}{2}
(D_{M}\Psi)^\dagger \Gamma^0 e^M_{A} \Gamma^{A} \Psi - M
\overline{\Psi} \Psi \right\} 
\nonumber \\[0.5em]
&& \mbox{}
+ S^{\rm UV}_{\rm fermion} + S^{\rm IR}_{\rm fermion}~,
\label{FermionAction}
\eeqa
where $\Gamma^{A} = (\gamma^\mu, -i \gamma_{5})$ are the flat space
Dirac gamma matrices in 5D space, $D_{M}$ is the covariant derivative
with respect to the gauge symmetry as well as general coordinate and
local Lorentz transformations~\footnote{For a diagonal metric of the
form $ds^2 = a(z)^2 \eta_{\mu\nu} dx^\mu dx^\nu + b(z)^2 dz^2$ the
spin connection in $D_{M}$ cancels out in the fermion action,
Eq.~(\ref{FermionAction}).}, and $S^{\rm UV}_{\rm fermion}$, $S^{\rm
IR}_{\rm fermion}$ contain possible fermion localized terms.  It is
convenient to express the 5D mass term in units of $1/L_{0}$ thus
defining a dimensionless parameter $c = M L_{0}$.  The KK
decomposition for each 4D chirality of the 5D fermion reads
($\gamma_{5} \Psi_{L,R} = \mp \Psi_{L,R}$)
\beqa
\Psi_{L,R}(x,z) &=& \sum_{n} \psi^n_{L,R}(x) f^n_{L,R}(z)~,
\label{FermionKK}
\eeqa
where $f^n_{L}(z)$ and $f^n_{R}(z)$ are related by $O_{c} f^n_{L} =
m_{n}f^n_{R}$, with 
\beqa
O_{c} = \partial_{z} - z^{-1} \left(2 - c \right) ~,
\label{FirstOrderOp}
\eeqa
and $m_{n}$ are the fermion KK masses.  The LH profile, $f^n_{L}(z)$,
obeys the second order differential equation $O_{-c} \, O_{c} f^n_{L}
= -m^2_{n} f^n_{L}$.  Explicitly, the solution for the LH modes reads
\beqa
f^n_L(z) &=& A^f_{n} \, a(z)^{-5/2} \left[ J_{c+1/2} ^{z} + B^f_{n} Y_{c+1/2}^z \right]~,
\label{Fermionf}
\eeqa
where the overall constants $A^f_{n}$ are determined by the
normalization condition Eq.~(\ref{GaugeNormalization}) with $f_{n}(z)
\rightarrow L^{1/2} a(z)^{3/2} f^n_L(z)$.

The b.c.'s for both chiralities are not independent but related by the
equations of motion.  4D chirality can be generated at the level of
zero modes by a choice of b.c.'s that forbid a zero mode for one of
the two chiralities.  For instance, a LH zero mode is obtained by
choosing b.c.'s for $ f^n_L(z)$ as in Eqs.~(\ref{bcUVGauge}) and
(\ref{bcIRGauge}) with $b_{\rm UV, IR} \rightarrow \pm (2 - c) +
b^f_{\rm UV, IR}$, and $b^f_{i}$ as in Eq.~(\ref{GaugeLocalizedTerms})
but with $\hat{m}_{i} = 0$ and $r_{i}$ denoting the coefficients of
possible fermion brane kinetic terms.  Neglecting brane kinetic terms,
the corresponding LH zero mode wave function is
\beqa
f^0_{L}(z) &=& \sqrt{\frac{(1-2c)}{L_{0} \left[ (L_{1}/L_{0})^{(1-2c)}
-1 \right]}} \, \left( \frac{z}{L_{0}} \right)^{2-c}~,
\label{FermionZeroMode}
\eeqa
which corresponds to a chiral fermion exponentially localized towards
the UV (IR) brane for $c > 1/2$ ($c < 1/2$), while the physical
profile is flat for $c = 1/2$.  The remaining constants for the
massive LH modes in Eq.~(\ref{Fermionf}) are given by $B^f_{n} =
-\tilde{J}^{\rm IR}_{c+1/2}/\tilde{Y}^{\rm IR}_{c+1/2}$, where
$\tilde{J}^{\rm IR}_{c+1/2}$ and $\tilde{Y}^{\rm IR}_{c+1/2}$ are
defined in Eq.~(\ref{Besseltilde}), taking $b_{i} = b^f_{i}$.  The
profiles for the associated RH chiralities, $f^n_R(z)$, are simply
obtained from Eq.~(\ref{Fermionf}) by the replacement $c+1/2 \to
c-1/2$ [but with exactly the same constants $A^f_{n}$ and $B^f_{n}$
that appear in $f^n_L(z)$].  The fermion KK masses are determined by
Eq.~(\ref{eigenGauge}) with $\alpha = c + 1/2$.

If instead the zero mode is right-handed, the corresponding profiles
are obtained with the replacement $c \rightarrow -c$ everywhere in
Eqs.~(\ref{Fermionf}) and (\ref{FermionZeroMode}) (and exchanging the
LH fields by the RH ones).

\subsubsection{Gravitons}

KK excitations of the graviton (spin-2 resonances) may give a striking
signature of the extra-dimensional structure.  We summarize briefly
the relevant results for the graviton KK wavefunctions.  The tensor
fluctuations $h_{\mu\nu}$ are introduced by making $\eta_{\mu\nu}
\rightarrow \eta_{\mu\nu} + h_{\mu\nu}(x,z)$ in Eq.~(\ref{metric}).
Starting from the Einstein-Hilbert action, linearizing in
$h_{\mu\nu}$, and choosing the transverse-traceless gauge,
$\partial^\mu h_{\mu\nu} = h^\alpha_{\alpha} = 0$, the KK expansion
reads
\beqa
h_{\mu\nu}(x,z) &=& \sum_{n} h^n_{\mu\nu}(x) f_n^G(z)~,
\label{GravitonKK}
\eeqa
where $\left[ a(z)^{-3} \partial_{z} a^3(z) \partial_{z} + m^2_{n}
\right]f_n^G(z) = 0$, so that [see the definitions of Eq.~(\ref{Bessel})]
\beqa
f_n^G(z) &=& A^G_{n} \, a(z)^{-2} \left[ J_2 ^{z} + B^G_{n} Y_{2}^z \right]~.
\eeqa
The normalization constants $A^G_{n}$ are obtained from
Eq.~(\ref{GaugeNormalization}) with $f_{n}(z) \rightarrow a(z)
f^G_n(z)$.  Localized curvature (``kinetic'')
terms~\cite{Davoudiasl:2003zt} can be introduced through the functions
of Eq.~(\ref{Besseltilde}), as was done for gauge and fermion fields,
so that $B^G_{n} = -\tilde{J}^{\rm IR}_{2}/\tilde{Y}^{\rm IR}_{2}$.
The graviton KK masses are determined by Eq.~(\ref{eigenGauge}) with
$\alpha = 2$.

\subsection{5D Propagators: the gauge field case}

Let us consider the Green's function of the quadratic operator for a
gauge boson.  We add to the action~(\ref{GaugeAction}) a gauge
fixing term~\cite{Randall:2001gb}
\beqa
{\cal L}_{g.f.} &=& -\frac{1}{2\xi} a(z) \left\{ \partial_{\mu} A^\mu - \xi a(z)^{-1} \partial_{z} \left[ a(z) A_5 \right] \right\}^2~,
\eeqa
where $\xi$ is the gauge fixing parameter.  Working in mixed
position/momentum space~\cite{ArkaniHamed:2001mi}, the propagator
takes the form
\beqa
-i P_{p}(z,z') \mathcal{P}_{\mu\nu} - i P_{\frac{p}{\sqrt{\xi}}}(z,z') \frac{p_{\mu} p_{\nu}}{p^2}~,
\eeqa
where $\mathcal{P}^{\mu\nu} = \eta^{\mu\nu}-p^\mu p^\nu/p^2$ is the
transverse projector, and $P_{p}(z,z')$ satisfies
\begin{equation}
O(p,z)P_p(z,z^\prime)=\delta(z-z^\prime)~,
\end{equation}
with $O(p,z)$ given by Eq.~(\ref{GaugeOp}).  $P_p(z,z^\prime)$
satisfies the same boundary conditions as the gauge field
wavefunctions $f_{n}(z)$, Eqs.~(\ref{bcUVGauge}) and
(\ref{bcIRGauge}); for the inclusion of brane localized terms,
see~\cite{Carena:2002dz}.  On the IR brane, we take $m_{\rm IR} \equiv
m$ in Eq.~(\ref{GaugeLocalizedTerms}).  For the UV b.c. we will choose
simple Neumann or Dirichlet (i.e. $b_{\rm UV} = 0$ or $b_{\rm UV}
\rightarrow \infty$), which imply, respectively, an unbroken or
spontaneously broken gauge symmetry at the UV brane:
\begin{equation}
\partial_z P^N_p(z,z^\prime) \Big|_{z=L_0}=0~ 
\qquad \qquad
P^D_p(z,z^\prime) \Big|_{z=L_0}=0~. 
\end{equation}
The two solutions can be written in terms of two functions, denoted
$K_m(p,z)$ and $S(p,z)$~\cite{Falkowski:2008yr}, that satisfy the
homogeneous differential equation
\begin{equation}
O(p,z)K_m(p,z)=O(p,z)S(p,z)=0~,
\label{DEKS}
\end{equation}
and boundary conditions
\begin{eqnarray}
K_m(p,L_0)&=&1~,
\quad \quad
L_{0} K^\prime_m(p,L_1)+ a_1 b_{\rm IR} K_m(p, L_1)=0~,
\label{bcsK}
\\
S(p,L_0)&=&0~,
\quad \quad
S^\prime(p, L_0)=1~,
\label{bcsS}
\end{eqnarray}
where a prime denotes derivative with respect to $z$.  The
Green's functions read
\begin{eqnarray}
P_p^N(z,z^\prime) &=& 
\frac{K_m(p,z_<) K_m(p,z_>)}{K^\prime_m(p,L_0)}  
- S(p, z_<) K_m(p, z_>)~, 
\label{PpN:general} \\
P_p^D(z,z^\prime) &=& 
- S(p, z_<) K_m(p, z_>)~, 
\label{PpD:general}
\end{eqnarray}
where $z_<$ and $z_>$ are the minimum and maximum of $z$ and
$z^\prime$, respectively.  The explicit solutions for $K_{m}$ and
$S$ are given in Eqs.~(\ref{solK}) and (\ref{solS}) of the
appendix.  The full propagators read
\begin{eqnarray}
P_p(z,z^\prime)&=&
\frac{\pi}{2} \frac{z_< z_>}{L_0}
\frac{(\tilde{Y}_1^{\rm UV} J_1^{z_<} - \tilde{J}_1^{\rm UV} Y_1^{z_<}) 
(\tilde{Y}_1^{\rm IR} J_1^{z_>} 
- \tilde{J}_1^{\rm IR} Y_1^{z_>})}{\tilde{J}_1^{\rm UV} \tilde{Y}_1^{\rm IR} 
- \tilde{Y}_1^{\rm UV} \tilde{J}_1^{\rm IR}}~, 
\label{AdSPropagator}
\end{eqnarray} 
where we used the notation for the Bessel functions of
Eqs.~(\ref{Bessel}) and (\ref{Besseltilde}) with $m_{n} \rightarrow
p$, and $b_{\rm UV} = 0$ for $P_p^N(z,z^\prime)$ while $b_{\rm UV}
\rightarrow \infty$ for $P_p^D(z,z^\prime)$.

We have split the Neumann propagator in Eq.~(\ref{PpN:general}) in two
pieces, one that vanishes at the UV brane and exactly coincides with
the Dirichlet propagator, Eq.~(\ref{PpD:general}), and another that is
the product of the holographic functions times the boundary field two
point function, to be introduced in Section~\ref{sec:holography}.
Alternatively, we can separate it into the contribution from the
massless zero mode (before EWSB) and that of the massive modes
\begin{equation}
P_p(z,z^\prime) = P_p^{(0)}(z,z^\prime) + \tilde{P}_p(z,z^\prime)~,
\label{Pp:zero:heavy}
\end{equation}
where the Green's function for the zero mode reads
\begin{equation}
P_p^{(0)} = \frac{1}{p^2 L_0 \log\frac{L_1}{L_0}}~,
\label{ZeroModePropagator}
\end{equation}
and, for simplicity, we set $r_{\rm UV} = r_{\rm IR} = 0$
(see~\cite{Davoudiasl:2002ua,Carena:2002dz,delAguila:2003bh} for a
thorough discussion of the phenomenological consequences of localized
gauge kinetic terms).  Propagator methods are also useful for
resumming the effects of the fermion KK tower (see e.g.
\cite{Carena:2004zn}). 

\subsection{Holographic method}
\label{sec:GeneralHolography}

Let us consider a bulk gauge boson $A_{M}$ with the following action
\begin{equation}
S_{\mathrm{gauge}}=\int \frac{d^4p}{(2\pi)^4} dz \sqrt{g} \left\{
-\frac{1}{4g_{5A}^2} (A_{MN})^2
+ \frac{1}{2} \, \textrm{v}^2(z)
(A_M)^2
\right\}~,
\end{equation}
where we allow for a bulk Higgs vev profile, $\textrm{v}(z)$, with
mass dimension $3/2$.~\footnote{For EWSB localized on the IR brane,
$\textrm{v}^2(z) \equiv \frac{1}{2} [v/a(L_{1})]^2 a(z)^{-1}
\delta(z-L_{1})$, which gives $m = \frac{1}{2} g^2_{5A}
[v/a(L_{1})]^2$ in Eqs.~(\ref{DEKS})-(\ref{bcsS}).  We introduce the
factors of $1/2$ and $1/a(L_{1})$ to match to the $SU(2) \times
U(1)_{Y}$ theory with $v \sim 174~{\rm GeV}$.} We will go to a gauge
with $A_5=0$ and assume that the gauge boson obeys Neumann b.c.'s on
the UV brane.  We write the 5D field as follows
\begin{equation}
A_\mu(p,z)=f_A(p,z) \bar{A}_\mu(p)~,
\end{equation}
with $f_A(p,L_0)=1$ so that $\bar{A}_\mu(p)$ is the boundary value of
the gauge field.  The IR boundary condition is the same as for the 5D
field.  We assume the boundary field satisfies the equation of motion
of a four-dimensional gauge field, which implies the following
equation for the holographic profile
\begin{equation}
O_{A} f_A(p,z)=0~,
\label{OpMA}
\end{equation}
where $O_{A} = O(p,z) - a(z)^3 g^2_{5A} \textrm{v}^2(z)$ with $O(p,z)$
as defined in Eq.~(\ref{GaugeOp}).  The effective action for the
boundary field can be obtained at tree level and to quadratic order in
the fields by inserting the equations of motion back in the action and
integrating over the extra dimension.  The bulk action vanishes due to
the equations of motion and the only remaining term is a boundary
piece that reads
\begin{equation}
S_{\mathrm{bound.}}= -\frac{1}{2}\int \frac{d^4p}{(2\pi)^4} 
\bar{A}_\mu \bar{\Pi}_A \bar{A}^\mu~,
\end{equation}
with the vacuum polarization function for the boundary field given by
\begin{equation}
\bar{\Pi}_A(p^2)= \frac{1}{g^2_{5A}} \partial_z f_A(p^2,z)\Big|_{z=L_0}~.
\label{GeneralPiA}
\end{equation}

Fermions can be treated in a similar way.  The action for a bulk
fermion is as in Eq.~(\ref{FermionAction}) but with an additional UV
localized term~\cite{Contino:2004vy}
\begin{equation}
\delta S_{\rm UV} =
\int_{\rm UV} \! d^4x \sqrt{g_{\rm ind}} \left( \pm \frac{1}{2} \right) \overline{\Psi} \Psi~,
\label{AdditionalFermionUVTerm}
\end{equation}
where $g_{\rm ind}$ is the induced metric, and the factor of $+1/2$
($-1/2$) is determined by the requirement that the LH (RH) chirality
be unconstrained on the UV brane.  We assume the LH component
satisfies Neumann b.c.'s on the UV brane and adopt a LH source
description
\begin{equation}
\Psi_{L,R}(p,z)=f_{L,R}(p,z) \psi_{L,R}(p)~,
\end{equation}
where $\not \! p \psi_{L,R}(p) = p \psi_{R,L}(p)$ and $f_{L,R}$
satisfy the equations
\begin{equation}
O_{-c} \, O_{c} f_L= - p^2 f_{L}~, 
\quad \quad \quad
O_{c} f_L = p f_R~,
\label{Hol:Fermion}
\end{equation}
where $O_{c}$ is given in Eq.~(\ref{FirstOrderOp}).  The boundary
conditions on the IR brane are as for the 5D field, while on the UV
brane we impose
\begin{equation}
f_L(p,L_0)=1~.
\label{BC:Hol:Fermion}
\end{equation}
The boundary action for $\psi_L$ is computed by replacing the
classical equations of motion and integrating over the extra
dimension.  Again, at the quadratic level there is no bulk
contribution and the boundary contribution,
Eq.~(\ref{AdditionalFermionUVTerm}), simply reads
\begin{equation}
S_{\mathrm{bound.}}= 
\int \frac{d^4p}{(2\pi)^4} 
\bar{\psi}_L \cancel{p} \Sigma(p) \psi_L~,
\end{equation}
where the kinetic function for the boundary field is
\begin{equation}
\Sigma(p)=\frac{f_R(p,L_0)}{p}~,
\end{equation}
where we have used $\Psi_R(p,L_0)=f_R(p,L_0) (\not \! p/p)\psi_L$.
Canonical normalization is obtained by the field redefinition
$\tilde{\psi}_{L} = \sqrt{\Sigma(p)} \, \psi_{L}$.  In the
limit of zero momentum we get $\tilde{\psi}_{L} \stackrel{p\rightarrow
0}{\longrightarrow} \psi_{L}/f^0_{L}(L_{0})$, where $f^0_{L}(z)$ is
the fermion zero-mode wavefunction, Eq.~(\ref{FermionZeroMode}), and
$\tilde{\psi}_{L}$ has 4D mass dimension, $[\tilde{\psi}_{L}] = 3/2$.

The explicit solutions for the gauge and fermion holographic profiles,
$f_{A}(p,z)$ and $f_{L}(p,z)$, are given in Eqs.~(\ref{solK}) and
(\ref{solfL}) of the appendix, respectively.

\subsection{Relation between the methods}

The three methods we have reviewed in the previous sections are
related to each other.  The 5D propagator can be written in terms of
the KK expansion as
\begin{equation}
P_p(z,z^\prime) = \sum_n \frac{f_n(z) f_n(z^\prime)}{p^2-m_n^2}~.
\label{PropKK}
\end{equation}
Similarly, the boundary kinetic function in the holographic method is
given by the inverse of the (Neumann) boundary to boundary propagator
\begin{equation}
\partial_z f_A(p,L_0) = \frac{1}{P^N_p(L_0,L_0)} = K'_m(p,L_{0})~,
\label{HolProp}
\end{equation}
where in the second equality we used Eq.~(\ref{PpN:general}) for the
case that the EWSB mass $m$ is IR brane localized, together with the
b.c.'s (\ref{bcsK}) and (\ref{bcsS}).  The holographic profile itself,
$f_A(p,z)$, is given by the amputated bulk to boundary propagator,
where amputation means dividing by the boundary to boundary propagator
\begin{equation}
f_A(p,z)= \frac{P^N_p(L_0,z)}{P^N_p(L_0,L_0)} = K_m(p,z)~.
\end{equation}
In particular, both the holographic functions and the 5D propagator
contain information on the whole spectrum, which can be extracted by
evaluating the residue of the corresponding functions on-shell (they
have poles at the corresponding masses of the physical particles).
Eq.~(\ref{PropKK}) shows that the 5D propagators resum
the contribution of the whole tower of KK modes and are therefore
rather efficient when computing indirect effects of the massive
modes. The holographic method also resums the effect of the whole
tower, although it does so in a different basis which can be very
useful in certain situations as we will review below.

\section{Low-energy effective Lagrangian}

\subsection{Integrating out gauge boson KK modes with 5D propagators}
\label{CustodialModel}

We start our discussion of electroweak tests of models with warped
extra dimensions by introducing the prototype of realistic
model~\cite{Agashe:2003zs,Agashe:2006at} and computing its low energy
effective Lagrangian.  In the following sections we show how to use
different techniques to obtain the same effective Lagrangian and study
the constraints on the model from electroweak precision data.  The
model has a bulk $SU(2)_L \times SU(2)_R \times U(1)_X$ gauge
symmetry, broken by boundary conditions to the SM gauge symmetry
$SU(2)_L \times U(1)_Y$ on the UV brane.  Separating the gauge fields
into zero modes and massive modes, the full covariant derivative can
be written as
\begin{equation}
D_\mu^\mathrm{full}=D_\mu-\mathrm{i} 
\Big[ g_{5\,L} 
\tilde{L}^a_{\mu} T^a_L +g_{5\,R} \tilde{R}^b_{\mu}
  T^b_R
+g_5^\prime Y \tilde{B}_\mu + g_{5\,Z^\prime} Q_{Z^\prime}
\tilde{Z}^\prime_\mu \Big]~,
\label{Dmu}
\end{equation}
where $D_\mu$ represents the SM covariant derivative (in 4D momentum
space) and we use tildes to denote the massive KK components of the 5D
fields.  Here $a=1,2,3$, $b=1,2$, $L_\mu^a$ and $R^a_\mu$ are the
gauge fields corresponding to $SU(2)_L$ and $SU(2)_R$, respectively,
and we have defined the hypercharge and $Z^\prime$ gauge bosons as
\begin{equation}
B_\mu=\frac{g_{5\,X} R^3_\mu
+g_{5\,R} X_\mu}{\sqrt{g^2_{5\,R}+g^2_{5\,X}}}~,
\qquad
Z^\prime_\mu=\frac{g_{5\,R} R^3_\mu
-g_{5\,X} X_\mu}{\sqrt{g^2_{5\,R}+g^2_{5\,X}}}~.
\label{BZp}
\end{equation}
We have denoted with $g_{5\,L}$, $g_{5\,R}$ and $g_{5\,X}$ the
five-dimensional gauge couplings of the three bulk gauge groups.  The
corresponding couplings of the hypercharge and $Z^\prime$ read
\begin{equation}
g_5^\prime=\frac{g_{5\,R}\,g_{5\,X}}{\sqrt{g^2_{5\,R}+g^2_{5\,X}}}~,
\quad \quad \quad
g_{5\,Z^\prime}=\sqrt{g^2_{5\,R}+g^2_{5\,X}}~,
\label{g5pg5Zp}
\end{equation} 
with charges
\begin{equation}
Y=T^3_R+Q_X~, \quad \quad \quad Q_{Z'}=\frac{g_{5\,R}^2 
T^3_R - g_{5\,X}^2 Q_X}
{g_{5\,R}^2+g_{5\,X}^2}~,
\label{YQZ}
\end{equation} 
so that the electric charge is $Q =T^3_L+ T^{3}_{R} + Q_X$.

Our goal in this section is to compute the low energy effective
Lagrangian for this model.  We will do so at tree level, assuming
there is a light Higgs in the spectrum.  The effective Lagrangian can
be obtained at tree level by solving the equations of motion of the
heavy particles and replacing the solutions back in the Lagrangian.
The 5D Lagrangian, omitting tensor indices, can be written as
\beqa
\mathcal{L}_{5} &=& \mathcal{L}_{SM} 
- \frac{1}{2} \tilde{L}^a O \tilde{L}^a 
- \frac{1}{2} \tilde{R}^b O \tilde{R}^b 
- \frac{1}{2} \tilde{B} O \tilde{B} 
- \frac{1}{2} \tilde{Z}^\prime O \tilde{Z}^\prime 
\nonumber \\
&& \hspace{8.5mm}
\mbox{} + g_{5\,L} 
\tilde{J}^{aL}
\tilde{L}^a 
+ g_5^\prime 
\tilde{J}^Y
\tilde{B}
+ g_{5R} 
\tilde{J}^{bR}
\tilde{R}^b 
+ g_{5Z^\prime} 
\tilde{J}^{Z^\prime}
\tilde{Z}^\prime+ \ldots~, 
\eeqa
where $O = O(p,z)$ was defined in Eq.~(\ref{GaugeOp}), and the dots
denote interaction terms with more than one heavy field which, in the
case of gauge bosons, do not give contributions to the effective
Lagrangian at leading order.  We have defined the effective currents
\beqa
\fl 
\tilde{J}^{aL}_{\mu} =
\big[a^3 (f_h^{0})^2 J^{aL}_{h\,\mu} + a^4 \sum_{\psi} (f_L^{0})^2
  J^{aL}_{\psi\,\mu} \big]~, 
\quad \tilde{J}^{bR}_{\mu} =
\big[a^3 (f_h^{0})^2 J^{bR}_{h\,\mu} + a^4 \sum_{\psi} (f_L^{0})^2
J^{bR}_{\psi\,\mu} \big]~, 
\label{HiggsCurrent_ab}
\\
\fl
\tilde{J}^Y_{\mu} = 
\big[a^3 (f_h^{0})^2 J^{Y}_{h\,\mu} + a^4 \sum_{\psi} (f_L^{0})^2
  J^{Y}_{\psi\,\mu} \big]~, 
\quad
\tilde{J}^{Z^\prime}_{\mu} = 
\big[a^3 (f_h^{0})^2 J^{Z^\prime}_{h\,\mu} + a^4 \sum_{\psi} (f_L^{0})^2
J^{Z^\prime}_{\psi\,\mu} \big]~, 
\label{HiggsCurrent_YZp}
\eeqa
where $f^{0}_{L}$ is the wave function of the fermion zero modes as
defined in Eq.~(\ref{FermionZeroMode}), the Higgs wavefunction
$f_h^{0}$ is normalized as in Eq.~(\ref{GaugeNormalization}) with
$f_{n}(z) \rightarrow L^{1/2} a(z) f^0_h(z)$~\footnote{\label{Higgs}For
instance, in models of Gauge-Higgs unification one has $f_h^{0}(z) =
\sqrt{2/[L_{0}(L^2_{1}/L^2_{0} - 1)]} \, a(z)^{-2}$.  For an IR brane
localized Higgs: $a^3(z)[f_h(z)]^2=\delta(z-L_1)$.}, and the fermionic
currents are
\begin{equation}
\fl
J^{aL,R\,\mu}_{\psi} \equiv  
\bar{\psi} \gamma^\mu T^a_{L,R} \psi~,
\quad \quad
J^{Y\,\mu}_{\psi} \equiv 
\bar{\psi} \gamma^\mu Y \psi~,
\quad \quad
J^{Z^\prime\,\mu}_{\psi} \equiv 
\bar{\psi} \gamma^\mu Q_{Z^\prime} \psi~,
\label{J:psi}
\end{equation}
while the Higgs currents are
\begin{equation}
\fl
J^{aL,R}_{h\,\mu} \equiv  
h^\dagger T^a_{L,R} i D_\mu h + \mathrm{h.c.}~,
\quad
J^{Y}_{h\,\mu} \equiv  
h^\dagger Y i D_\mu h + \mathrm{h.c.}~,
\quad
J^{Z^\prime}_{h\,\mu} \equiv  
h^\dagger Q_{Z^\prime} i D_\mu h + \mathrm{h.c.}
\label{J:h}
\end{equation}

We can now integrate out the heavy fields by replacing back in the
Lagrangian the solution of the classical equations of motion, which
can be written in terms of the Green's function of the corresponding
differential operator [see
Eqs.~(\ref{PpN:general})-(\ref{ZeroModePropagator})] as
\begin{equation}
\fl
\tilde{L}^a(z) =g_{5\,L} \int_{L_0}^{L_1} \! dz^\prime \, \tilde{P}_p^N(z,z^\prime) 
\tilde{J}^{aL}~, \qquad \quad
\tilde{B}(z) = g_5^\prime \int_{L_0}^{L_1} \! dz^\prime \, \tilde{P}_p^N(z,z^\prime) 
\tilde{J}^Y~, 
\end{equation}
for the fields that are unbroken on the UV brane, and
\begin{equation}
\fl
\tilde{R}^b(z) =g_{5\,R} \int_{L_0}^{L_1} \! dz^\prime \, \tilde{P}_p^D(z,z^\prime) 
\tilde{J}^{bR}~, \qquad \quad
\tilde{Z}^\prime(z) 
= g_{5\,Z^\prime} \int_{L_0}^{L_1} \! dz^\prime \, \tilde{P}_p^D(z,z^\prime) 
\tilde{J}^{Z^\prime}~,
\end{equation}
for those that vanish on the UV brane.  The Neumann (with the zero
mode subtracted) or Dirichlet propagator takes care of the
corresponding boundary conditions on the UV brane.  Also, the
propagators are computed in the EW preserving vacuum [e.g., for an IR
localized Higgs, $m = 0$ in Eq.~(\ref{AdSPropagator})].  The resulting
effective Lagrangian can be put in the standard basis
of~\cite{Buchmuller:1985jz} by using the equations of motion of the SM
fields.  It reads
\begin{equation}
\fl
\mathcal{L}_{\mathrm{eff}}=\mathcal{L}_{\mathrm{SM}}
+ \alpha_h \mathcal{O}_h 
+ \sum_{\psi_L} \alpha^t_{h\psi_L} \mathcal{O}^t_{h\psi_L}
+ \sum_{\psi} \alpha^s_{h\psi} \mathcal{O}^s_{h\psi}
+ \sum_{\psi_L,\psi_L^\prime} \alpha^t_{\psi_L\psi_L^\prime} 
\mathcal{O}^t_{\psi_L \psi_L^\prime}
+ \sum_{\psi,\psi^\prime} \alpha^s_{\psi\psi^\prime} 
\mathcal{O}^s_{\psi \psi^\prime},
\label{Leff:mostgeneral}
\end{equation}
where $\psi_L$ stands for any of the SM LH doublets, $\psi$ for any of
the SM fermion fields, the gauge fields in $\mathcal{L}_{\mathrm{SM}}$
are assumed to be canonically normalized, and the different operators
are defined as follows ($\mathcal{O}_{WB}$ is not induced at tree
level in this basis):
\begin{itemize}
\item Oblique operators
\begin{equation}
\mathcal{O}_h=|h^\dagger D_\mu h|^2~.
\qquad \hspace{2.5mm}
\mathcal{O}_{WB} = (h^\dagger \sigma^a h) W^a_{\mu\nu} B^{\mu\nu}~.
\label{TOp}
\end{equation}
\item Two-fermion operators
\begin{equation}
\fl
\mathcal{O}_{h\psi}^s = 
\mathrm{i} (h^\dagger D_\mu h)
(\bar{\psi}\gamma^\mu \psi) + \mathrm{h.c.}~, 
\qquad
\mathcal{O}_{h\psi_L}^t = 
\mathrm{i} (h^\dagger \sigma^a D_\mu h)
(\bar{\psi}_L\gamma^\mu \sigma^a \psi_L) + \mathrm{h.c.}
\label{hpsiOps}
\end{equation}
\item Four-fermion operators
\begin{equation}
\fl 
\mathcal{O}_{\psi \psi^\prime}^s =
\frac{1}{1+\delta_{\psi \psi^\prime}}
(\bar{\psi}\gamma^\mu \psi)(\bar{
\psi}^\prime \gamma_\mu \psi^\prime)~,
\quad
\mathcal{O}_{\psi_L \psi^\prime_L}^t =
\frac{1}{1+\delta_{\psi_L \psi^\prime_L}}
(\bar{\psi}_L \gamma^\mu \sigma^a \psi_L)
(\bar{\psi}_L^\prime \gamma_\mu \sigma^a \psi_L^\prime)~. 
\label{fourFOps}
\end{equation}
\end{itemize}
The coefficients $\alpha_i$, which encode the dependence on the
model parameters, read
\begin{eqnarray}
\alpha_h&=&\frac{\bar{g}^{\prime\, 2}}{2} [\alpha^{N} -\alpha^{D}]~, 
\label{alpha_h}
\\ 
\alpha^t_{h\psi_L}&=&
\frac{\bar{g}_L^2}{4} \, \beta_{\psi_L}^{N},
\label{alphat_hl}
\\
\alpha^s_{h \psi}&=& 
\frac{\bar{g}^{\prime \,2}}{2} Y_{\psi} \, \beta^{N}_\psi
+\frac{\bar{g}_R^2 T^3_R(\psi) - \bar{g}^{\prime\,2}Y_{\psi}}{2}
\, \beta^{D}_{\psi}~,
\label{alphas_hl}
\\
\alpha^t_{\psi_L \psi^\prime_L}&=&\frac{\bar{g}_L^2}{4} 
\, \gamma_{\psi_L \psi^\prime_L}^{N}~,
\label{alphat_ll}
\\
\alpha^s_{\psi \psi^\prime}&=& \bar{g}^{\prime\,2} Y_\psi Y_{\psi^\prime} 
\gamma_{\psi \psi^\prime}^{N}
+
\frac{
[\bar{g}_R^2T^3_R(\psi) -\bar{g}^{\prime\,2} Y_\psi]
[\bar{g}_R^2T^3_R(\psi^\prime) -\bar{g}^{\prime\,2} Y_{\psi^\prime}]
}{\bar{g}_R^2-\bar{g}^{\prime\,2}}
\, \gamma_{\psi \psi^\prime}^{D}~,
\label{alphas_ll}
\end{eqnarray}
where $\bar{g}^2_{L,R} = g^2_{5\,L,R}/L$ and $\bar{g}^{\prime\,2} =
g^{\prime\,2}_{5}/L$, with $L$ the volume factor (\ref{volume}), while
$Y_\psi$ and $T^3_R(\psi)$ are the hypercharge and third component of
$SU(2)_R$ isospin for the field $\psi$, respectively (for the SM
fermions, $Y_{q} = 1/6$, $Y_{u} = 2/3$, $Y_{d} = -1/3$, $Y_{l} =
-1/2$, $Y_{e} = -1$).  The parameters $\alpha^{N,D}, \beta^{N,D}_\psi,
\gamma^{N,D}_{\psi\psi^\prime}$ are defined as
\begin{eqnarray}
\alpha^{N,D} &=& 
L
\int_{L_0}^{L_1} dz\, dz^\prime 
a^3(z) [f^0_h(z)]^2
\tilde{P}_p^{N,D}(z,z^\prime)
a^3(z) [f^0_h(z)]^2~, 
\label{alpha}
\\
\beta_\psi^{N,D} &=& 
L
\int_{L_0}^{L_1} dz\, dz^\prime 
a^4(z) [f^0_\psi(z)]^2
\tilde{P}_p^{N,D}(z,z^\prime)
a^3(z) [f^0_h(z)]^2~, 
\label{beta}
\\
\gamma_{\psi\psi^\prime}^{N,D}
 &=& 
L
\int_{L_0}^{L_1} dz\, dz^\prime 
a^4(z) [f^0_\psi(z)]^2
\tilde{P}_p^{N,D}(z,z^\prime)
a^4(z) [f^0_{\psi^\prime}(z)]^2~.
\label{gamma}
\end{eqnarray}
Since we are interested in the contribution of dimension-6 operators,
we can evaluate the 5D propagators at zero momentum,
\beqa
\tilde{P}_0^N(z,z^\prime) &=&
\frac{
z^2_{<} \left(1+2\log\frac{L_0}{z_{<}}\right)
+z^{2}_{>} \left(1+2\log\frac{L_1}{z_{>}}\right)
-\frac{L_1^2}{\log\frac{L_1}{L_0}}
}{4L_0\log\frac{L_1}{L_0}}~,
\\
\tilde{P}^D_0(z,z^\prime) &=& \frac{L_0}{2}\left(1-\frac{z_<^2}{L_0^2}\right).
\label{PD0}
\eeqa
Note that, although we have put a tilde on the Dirichlet propagator,
there is no zero mode subtraction in that case.

The above results provide the dimension-6 effective Lagrangian for
general models in warped extra dimensions with custodial symmetry and
a light Higgs, after integration of the gauge boson heavy
modes.~\footnote{The integration of fermion heavy modes can be
performed trivially from the general results
in~\cite{delAguila:2000aa}, see for instance~\cite{delAguila:2000kb}.}
Simple limits can be easily obtained from this general Lagrangian.
For instance, if we consider that the bulk symmetry is just the SM
one, we obtain, by setting to zero the Dirichlet coefficients, the
following effective Lagrangian
\begin{eqnarray}
\fl \hspace{10mm}
\mathcal{L}_{\mathrm{eff}} &=&
\mathcal{L}_{SM} +
\frac{\bar{g}^2}{2} \left [
\alpha^N J^{aL\,\mu}_h J^{aL}_{h\,\mu} 
+ 2 \sum_{\psi} \beta_{\psi}^N  J^{aL\,\mu}_h J^{aL}_{\psi\,\mu}
+ \sum_{\psi,\psi'} \gamma_{\psi\psi'}^N  J^{aL\,\mu}_\psi J^{aL}_{\psi'\,\mu}
\right] 
\nonumber  \\
\fl \hspace{10mm}
&& \mbox{ }\phantom{mm} +
\frac{\bar{g}^{\prime\,2}}{2} \left [
\alpha^N J^{Y\,\mu}_h J^{Y}_{h\,\mu} 
+ 2 \sum_{\psi} \beta_{\psi}^N  J^{Y\,\mu}_h J^{Y}_{\psi\,\mu}
+ \sum_{\psi,\psi'} \gamma_{\psi\psi'}^N  J^{Y\,\mu}_\psi J^{Y}_{\psi'\,\mu}
\right]+ \ldots~. 
\label{Leff:general}
\end{eqnarray}
Note that the Dirichlet terms involving fermions vanish in the limit
of UV localized fermions.  Thus, the above Lagrangian, with the
replacement $\alpha^N \to \alpha^N-\alpha^D$, is also the effective
Lagrangian of models with custodial symmetry and UV localized
fermions. 

In the case that the Higgs field is localized on the IR brane,
$a^3(z)[f_h(z)]^2=\delta(z-L_1)$, these coefficients are explicitly
given by
\begin{eqnarray}
\alpha^N &=&
\frac{L_1^2}{4}
\left[-2 \log\frac{L_1}{L_0}+2-\frac{1}{\log\frac{L_1}{L_0}}\right]~,
\label{alpha:localized}
\\ 
\beta_\psi^N  &=&
\frac{L_1^2}{4} \left[1-\frac{1}{\log\frac{L_1}{L_0}}
+g_2(c_\psi) \left(1-2\log\frac{L_1}{L_0}\right)-2 \tilde{g}_2(c_\psi)
\right]~,
\label{beta:localized}
\end{eqnarray}
where $c_\psi$ is the bulk mass parameter for the fermion $\psi$ and
the auxiliary functions $g_n(c)$ and $\tilde{g}_n(c)$ are defined in
Eqs.~(\ref{gn}) and (\ref{gtn}) of the appendix.  Finally $\gamma^N$
is a complicated function of $c_\psi$ and $c_{\psi^\prime}$.  In the
limit $c \to \infty$ (UV localized fermions) it simplifies to
\begin{equation}
\gamma^N(c \to \infty) =
-\frac{L_1^2}{4 \log \frac{L_1}{L_0}}~.
\label{gamma:localized:UV}
\end{equation}
Also, for future reference, when the Higgs is localized on the IR
brane
\begin{eqnarray}
\alpha^D =
-\frac{L_1^2}{2} \log\frac{L_1}{L_0}~,
\qquad
\beta_\psi^D  =
-\frac{L_1^2}{2} \log\frac{L_1}{L_0} \, g_2(c_\psi)~.
\label{Dirichlet:localized}
\end{eqnarray}
%

\subsubsection{Universal case \label{sec:universal}} 

An assumption that significantly simplifies the analysis of EWPT is
that of universal new physics~\cite{Barbieri:2004qk}.  Models of
universal new physics are those for which a combination of gauge
bosons exists (so called interpolating fields) $\bar{W}^a$, with
$a=1,2,3$, and $\bar{B}$ such that the only couplings of fermions
(excluding Yukawa couplings) are of the form
\begin{equation}
\mathcal{L}_{\mathrm{fermions}}^{\mathrm{Univ.}}
= \bar{W}^a_\mu J^{aL\,\mu}_f + \bar{B}_\mu J^{Y\,\mu}_f + \cdots~,
\label{UniversalFermionGauge}
\end{equation}
where the dots denote kinetic and Yukawa terms for the fermions.  
In
particular, the interpolating fields $\bar{W}^a$ and $\bar{B}$ are
not canonically normalized.  
The SM fermionic currents are
defined by [see Eq.~(\ref{J:psi})]
\begin{equation}
J^{aL\,\mu}_{f} \equiv  
\sum_{\psi} J^{aL\,\mu}_{\psi}~,
\quad \quad
J^{Y\,\mu}_{f} \equiv 
\sum_{\psi} J^{Y\,\mu}_{\psi}~.
\label{J:f}
\end{equation}
In this case, all relevant contributions to EWPT can be encoded in the
vacuum polarizations of the interpolating fields (oblique
corrections).  Assuming unbroken QED, the quadratic Lagrangian for the
interpolating fields, together with the gauge-fermion interacions, can
be written as
\begin{eqnarray}
\mathcal{L}_{\mathrm{Oblique}}&=&
-\mathcal{P}^{\mu\nu}\left[
\bar{W}^+_\mu \bar{\Pi}_{+-}(p^2) \bar{W}^-_\nu 
+\frac{1}{2} \bar{W}^3_\mu \, \bar{\Pi}_{33}(p^2) \bar{W}^3_\nu 
\right. \nonumber \\
&&\phantom{-\mathcal{P}^{\mu\nu}}\left.
\mbox{}+\frac{1}{2} \bar{B}_\mu \bar{\Pi}_{BB}(p^2) \bar{B}_{\nu}
+ \bar{W}^3_\mu \, \bar{\Pi}_{3B}(p^2) B_\nu \right] 
+\mathcal{L}_{\mathrm{fermions}}^{\mathrm{Univ.}}~,
\label{L:Oblique}
\end{eqnarray}
where $\mathcal{P}^{\mu\nu}$ is the transverse projector and we have
defined $\bar{W}^{\pm} \equiv (\bar{W}^1 \mp i \bar{W}^2)/\sqrt{2}$.

In the case that the coefficients $\beta^{N}_{\psi} \equiv \beta$ and
$\gamma^{N}_{\psi\psi'} \equiv \gamma$ in Eq.~(\ref{Leff:general}) are
independent of the fermion type, $\psi$, the effective Lagrangian
takes the form
\begin{eqnarray}
\mathcal{L}_{\mathrm{eff}} =
\mathcal{L}_{SM} 
&+&
\frac{\bar{g}^2}{2} \left [
\alpha J^{aL\,\mu}_h J^{aL}_{h\,\mu} 
+ 2 \beta  J^{aL\,\mu}_h J^{aL}_{f\,\mu}
+ \gamma  J^{aL\,\mu}_f J^{aL}_{f\,\mu}
\right] 
\nonumber  \\
&+&
\frac{\bar{g}^{\prime\,2}}{2} \left [
\alpha J^{Y\,\mu}_h J^{Y}_{h\,\mu} 
+ 2 \beta  J^{Y\,\mu}_h J^{Y}_{f\,\mu}
+ \gamma   J^{Y\,\mu}_f J^{Y}_{f\,\mu}
\right]+ \ldots~. 
\label{Leff:universal}
\end{eqnarray}
This effective Lagrangian is not in oblique form, since it includes
corrections to the fermion gauge couplings and four-fermion
interactions, proportional to $\beta$ and $\gamma$, respectively.  The
corrections are however universal and can be written purely in terms
of oblique corrections.  This can be done in two equivalent ways,
either by using the classical equations of motion of the SM fields as
determined by $\mathcal{L}_{SM}$ or by doing field redefinitions that
eliminate the corrections involving fermions.  The latter approach
consists of performing a shift of the gauge fields, proportional to
the fermion currents that eliminates the four-fermion terms, followed
by a gauge field rescaling that puts the fermion-gauge interactions in
the form of Eq.~(\ref{UniversalFermionGauge}).  Replacing a
Higgs vev $\langle h \rangle = (0,v)$ in the Higgs currents
(\ref{J:h}), the field redefinitions are, to first order in $\alpha$,
$\beta$ and $\gamma$:
\begin{eqnarray}
\fl \hspace{10mm}
W^a &=&
\frac{1}{\bar{g}}\bar{W}^a \left[ 1 - \frac{\bar{g}^2v^2}{2} \beta  
- \frac{\gamma}{2}
  \Pi_{aa}^{\rm SM} \right] - \frac{\bar{g}}{2} \gamma J_f^{aL}
+\delta^{a3} \frac{1}{\bar{g}^\prime}\bar{B} \left[
\frac{\bar{g}\bar{g}^\prime v^2}{2}\beta -
\frac{\gamma}{2}\Pi^{\rm SM}_{3B} \right]~, 
\label{DefW}
\\
\fl \hspace{10mm}
B &=&
\frac{1}{\bar{g}^\prime}\bar{B} 
\left[ 1 - \frac{\bar{g}^{\prime\,2}v^2}{2} \beta 
- \frac{\gamma}{2}
  \Pi_{BB}^{\rm SM} \right] - \frac{\bar{g}^\prime}{2} \gamma J_f^{Y}
+\frac{1}{\bar{g}}\bar{W}^3 \left[
\frac{\bar{g}\bar{g}^\prime v^2}{2}\beta 
- \frac{\gamma}{2}\Pi^{\rm SM}_{3B} \right]~,
\label{DefB}
\end{eqnarray}
where the vacuum polarization functions in the SM limit are defined by
\begin{equation}
\Pi^{\rm SM}_{aa} = p^2 - \frac{\bar{g}^2 v^2}{2}~, \qquad
\Pi^{\rm SM}_{BB} = p^2 - \frac{\bar{g}^{\prime\,2} v^2}{2}~, \qquad
\Pi^{\rm SM}_{3B} = \frac{\bar{g} \bar{g}^\prime  v^2}{2}~.
\end{equation}
These field redefinitions lead to an oblique form for the effective
Lagrangian written in terms of the interpolating fields $\bar{W}^a,
\bar{B}$, as in Eq.~(\ref{L:Oblique}), with the following
vacuum polarization functions
\begin{eqnarray}
\fl
\bar{g}^2\bar{\Pi}_{aa}
&=&
\Pi_{aa}^{SM}(1-\bar{g}^2v^2 \beta - \gamma \Pi_{aa}^{SM})
-\frac{\bar{g}^2v^4}{4} 
\left( \bar{g}^2 + \delta_{a3} \bar{g}^{\prime\,2} \right) \alpha
+ \delta_{a3} \Pi^{SM}_{3B} (\bar{g}\bar{g}^\prime v^2
\beta - \gamma
\Pi^{SM}_{3B}) 
\nonumber \\
\fl
&=&
-\frac{\bar{g}^2v^2}{2} \left[
1 - \left( \frac{\bar{g}^2}{\bar{g}^{\prime 2}} + \delta_{a3} \right)\hat{T}^{\rm Op} \right] 
+ \left(1 - \hat{S}^{\rm Op} \right) p^2 + \frac{2}{\bar{g}^2v^2} W^{\rm Op} p^4 +
\ldots~, 
\label{Pi:bar:aa}
\\
\fl
\bar{g}^{\prime\,2}\bar{\Pi}_{BB}
&=&
\Pi_{BB}^{SM}(1-\bar{g}^{\prime\,2}v^2 \beta 
- \gamma \Pi_{BB}^{SM})
-\frac{\bar{g}^{\prime\,2}v^4}{4} 
\left( \bar{g}^2 + \bar{g}^{\prime\,2} \right) \alpha
+\Pi^{SM}_{3B} (\bar{g}\bar{g}^\prime v^2 \beta - \gamma
\Pi^{SM}_{3B})  
\nonumber \\
\fl
&=&
-\frac{\bar{g}^{\prime\,2}v^2}{2}\left[
1 - \left( \frac{\bar{g}^2}{\bar{g}^{\prime 2}} + 1 \right)\hat{T}^{\rm Op} \right] 
+ \left(1 - \frac{\bar{g}^{\prime 2}}{\bar{g}^2} \hat{S}^{\rm Op} \right) p^2 + 
\frac{2}{\bar{g}^2v^2} W^{\rm Op} p^4 + \ldots~, 
\label{Pi:bar:BB}
\\
\fl
\bar{g}\bar{g}^\prime \bar{\Pi}_{3B}&=&
\Pi_{3B}^{SM}
\left[1-
\frac{v^2}{2} \left( \bar{g}^{2} + \bar{g}^{\prime\,2} \right) \beta
- \gamma \Pi_{33BB} \right]
+\frac{\bar{g}\bar{g}^\prime v^2}{2} \left[
\frac{v^2}{2} \left( \bar{g}^2 + \bar{g}^{\prime\,2}\right)\alpha
+ \beta \, \Pi_{33BB}
\right]
\nonumber \\
\fl
&=&
\frac{\bar{g}\bar{g}^\prime v^2}{2}\left[
1 - \left( \frac{\bar{g}^2}{\bar{g}^{\prime 2}} + 1 \right)\hat{T}^{\rm Op} \right] 
+ \frac{\bar{g}^\prime}{\bar{g}} \hat{S}^{\rm Op} \, p^2 
+\ldots~, 
\label{Pi:bar:3B}
\end{eqnarray}
where $\Pi_{33BB} \equiv \Pi^{SM}_{33} + \Pi^{SM}_{BB}$, and in the
second equality of each definition we expanded for small $p^2$
and wrote the result in terms of
\begin{eqnarray}
\fl
\hat{T}^{\rm Op} &\equiv 
\frac{\bar{g}^{\prime\,2}v^2}{2} (-\alpha+2\beta -
\gamma)~,
\qquad 
\hat{S}^{\rm Op} & \equiv
\bar{g}^{2} v^2 (\beta - \gamma)~,
\qquad
W^{\rm Op} \equiv  
- \frac{\bar{g}^2v^2}{2} \gamma~,
\label{That:Y:universal}
\end{eqnarray}
which will be identified in Section~\ref{sec:EWPT} as the oblique
parameters of~\cite{Barbieri:2004qk}.  Recall that these same vacuum
polarization functions represent also the effective Lagrangian of a
custodially symmetric model with fermions localized on the UV brane if
we make the replacement $\alpha \to \alpha^N-\alpha^D$. 

\subsection{Holographic method 
\label{sec:holography}}

In this section we rederive the effective Lagrangian of the model with
custodial symmetry using the holographic method (for further
discussions of holography, see the review
in~\cite{Gherghetta:2006ha}).  We consider first the case of UV
localized fermions, which is the situation in which the holographic
method is the most efficient, as it produces with minimal effort the
effective Lagrangian in oblique form.  In
Subsection~\ref{sec:holography:bulk:fermions} we discuss how the
formalism changes when fermions are allowed to propagate in the bulk.

In the presence of a bulk EWSB vev, $\textrm{v}(z)$, the gauge part of
the action reads
\begin{eqnarray}
\fl
S^{\rm Cust.}_{\mathrm{gauge}}&=&\int \! \frac{d^4p}{(2\pi)^4} dz \sqrt{g} \left\{
-\frac{1}{4g_{5L}^2} (L^a_{MN})^2
-\frac{1}{4g_{5R}^2} (R^a_{MN})^2
-\frac{1}{4g_{5X}^{2}} (X_{MN})^2
\right.
\nonumber \\
\fl
&& \hspace{2.7cm} \left. \mbox{}
+ \frac{1}{2} \, \textrm{v}(z)^2 
(L^a_M-R^a_M)^2
\right\} 
\nonumber \\
\fl
&=&\int \! \frac{d^4p}{(2\pi)^4} dz \sqrt{g} \left\{
-\frac{1}{4g_{5Z}^2} (V^a_{MN})^2
-\frac{1}{4g_{5Z}^2} (A^a_{MN})^2
-\frac{1}{4g_{5X}^{2}} (X_{MN})^2
+\frac{1}{2} \textrm{v}(z)^2 A_M^2
\right\}
\nonumber \\
\fl
& \stackrel{\rm Hol}{=}&
-\frac{1}{2}\int \! \frac{d^4p}{(2\pi)^4} dz \sqrt{g} \left\{
 \bar{A}^a \Pi_A(p^2) \bar{A}^a
+ \bar{V}^a \Pi_V(p^2) \bar{V}^a
+ \bar{X} \Pi_X(p^2) \bar{X}
\right\}~,
\end{eqnarray} 
where $g_{5Z}^2 \equiv g_{5L}^2 + g_{5R}^2$, and in the second
equality we defined the axial and vector combinations
\begin{equation}
A_M^a=L_M^a-R_M^a~, \qquad \qquad
V_M^a=\frac{g_{5R}}{g_{5L}}L_M^a + \frac{g_{5L}}{g_{5R}} R_M^a~.
\end{equation}
In the last line we used the holographic method (indicated by the
$\stackrel{\rm Hol}{=}$ symbol) to obtain the effective action for the
fields evaluated on the UV brane:
\begin{equation}
\bar{A}^a=A^a(z=L_0)~,\qquad
\bar{V}^a=V^a(z=L_0)~,\qquad
\bar{X}=X(z=L_0)~.
\end{equation}
The corresponding vacuum polarizations read
\begin{equation}
\Pi_{A,V} = \frac{1}{g_{5Z}^2} \partial_z f_{A,V}(p^2,z=L_0)~,
\qquad
\Pi_X = \frac{1}{g_{5X}^2} \partial_z f_X(p^2,z=L_0)~,
\end{equation}
where $f_{A}$, $f_{V}$ and $f_{X}$ satisfy Eq.~(\ref{OpMA}) with mass
terms $M^2_{A} = g^2_{5Z} \textrm{v}^2$, $M^2_{V} = 0$ and $M^2_{X} =
0$, respectively.  By assumption, the fermionic action is already
localized on the UV brane.  In particular, the coupling between the
fermions and gauge bosons is universal, as in Eq.~
(\ref{UniversalFermionGauge}), with interpolating fields that are the
5D SM gauge boson fields evaluated at the UV brane.  Replacing the
boundary values of the axial, vector and $X$ fields in terms of the SM
ones~\footnote{Note the difference with respect to Eqs.~(\ref{Dmu})
and (\ref{BZp}) due to the non-canonical normalization of the fields
in this section.}
\begin{equation}
\bar{A}^a=\bar{W}_L^a -\delta^{a3}\bar{B}~, \qquad
\bar{V}^a=\frac{g_R}{g_L} \bar{W}_L^a
+ \delta^{a3}\frac{g_L}{g_R} \bar{B}~, \qquad
\bar{X}=\bar{B}~,
\end{equation}
we obtain an effective Lagrangian in the oblique form~(\ref{L:Oblique}),
with vacuum polarizations
\begin{eqnarray}
\bar{\Pi}_{aa}&=& \Pi_A + \frac{g_{5R}^2}{g_{5L}^2} \Pi_V~,
\label{Piaa_Hol}
\\
\bar{\Pi}_{BB}&=& \Pi_A + \frac{g_{5L}^2}{g_{5R}^2} \Pi_V
+\Pi_X~,
\label{PiBB_Hol}
\\
\bar{\Pi}_{3B}&=& -\Pi_A + \Pi_V~.
\label{Pi3B_Hol}
\end{eqnarray}
Assuming that $\textrm{v}(z)^2 = (v^2/2) [f^0_h(z)]^2$, where the
Higgs profile $f^0_h(z)$ is normalized as specified after
Eq.~(\ref{HiggsCurrent_YZp}), one can check that these vacuum
polarization functions agree with the ones obtained using 5D
propagators, Eqs.~(\ref{Pi:bar:aa})-(\ref{Pi:bar:3B}), up to a
redefinition of the vev, $v^2 \rightarrow v^2[1 - \frac{1}{4}
(g^2_{5L} + g^2_{5R}) (v/a(L_{1}))^2 L_{0}]$, when expanded to the
same order in $v^2$ and with the replacement $\alpha \to \alpha^N
-\alpha^D$ to take into account the modifications due to the custodial
symmetry.  Note also that in the holographic formalism the custodial
symmetry is explicit from Eq.~(\ref{Piaa_Hol}), i.e. a symmetry
between the charged and neutral $W$'s.

\subsubsection{Holography with bulk fermions} 
\label{sec:holography:bulk:fermions}

\begin{figure}[ht]
\begin{center}
\includegraphics[width=.25\textwidth]{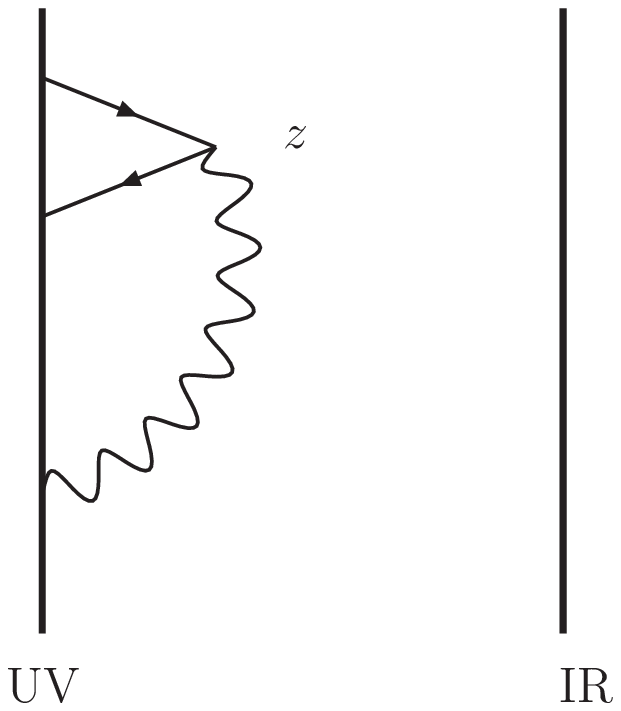}
\hspace{2cm}
\includegraphics[width=.25\textwidth]{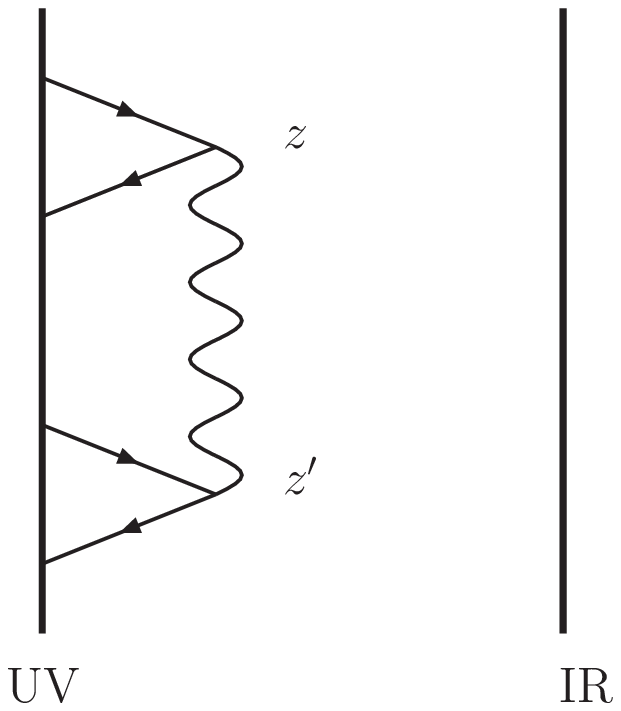}
\caption{ Vertex and four-fermion interaction contributions to the
boundary action.  The four-fermion interaction, with points at $z$ and
$z^\prime$, is to be computed with the Dirichlet propagator,
Eq.~(\ref{PpD:general}).
\label{DiagramsHolography}}
\end{center}
\end{figure}

As we showed in the previous section, the holographic method is
particularly suitable to compute the effective Lagrangian in the case
of UV boundary localized fermions.  The holographic method can still
be used when the fermions propagate in the bulk although obtaining the
effective Lagrangian requires more work.  In this section we discuss
the new steps one has to perform in the simple case of the SM bulk
gauge symmetry, which can be obtained from our custodial model by
setting $R^b_M=0$, $R^3_M=X_M=B_M$, and identifying the $U(1)_{Y}$
gauge coupling, $g_{5}^\prime$, as in Eq.~(\ref{g5pg5Zp}).
The gauge part
of the Lagrangian then reads
\beqa
S_{\mathrm{gauge}} &=& 
\int \! \frac{d^4p}{(2\pi)^4} dz \sqrt{g} \left\{ -\frac{1}{4g^2_{5}} (W^{b}_{MN})^2 -
\frac{1}{4g^{2}_{5Z}} \left[(A_{MN})^2 + (V_{MN})^2\right] \right.
\nonumber \\
&& \hspace{2.6cm}
\left. \mbox{} + \frac{1}{2} \, \textrm{v}(z)^2 \left[ (W^{b}_{M})^2 + (A_{M})^2 \right] \right\} 
\nonumber \\
\fl
&\stackrel{\rm Hol}{=}&
-\frac{1}{2}
\int \! \frac{d^4p}{(2\pi)^4} \left\{ 
\bar{W}^{b}_{\mu} \, \Pi_{W} \bar{W}^{\mu}_{b} 
+\bar{A}_{\mu} \Pi_{A} \bar{A}^{\mu} 
+ \bar{V}_{\mu} \Pi_{V} \bar{V}^{\mu} 
\right\}~,
\label{SMgaugeHol}
\eeqa
where 
\begin{equation}
\Pi_{W,A,V} = \Pi(g_5^2 \textrm{v}^2,g_{5Z}^2 \textrm{v}^2,0)~,
\label{PiWAV}
\end{equation}
and we indicate inside the parenthesis the masses to be used in
Eqs.~(\ref{OpMA}) and (\ref{GeneralPiA}).  Now we have $g^2_{5Z}
\equiv g^2_{5} + g^{\prime 2}_{5}$ and the following relation between
the axial and vector fields, and the SM neutral ones:
\begin{equation}
A_{M} = W^3_{M} - B_{M}~, \qquad  V_{M} =
\frac{g^\prime_{5}}{g_{5}} \, W^3_{M} + \frac{g_{5}}{g^\prime_{5}} \, B_{M}~,
\label{AVtoW3B}
\end{equation}
so that the vacuum polarization functions for the SM boundary fields
read
\begin{eqnarray}
\Pi_{+-}&=& \Pi_W~, \qquad \qquad \qquad \hspace{-1mm}
\Pi_{3B}= -\Pi_A + \Pi_V~, 
\label{Pi:bar:aa:hol:UV}
\\ 
\Pi_{33}&=& \Pi_A + \frac{g_{5}^{\prime\,2}}{g_5^2} \Pi_V~, 
\qquad
\Pi_{BB}= \Pi_A + \frac{g_5^2}{g_{5}^{\prime\,2}}\Pi_V~. 
\label{Pi:bar:3B:hol:UV}
\end{eqnarray}
As in the previous section, these would be the vacuum polarization
functions for the interpolating fields for UV localized fermion
fields.  When the fermion fields live in the bulk, however, these
terms do not give the full contribution to the boundary effective
Lagrangian.  Indeed there are both vertex and four-fermion interaction
corrections as shown in Fig.~\ref{DiagramsHolography}.
The left panel in the figure represents the vertex corrections.  The
contribution to the effective Lagrangian is obtained by matching the
corresponding amputated three point function in the full and effective
theories.  The diagram in the full theory requires bulk to boundary
propagators for the gauge boson and fermions, and amputation means
that the external legs are divided by the corresponding boundary to
boundary propagators [$=g^2_{5}/K^\prime_{m}(p^2,L_{0})$ for the case
of an IR brane localized EWSB mass $m$, with $g_{5}$ and $m$ the
parameters of the corresponding gauge boson; see Eqs.~(\ref{OpMA}),
(\ref{GeneralPiA}) and (\ref{HolProp})].  Assuming fermion
localization universality,\footnote{Eqs.~(\ref{SMgaugeHol}),
(\ref{SintHol}) and (\ref{SM4FermionHol}) below are trivially
generalized to the non-universal case by expressing the results in
terms of the individual fermion currents of Eqs.~(\ref{J:f}), and
using the corresponding fermion profiles in Eqs.~(\ref{bulkcouplings})
and (\ref{gamma}).} the resulting extra contribution to the boundary
action reads
\begin{equation}
S_{\rm Vertex} 
= 
\int\! \frac{d^4p}{(2\pi)^4} \,
\left\{ 
\bar{g}_{W} \bar{W}^{b}_{\mu} \tilde{J}^{b L \, \mu}_{f} 
+ \bar{g}_{A} \bar{A}_{\mu} \tilde{J}^\mu_{Z} 
+ s_{\theta} c_{\theta} \, \bar{g}_{V} \bar{V}_{\mu} \tilde{J}^\mu_{Q} 
\right\}~, 
\label{SintHol}
\end{equation}
where $c_{\theta} = g_{5}/g_{5Z}$, $s_{\theta} = g^\prime_{5}/g_{5Z}$,
$\tilde{J}^\mu_{Z} = c^2_{\theta} \tilde{J}^{3 L \mu}_{f} -
s^2_{\theta} \tilde{J}^{Y \mu}_{f}$ and $\tilde{J}^\mu_{Q} =
\tilde{J}^{3 L \mu}_{f} + \tilde{J}^{Y \mu}_{f}$, with $\tilde{J}^{a L
\mu}_{f}$ and $\tilde{J}^{Y \mu}_{f}$ as defined in Eqs.~(\ref{J:f})
and (\ref{J:psi}), but with $\psi \rightarrow \tilde{\psi}$
representing the canonically normalized boundary fermion fields (see
end of Subsection~\ref{sec:GeneralHolography}).  We also defined
\begin{equation}
\bar{g}_{W,V,A}(p^2)= 
\int^{L_{1}}_{L_{0}} \! dz \, a^4(z) [f^0_{\psi}(z)]^2 \, f_{W,V,A}(p^2,z)~,
\label{bulkcouplings} 
\end{equation}
where $f_{W,V,A}$ are the gauge boson ``holographic'' profiles
associated with Eq.~(\ref{PiWAV}).  Since we are interested in the
couplings of the fermion zero modes, we have evaluated the
corresponding fermion profiles at $p^2_{\psi}=0$
in~(\ref{bulkcouplings}) [neglecting the small fermion masses from
EWSB], so that the zero-mode wavefunctions,
Eq.~(\ref{FermionZeroMode}), appear inside the integral, and there is
dependence on a single momentum scale, associated with the gauge
boson.

Bulk fermions also imply that four-fermion interactions are generated
in the boundary action, due to the diagram in the right panel of
Fig.~\ref{DiagramsHolography}.  Again the coefficient in the effective
theory is obtained by matching the amputated four-point function in
the full and effective theories.  On the full theory side, we use the
Neumann bulk to bulk propagator and integrate over both
interaction points.  On the effective theory side, we have to use the
boundary to boundary propagator using the couplings we have computed
in Eq.~(\ref{bulkcouplings}).  This latter term exactly coincides with
the $K(z)K(z^\prime)/K^\prime(L_0)$ part of full propagator,
Eq.~(\ref{PpN:general}).  The difference of the two therefore
corresponds to computing the diagram on the right panel of
Fig.~\ref{DiagramsHolography} using the Dirichlet bulk to bulk
propagator instead of the full one.  The induced 4-fermion operators
are
\beqa
\fl
S_{\rm 4-fermion} 
= 
\int\! \frac{d^4p}{(2\pi)^4} \,
\left\{ 
\frac{g^2_{5}}{2L} \, \gamma^D_{W} \tilde{J}^{b L}_{f \mu} \tilde{J}^{b L \, \mu}_{f} 
+\frac{g^2_{5Z}}{2L} \, \gamma^D_{A} \tilde{J}_{Z \mu} \tilde{J}^{\mu}_{Z} 
+ s^2_{\theta} c^2_{\theta} \frac{g^2_{5Z}}{2L} \, \gamma^D_{V} \tilde{J}_{Q \mu} \tilde{J}^{\mu}_{Q} 
\right\}~,
\label{SM4FermionHol}
\eeqa
where $g^2_{5Z}$ was defined after Eq.~(\ref{PiWAV}), $L$ is the
volume factor Eq.~(\ref{volume}), and $\gamma^D_{W,A,V}$ were defined
in Eq.~(\ref{gamma}), with the Dirichlet propagator corresponding to
$W$, $A$ and $V$ (i.e. gauge boson squared masses $g^2_{5}
\textrm{v}^2$, $g^2_{5Z} \textrm{v}^2$ and $0$, respectively).  If we
are interested in the contribution to the effective Lagrangian up to
operators of dimension 6, we can evaluate the corresponding propagator
at zero momentum and neglect EWSB effects, i.e. use Eq.~(\ref{PD0}).
In that case, $\gamma^D_{W} = \gamma^D_{A} = \gamma^D_{V} =
\gamma^{\rm Hol.}$, and assuming universality we have
\begin{eqnarray}
\fl
\gamma^{\rm Hol.} 
&=&
\frac{L_1^2 \log \frac{L_1}{L_0}}{4}
\frac{
(3-2c)\left(\frac{L_0}{L_1}\right)^{2}
-(2c-1)^2 \left(\frac{L_0}{L_1}\right)^{4c-2}
+8(c-1)\left(\frac{L_0}{L_1}\right)^{2c+1}}
{(2c^2-5c+3)\left[1-\left(\frac{L_0}{L_1}\right)^{2c-1}\right]^2}~.
\end{eqnarray}

The effective Lagrangian given by Eqs.~(\ref{SMgaugeHol}),
(\ref{SintHol}) and (\ref{SM4FermionHol}) is not in the oblique form
but can be rewritten in such a form due to the assumption of
universality of fermion localization.  The simplest way to proceed is
to first shift the gauge fields as
\beqa
\fl
\bar{W}^b \rightarrow \bar{W}^b - \frac{g^2_{5}}{2L\bar{g}_{W}} \gamma^D_{W} \tilde{J}^{b L}_{f}~,
\quad
\bar{A} \rightarrow \bar{A} - \frac{g^2_{5Z}}{2L\bar{g}_{A}} \gamma^D_{A} \tilde{J}_{Z}~,
\quad
\bar{V} \rightarrow \bar{V} - \frac{s_{\theta} c_{\theta} g^2_{5Z}}{2L\bar{g}_{V}} \gamma^D_{V} \tilde{J}_{Q}~,
\nonumber
\eeqa
so as to eliminate the four-fermion interactions (to first order in
the $\gamma^D_{i}$), followed by a rescaling $\bar{W}^b \rightarrow
\bar{W}^b/\bar{\bar{g}}_{W}$, $\bar{A} \rightarrow
\bar{A}/\bar{\bar{g}}_{A}$ and $\bar{V} \rightarrow
\bar{V}/\bar{\bar{g}}_{V}$, with
\beqa
\bar{\bar{g}}_{i}(p^2) = \bar{g}_{i}(p^2) \left[ 1 + \frac{g^2_{5i}}{2L\bar{g}^2_{i}(p^2)} \gamma_{i} \Pi_{i} (p^2) \right]~,
\qquad \quad
i = W,A,V,
\eeqa
where $g^2_{5i} = g^2_{5}$ for $i=W$, $g^2_{5i} = g^2_{5Z}$ for $i =
A,V$, $\bar{g}_{i}$ are given in Eq.~(\ref{bulkcouplings}), and
$\Pi_{i}$ are given in Eq.~(\ref{PiWAV}), for $i=W,A,V$.  Expressing
the resulting Lagrangian in terms of $W^a$ and $B$, as in
Eq.~(\ref{AVtoW3B}) one finds that the fermion couplings to the new
gauge bosons are universal as in~(\ref{UniversalFermionGauge}), while
the vacuum polarizations are
\begin{eqnarray}
\bar{\Pi}_{+-}
&=& \frac{1}{\bar{\bar{g}}_W^2}
\Pi_W~, 
\qquad \qquad \qquad \hspace{2.5mm}
\bar{\Pi}_{33} = 
\frac{1}{\bar{\bar{g}}_A^2}\Pi_A + 
\frac{1}{\bar{\bar{g}}_V^2}
\frac{g_{5}^{\prime\,2}}{g_5^2} \Pi_V~, 
\label{Pi:bar:aa:hol}\\ 
\bar{\Pi}_{3B} &=& -\frac{1}{\bar{\bar{g}}_A^2}\Pi_A +
\frac{1}{\bar{\bar{g}}_V^2} \Pi_V~, 
\qquad
\bar{\Pi}_{BB} = 
\frac{1}{\bar{\bar{g}}_A^2}\Pi_A +
\frac{1}{\bar{\bar{g}}_V^2} \frac{g_5^2}{g_{5}^{\prime\,2}}\Pi_V~,
\label{Pi:bar:3B:hol}
\end{eqnarray}
with $\Pi_{W,A,V}$ as defined in Eq.~(\ref{PiWAV}).  Recall that all
these quantities depend on $p^2$, including the $\bar{\bar{g}}_{i}$,
even though we do not explicitly indicate so.  Expanding in the EWSB
masses to the corresponding order in $v^2$, one can check that these
vacuum polarizations agree exactly with the ones obtained with the
method of propagators.  Notice, however, that the above formulas hold
for arbitrary $v$.  Also, in the limit of UV localized fermions, $c
\rightarrow +\infty$, one has $\gamma^{\rm Hol.} \rightarrow 0$ and
$\bar{g}_{i} \rightarrow 1$, so that
Eqs.~(\ref{Pi:bar:aa:hol})-(\ref{Pi:bar:3B:hol}) reduce to
Eqs.~(\ref{Pi:bar:aa:hol:UV}) and (\ref{Pi:bar:3B:hol:UV}).

\section{Electroweak Precision Tests}
\label{sec:EWPT}

Very precise data from low energy neutrino and electron scattering,
LEPI and SLC data at the $Z$ pole, LEPII above the $Z$ pole and the
Tevatron impose stringent constraints on any physics beyond the
SM~\cite{Amsler:2008zzb}, commonly called electroweak precision tests
(EWPT).  These constraints can be computed on each model of new
physics by carefully considering the contribution to all the
(pseudo)observables that constitute the EWPT. Under the assumption of
linearly realized EWSB with a light Higgs and a large enough mass gap
with the new physics, one can use the SM effective Lagrangian up to
dimension 6 of reference~\cite{Buchmuller:1985jz} to easily constrain
large classes of new models.  Not all dimension 6 operators are
relevant for EWPT. In general, operators that violate CP or flavour
symmetries (except for the third family), operators that only involve
quarks or gluons and operators that just renormalize the SM operators
(\textit{i.e.} terms of the form $h^\dagger h\,
\mathcal{O}^\mathrm{SM}$, with $\mathcal{O}^\mathrm{SM}$ a dimension 4
operator that is already present in the SM Lagrangian) are irrelevant
for EWPT. The relevant ones were classified in Ref.~\cite{Han:2004az}
(see also~\cite{Cacciapaglia:2006pk}) and their effects on the EWPT
computed (\cite{Han:2004az} gives the $\chi^2$ as a function of the SM
input parameters and the coefficients of the relevant dimension-6
operators).  Using the results in Ref.~\cite{Han:2004az} and our
calculation of the effective Lagrangian for a general model with
custodial symmetry, Eqs.~(\ref{Leff:mostgeneral}) and
(\ref{alpha_h})-(\ref{PD0}), one can obtain the constraints in any
model of warped extra dimensions with a light Higgs.  However, 
the prototype of
realistic warped model, with light fermions localized close to the UV
brane, corresponds to universal new physics (except for the third
generation) and the corresponding computation of EWPT can be more
simply done following~\cite{Barbieri:2004qk}.

\subsection{Tree-level effects}

As we discussed in Section~\ref{sec:universal}, EW precision
constraints can be easily implemented in models of universal new
physics~\cite{Barbieri:2004qk}.  In this case, the relevant Lagrangian
is given by Eqs.~(\ref{UniversalFermionGauge}) and (\ref{L:Oblique}).
If we further assume there is a mass gap with the new physics that
allows us to reliably expand the vacuum polarizations as
\begin{equation}
\bar{\Pi}(p^2)= \bar{\Pi}(0)+p^2 \bar{\Pi}^\prime(0)+
\frac{(p^2)^2}{2} \bar{\Pi}^{\prime\prime}(0) + \cdots~,
\label{Oblique:expansion}
\end{equation}
where the prime denotes derivative with respect to $p^2$ (higher
derivative terms give contributions of mass dimension larger than 6),
we can parametrize all relevant EWPT in terms of four oblique
parameters:~\footnote[1]{The $Z$-pole observables plus the $W$ mass
depend only on the three linear combinations introduced in
\cite{Csaki:2002gy,Carena:2002dz}.  These are $\alpha T_{\rm eff} =
\hat{T} - \left[ W + (s^2_{W}/c^2_{W}) Y \right]$, $\alpha S_{\rm eff}
= 4 s^2_{W} ( \hat{S} - W - Y)$ and $\alpha U_{\rm eff} = 4 s^2_{W}
(\hat{U} - W)$~\cite{Barbieri:2004qk}.  }
\begin{eqnarray}
\hat{T}&= 1-\frac{\bar{\Pi}_{33}(0)}{\bar{\Pi}_{+-}(0)}~, \qquad
&W=\frac{g^2 M_W^2}{2} \bar{\Pi}_{33}^{\prime\prime}(0)~, 
\label{TW} \\
\hat{S}&= g^2 \bar{\Pi}_{3B}^\prime(0), 
\qquad
&Y=\frac{g^{\prime\,2} M_W^2}{2} \bar{\Pi}_{BB}^{\prime\prime}(0)~.
\label{SY}
\end{eqnarray}
Here $g$, $g^\prime$ and $M_W$ are fixed by the conditions
\beqa
\frac{1}{g^2} \,\,=\,\, \bar{\Pi}^\prime_{11}(0)~, 
\hspace{1cm}
\frac{1}{g^{\prime\,2}} \,\,=\,\, \bar{\Pi}^\prime_{BB}(0)~, 
\hspace{1cm}
-\frac{M^2_{W}}{g^2} \,\,=\,\, \bar{\Pi}_{+-}(0)~,
\label{Oblique:normalization}
\eeqa
which, to the order we are interested in, can be identified with the
experimentally measured weak gauge couplings and the $W$ mass.  Some
of these parameters are related to the
Peskin-Takeuchi~\cite{Peskin:1990zt} parameters as $\alpha S=4s_W^2
\hat{S}$ and $\alpha T=\hat{T}$, where $\alpha$ is the electromagnetic
fine structure constant.

If EWSB can be treated perturbatively, we can use the general results
for the vacuum polarizations~(\ref{Pi:bar:aa})-(\ref{Pi:bar:3B}) to
compute these oblique parameters.  For arbitrary $v$ one can use
either~(\ref{Piaa_Hol})-(\ref{Pi3B_Hol}) for custodially symmetric
models with UV localized fermions,
or~(\ref{Pi:bar:aa:hol})-(\ref{Pi:bar:3B:hol}) for models without
custodial symmetry, but arbitrary (though universal) fermion
localization.  In the case that EWSB is a perturbation, the
normalization conditions, Eq.~(\ref{Oblique:normalization}) give
\begin{eqnarray}
\fl
\bar{g}^2 = g^2 \left[1-\hat{S}^{\rm Op} \right]~, 
\qquad
\bar{g}^{\prime\,2} = g^{\prime\,2} \left[1-\frac{g^{\prime 2}}{g^2} \hat{S}^{\rm Op} \right]~, 
\qquad
\bar{v}^2 = v^2 \left[1+\frac{g^2}{g^{\prime 2}} \hat{T}^{\rm Op} \right]~,
\end{eqnarray}
where $\hat{T}^{\rm Op}$ and $\hat{S}^{\rm Op}$ were defined in
Eq.~(\ref{That:Y:universal}).  It is then straightforward to check
that $\hat{T} = \hat{T}^{\rm Op}$, $\hat{S} = \hat{S}^{\rm Op}$ and $W
= Y = W^{\rm Op}$.  In the particular case that the fermions are UV
localized, and in the absence of custodial symmetry,
the oblique parameters are explicitly given by
\begin{eqnarray}
\hat{T}_{UV}&= \frac{g^{\prime\,2}}{2g^2} (m_W L_1)^2
\log\frac{L_1}{L_0}~, \qquad 
& \hspace{-1.5mm}
W_{UV}=\frac{(m_WL_1)^2}{4 \log\frac{L_1}{L_0}}~,
\label{UVTW}
\\
\hat{S}_{UV}&= \frac{(m_W L_1)^2}{2}~, \qquad
&Y_{UV}=\frac{(m_WL_1)^2}{4 \log\frac{L_1}{L_0}}~.
\label{UVST}
\end{eqnarray}
Note that for UV localized fermions $\hat{T}$ is volume enhanced,
$\hat{S}$ is neither enhanced nor suppressed and $W$ and $Y$ are
volume suppressed.  Thus the most constrained experimentally is
$\hat{T}$, followed by $\hat{S}$, while $W$ and $Y$ are very mildly
constrained.  In models with custodial symmetry and UV localized
fermions, setting $\alpha \rightarrow \alpha^N - \alpha^D$ as
discussed at the end of Subsection~\ref{sec:holography}, one finds
$\hat{T}_{UV}^{\rm Cust.} = 0$, so that the most stringent and robust
constraints arise from $\hat{S}$.  This is true only for UV localized
fermion fields and is not maintained in general for bulk fermion
fields.

Light fermions localized on the UV brane are a good approximation in
models with a natural realization of flavour.  However , the large
mass of the top implies that neither of its chirality components can
be too far from the IR brane, and in particular that $b_{L}$ cannot be
UV localized.  The resulting corrections to the bottom couplings can
be computed with identical results using any of the methods we have
discussed in Section~\ref{sec:methods}.  Here we will use our general
result for the effective Lagrangian of models with custodial symmetry,
Eq.~(\ref{Leff:mostgeneral}).  The part of the effective Lagrangian
that we are interested in reads
\beqa
\mathcal{L}_{\mathrm{eff}} &=& 
\mathcal{L}_{\mathrm{SM}}+\alpha^t_{hq} \mathcal{O}^t_{hq}
+\alpha^s_{hq} \mathcal{O}^s_{hq}
+\ldots
\nonumber \\
&=& \frac{\bar{g}}{2c_W} Z_\mu \bar{b}_L \gamma^\mu b_L
\left[-1+\frac{2s_W^2}{3} - 2v^2(\alpha^t_{hq}+\alpha^s_{hq})\right]
+\cdots
\eeqa
Replacing the values of $\alpha^{t,s}_{hq}$ of Eqs.~(\ref{alphat_hl})
and (\ref{alphas_hl}) we obtain for the correction of the
$Z\bar{b}_L b_L$ coupling
\begin{eqnarray}
\frac{\delta g_{b_L}}{g_{b_L}} &=& \frac{2 v^2 (\alpha^t_{hq}+
\alpha^s_{hq})}{1-2s_W^2/3} 
\nonumber \\
&=& - v^2 \frac{
\big[\bar{g}_L^2 T^3_L -\bar{g}_R^2 T^3_R \big] \beta^D
+\big[\bar{g}_L^2 T^3_L - \bar{g}^{\prime\,2} Y \big] 
(\beta^N - \beta^N_{\mathrm{UV}}-\beta^D)
}{1-2s_W^2/3},
\label{deltagZbb:general}
\end{eqnarray}
where the quantum numbers and $\beta^{N,D}$ refer to $b_L$, and we
have explicitly subtracted the global effect that we are
paremeterizing in terms of oblique corrections with the term
$\beta^N_{\mathrm{UV}}$ (we are assuming UV localized light fermions).
We have separated the correction in two terms.  The first term,
proportional to $\beta^D$, corresponds to the coupling evaluated at
zero momentum and, as was the case of $\hat{T}$, is volume enhanced
[see Eq.~(\ref{alpha:localized})].  The term proportional to $(\beta^N
- \beta^N_{\mathrm{UV}}-\beta^D)$ is a correction that arises when the
external gauge boson line is evaluated at $p^2 = m^2_{Z}$, and is
neither (volume) enhanced nor suppressed [see~(\ref{beta:localized})
and (\ref{gamma:localized:UV})].  This separation is clear if one
computes the coupling using the holographic method and the expansion
in Eq.  (A.9).  The volume enhanced term exactly vanishes if we take
either $T^3_L=T^3_R=0$ or $T^3_L=T^3_R$ and $\bar{g}_L=\bar{g}_R$.
The vanishing of the coupling at zero momentum is guaranteed by a
subgroup of the custodial symmetry, as first discussed
in~\cite{Agashe:2006at} 
(and applies to any fermion satisfying
$T^3_R=T^3_L=0$ or $T^3_R=T^3_L$ and $g_L=g_R$).  
However, the on-shell coupling does not
vanish, reading for $T^3_L=T^3_R$ and $\bar{g}_L=\bar{g}_R$,
\begin{equation}
\frac{\delta g_{b_L}}{g_{b_L}}=
-\frac{v^2}{(1-2s_W^2/3)}
\Big[\bar{g}^2_L T^3_L -\bar{g}^{\prime\,2} Y](\beta^N -\beta^D- \beta_{\mathrm{UV}}^N)~.
\end{equation}
Assuming a boundary Higgs, we obtain using
Eqs.~(\ref{alpha:localized}), (\ref{beta:localized}) and
(\ref{Dirichlet:localized}),
\begin{equation}
\frac{\delta g_{b_L}}{g_{b_L}}=
-\frac{v^2 L^2_{1}}{4(1-2s_W^2/3)}
\Big[\bar{g}^2_L T^3_L -\bar{g}^{\prime\,2} Y] [g_2(c_{b_L})-2\tilde{g}_2(c_{b_L})]~.
\end{equation}
where $g_2(c)$ and $\tilde{g}_2(c)$ are defined in Eqs.~(\ref{gn}) and
(\ref{gtn}) of the appendix.

\subsection{Loop effects}

So far we have concentrated on tree-level effects.  Due to the
summation over the KK towers, loop effects can generically be expected
to be relevant.  In generic theories such effects are UV sensitive and
can only be estimated, e.g.~based on NDA in higher
dimensions~\cite{Chacko:1999hg}.  However, in theories with custodial
symmetry~\cite{Agashe:2003zs} (and custodial protection of some
fermion couplings to the $Z$~\cite{Agashe:2006at}) some of the EW
(pseudo)observables, namely $T$ and $\delta g_{b_{L}}$, are
calculable, since the symmetries forbid the associated counterterms.
Given that the corresponding tree-level contributions are small, it is
pertinent to assess more carefully the importance of such loop
effects.

The one-loop contribution to the $T$-parameter due to the top KK tower
was computed in certain scenarios in~\cite{Carena:2006bn} (see
also~\cite{Agashe:2005dk}).  In particular, it was observed that when
the fermions are assigned to representations [which involve
bi-doublets under $SU(2)_{L} \times SU(2)_{R}$] that protect the
$g_{b_{L}}$ coupling from large tree-level contributions, typically
the contribution to $T$ from the full tower associated with the top
quark decreases with respect to the SM one (i.e. the massive KK tower
gives a \textit{negative} $\Delta T$).  This can be understood as a
consequence of the quantum number assignments plus the boundary
conditions necessary to preserve the custodial symmetry on the IR
brane (and requiring also that the zero-mode sector be the SM one).
Given that there is a sizable positive tree-level contribution to the
$S$ parameter, an EW fit based on the oblique parameters can put
rather severe bounds on the scale of new physics in these scenarios.
Interestingly, there are well-defined regions of parameter space where
some of the top KK excitations become light and make $\Delta T$
positive, thus also allowing for lighter KK gauge
excitations~\cite{Carena:2006bn}.  This can have important
implications for the phenomenology of fermion KK modes as well as that
of the KK gauge bosons, which can decay into the light KK fermions
with significant branching fractions (as studied in the context of a
Gauge-Higgs unification scenario in~\cite{Carena:2007tn}).  We should
remark that the one-loop contribution to $T$ due to gauge fields has
not been computed, although it is expected to be smaller than the one
from the top sector since the KK gluons are heavier and their
couplings are controlled by the weak gauge couplings as opposed to the
top Yukawa coupling.

Also, the one-loop contributions to $\delta g_{b_{L}}$ were computed
in~\cite{Carena:2007ua}.  These tend to be small, but not necessarily
negligible.  As we will see in the next section, such loop
contributions can be important in relaxing the constraints from EW
precision measurements on the scenarios we consider.

\subsection{Summary of EW constraints in models with custodial symmetry}

With the low-energy effective Lagrangian presented in the previous
sections one can assess the indirect constraints from precision
measurements on models with custodial symmetry in warped spaces, under
the assumption that there exists a light Higgs degree of freedom.  In
the general case that the fermions propagate in the bulk, one should
perform a global fit to the EW observables based on the dimension-6
Lagrangian of Eq.~(\ref{Leff:mostgeneral}).  Such an analysis was
performed in~\cite{Carena:2007ua}.  It was found that, depending on
the SM fermion $SU(2)_{R}$ quantum number assignments, the bounds from
the EW constraints on the gauge boson masses are typically in the
$2.5$-$3.5~{\rm TeV}$ range (neglecting brane kinetic terms), although
in certain models they could be as low as $1.5~{\rm TeV}$.

Due to the $T^3_{R}$ charges in Eqs.~(\ref{alphas_hl}) and
(\ref{alphas_ll}), the low-energy effects of the heavy physics are in
general not universal, even when all fermions share the same
localization profiles.  However, the RS interpretation of the flavor
structure as arising from ``anarchy'' of the Yukawa couplings (see
Section~\ref{flavor} below) requires that the light fermion families
be localized close to the UV brane, in which case the low-energy
corrections become of the universal type (the effects of the massive
$SU(2)_{R}$ gauge bosons become exponentially suppressed and the
$c$-dependence of Eqs.~(\ref{beta}) and (\ref{gamma}) disappears).  It
is then simpler to use the fits based on the oblique parameters,
Eqs.~(\ref{TW}) and (\ref{SY}).  The dependence on the model
parameters is given in Eq.~(\ref{That:Y:universal}).  Furthermore, for
UV localized fermions, $W$ and $Y$ are volume suppressed and can be
neglected.  However, in general the corrections to the
$Z\bar{b}_{L}b_{L}$ coupling, which at tree-level are given by
Eq.~(\ref{deltagZbb:general}), need to be taken into account.  When
$g_L=g_R$ and $T^3_L=T^3_R$ the tree-level contributions to $\delta
g_{b_{L}}$ are relatively small due to a custodial
protection~\cite{Agashe:2006at}, but there can be non-negligible
loop-level effects.  Thus, a fit to $S$, $T$ and $\delta g_{b_{L}}$ is
typically appropriate.

Using the code of Ref.~\cite{Han:2004az}, we obtain the
$1\sigma~(\Delta \chi^2 = 1)$ intervals
\beqa
\fl
S = -0.03 \pm 0.09~,
\qquad \quad
T = 0.03 \pm 0.09~,
\qquad \quad
\frac{\delta g_{b_{L}}}{g_{b_{L}}} = (-0.4 \pm 1.4) \times 10^{-3}~.
\label{fit}
\eeqa
Here we have used the combined CDF and D$\slash{\!\!\!0}$ top mass
measurement of March-2009 (using up to $3.6~{\rm fb}^{-1}$ of data per
experiment): $m_{t} = 173 \pm 1.3~{\rm GeV}/c^2$~\cite{:2009ec}, as
well as the combined CDF/D$\slash{\!\!\!0}$ $W$ mass measurement of
Aug.-2008: $M_{W} = 80.432 \pm 0.039~{\rm GeV}/c^2$~\cite{:2008ut}.
We include the $Z$-pole observables, the low-energy measurements
(except for NuTeV), and LEPII data.  We also used a Higgs mass of
$m_{h_{\rm ref.}} = 117~{\rm GeV}$.

In models without custodial protection the main constraint comes from
$T = \hat{T}/\alpha$.  Requiring $T^{\rm tree} < 0.21$ (at $2\sigma$),
assuming a light Higgs as in the fit of Eq.~(\ref{fit}), and using
Eq.~(\ref{UVTW}), gives a strong lower bound of $1/L_{1} \sim 4.4~{\rm
TeV}$ (hence gauge boson masses $M_{\rm KK} > 2.45/L_{1} \approx
11~{\rm TeV}$).  Such a constraint can be somewhat relaxed for a
heavier Higgs (and for sizable gauge and/or fermion brane kinetic
terms~\cite{Carena:2002dz}).  The situation is dramatically improved
in models with custodial protection, since $T^{\rm tree}$ vanishes in
this case (for UV localized fermions).  A bound based on $S < 0.15$
(at $2\sigma$) alone results in $1/L_{1} \sim 1.5~{\rm TeV}$ ($M_{\rm
KK} > 3.7~{\rm TeV}$).  However, falling inside the $95\%$ C.L.
ellipsoid ($\Delta \chi^2 = 7.81$ for three parameters: $S$, $T$ and
$\delta g_{b_{L}}$) can allow for $S \sim 0.23$ (if $T \sim 0.25$ and
$\delta g_{b_{L}}/g_{b_{L}} \sim - 0.8 \times 10^{-3}$).  In this
case, and in the absence of additional contributions to $S$, one can
have $1/L_{1} \sim 1.25~{\rm TeV}$ ($M_{\rm KK} > 3~{\rm TeV}$).  As
mentioned in the previous subsection, loop level contributions to $T$
and $\delta g_{b_{L}}$ can be relevant (and depend on additional model
parameters, although typically $\Delta T^{\rm loop}$ and $\delta
g_{b_{L}}^{\rm loop}$ are correlated).  Requiring that these
additional contributions optimize the EW fit usually selects
well-defined regions of parameter space (e.g. localization of the
third family quarks, with important consequences for their KK spectrum
which can include fermion states lighter than the gauge KK modes).

In summary, we see that models with custodial protection and an
implementation of the flavor structure through fermion localization
can be consistent with gauge boson KK excitation of about $3~{\rm
TeV}$.  EW constraints place similar bounds on models of Gauge-Higgs
Unification~\cite{Agashe:2004rs,Agashe:2005dk}, where an interesting
connection between EWSB and EWPT arises, via the appearance of
relatively light KK fermion states~\cite{Medina:2007hz,Panico:2008bx}.
We will take
these scales as a guide to the study of the collider phenomenology, to
be undertaken in the second part of this review.

Some of the above bounds could be somewhat relaxed in the presence of
moderate IR brane kinetic terms.  Also, playing with the localization
parameters of LH versus RH fermions it may be possible to lower the
scale of new physics somewhat more~\cite{Carena:2007ua}.  Another 
possibility compatible with a lower scale of new physics is to consider a 
different background, e.g. soft-wall models (see for
instance~\cite{Falkowski:2008fz}-\cite{Gherghetta:2009qs}).

Higgsless models~\cite{Csaki:2003dt}, which have not been covered in
this review due to lack of space, require a much lower scale that can
be made compatible with EWPT at tree level by localizing the light
fermions close to $c=1/2$~\cite{Cacciapaglia:2004rb} for which all
corrections become small, and imposing the custodial protection of the
$T$ parameter and the $Z\bar{b}b$ coupling~\cite{Cacciapaglia:2006gp}
(for a review, see~\cite{Grojean:2006wq} or Section~3 of
\cite{Csaki:2005vy}).
  
\section{A few remarks on flavour (constraints)}
\label{flavor}

As remarked in the introduction, the RS framework with bulk fields
provides a theory of flavor.  The main idea~\cite{Grossman:1999ra}
arises from the $c$-dependence of the fermion zero-mode
wavefunctions~(\ref{FermionZeroMode}): for $c>1/2$, the profiles are
localized near the UV brane and therefore the overlap with an IR
localized Higgs field (so as to address the hierarchy problem) is
exponentially suppressed.  This allows one to easily generate
hierarchical effective 4D Yukawa couplings, even if the microscopic 5D
Yukawa couplings exhibit no special structure, an assumption usually
called ``anarchy'' (in other scenarios, such as gauge-Higgs
unification, the fermion-Higgs interactions depend on localized
fermion mixing mass parameters to which the anarchy assumption can be
applied).  In warped spaces, with their associated $z$-dependence of
the cutoff scale, (light) fermion localization near the UV brane also
provides an effective cutoff on non-renormalizable flavour-violating
operators that is sufficiently high to significantly suppress their
effects, even with the new KK physics entering at a few TeV.

The consequences of ``anarchy'' for flavour physics were considered
originally in \cite{Huber:2000ie}-\cite{Agashe:2006iy} and more recently   
in~\cite{Csaki:2008zd}-\cite{Azatov:2009na}.
The largest effects arise at tree-level from flavour changing KK gluon
couplings.  Although most of the predictions are consistent with
current flavour constraints for the scales determined by the EWPT
--and often lead to interesting expectations for indirectly observing
the new physics in the near future-- there are some observables that
can impose more severe bounds.  Most notably CP violation in the kaon
system can lead to a bound of ${\cal O}(10)~{\rm
TeV}$ on the KK scale~\cite{Csaki:2008zd}.

Thus, a strict application of the anarchy assumption can result in a
KK scale beyond the reach of the LHC. This may suggest some non-trivial
structure (for instance flavor symmetries), see
e.g.~\cite{Cacciapaglia:2007fw}-\cite{Chen:2009gy}, or perhaps that
these models exhibit a moderate amount of fine-tuning (hopefully in
the flavour sector, so that discovering the new physics at the LHC
remains a possibility).  Given the model dependence of such
conclusions, we will take the more robust bounds from EW precision
measurements as a starting point to summarize the LHC collider
phenomenology of warped scenarios.


\title{Part II \,\, Collider Phenomenology of Warped Models}
\vspace{2mm}
\noindent
\textbf{\it Contributed by H.~Davoudiasl and S.~Gopalakrishna}
\vspace{4mm}


In our presentation of the collider phenomenology of warped models~
\footnote{In what follows, we focus on KK signals of compactification.  However, 
an embedding of the RS model in string theory (UV completion) 
could result in the appearance of additional signals; see for example 
Refs.~\cite{Hassanain:2009at,Perelstein:2009qi}.}, we
will use a different notation from that of the preceding discussion.
This is largely done to match the conventions used in several of the
papers that are cited in the following review.  In our notation,
$k=1/L_0$ and the size of the extra dimension is denoted by $\pi r_c$.
Also, note that, in most of what follows, the LHC center of mass
energy $\sqrt{s}=14$~TeV is assumed, unless otherwise specified.
Obviously, given the parton distribution function (PDF) dependence of
cross sections for production of TeV-scale KK modes, many of the
conclusions presented below have to be revisited for smaller values of
$\sqrt{s}$.

\section{KK Gravitons}

We begin our discussion of KK phenomenology with the graviton sector, since KK gravitons were the first
tower of states whose collider phenomenology was studied \cite{Davoudiasl:1999jd}
and constitute a key and common signature of warped models.
As noted previously, in the original RS model the entire SM content was assumed to be confined
to the IR-brane.  The most distinct signature of this setup was then the tower of graviton KK modes.
These states would appear as a series of spin-2 resonances close to the weak
scale, which sets the scale of their masses and couplings to the SM.  Specifically,
the KK graviton masses are given by
\beq
m^{G}_n = x_n^G k e^{- k r_c \pi},
\label{mGn}
\eeq
where $x_n=3.83, 7.02,\ldots$, for $n=1,2, \dots$, are given by the roots of $J_1$ Bessel function, to a very good
approximation \cite{Davoudiasl:1999jd}; the zero mode, for $n=0$, is the massless 4D graviton.
The interactions of the graviton KK modes with the
SM fields are given by
\beq
{\cal L} = - {1\over\mP}T^{\alpha\beta}(x)h^{(0)}_{\alpha\beta}(x)-
{1\over\Lambda_\pi}T^{\alpha\beta}(x)\sum_{n=1}^\infty
h^{(n)}_{\alpha\beta}(x)\, ,
\label{gravSM}
\eeq
with $\mP \sim 10^{18}$~GeV the reduced Planck mass, $T_{\alpha \beta}$ the SM energy-momentum
tensor, $h^{(n)}_{\alpha\beta}(x)$ KK modes of the graviton, and
$\Lambda_\pi \equiv e^{- k r_c \pi} \mP$ \cite{Davoudiasl:1999jd}.

\begin{figure}[htbp]
\centerline{
\includegraphics[width=10cm,angle=0]{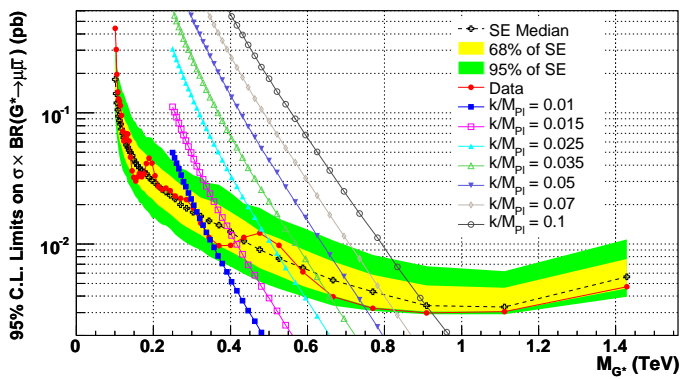}}
\vspace*{0.1cm}
\caption{CDF limits ($95\%$ confidence level) \cite{CDF_RS} on the product of cross section 
and dimuon branching fraction,  
for the lightest KK graviton in the original RS model. Theoretical cross sections and expected limits from simulated experiments (SE) are also shown.}
\label{CDF:RS}
\end{figure}

\begin{figure}[htbp]
\centerline{
\includegraphics[width=7cm,angle=-90]{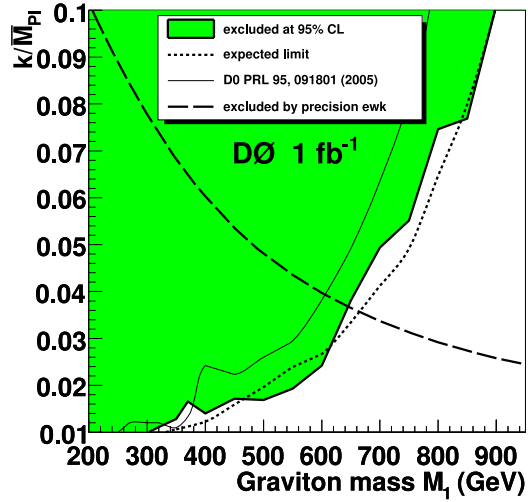}}
\vspace*{0.1cm}
\caption{D0 bounds \cite{D0_RS} 
on the lightest KK graviton mass, as a function of $k/\mP$, in the original RS model.}
\label{D0:RS}
\end{figure}

The best direct limits on the original RS model are from the Tevatron experiments.
We present recent limits from the CDF \cite{CDF_RS} and D0 \cite{D0_RS} experiments
in Figs.~\ref{CDF:RS} and \ref{D0:RS}.  The graviton KK phenomenology of the original
RS model can be described by the KK mass and the ratio $k/\mP$.  Calculations in this
background are reliable as long as $k/M_5$ is not $\ord{1}$ or larger (this is
a classical argument that can be somewhat modified if we consider quantum
effects \cite{KKgraviton2}).  For this reason, $k/\mP\lsim 1$
is often considered in phenomenological and experimental studies, as seen for example
in these figures.  We see that the direct bounds on the mass of the lightest KK graviton
range over 300-900~GeV for $0.01 \leq k/\mP \leq 0.1$.  The LHC reach for the first
graviton KK mode in this model has been
calculated by ATLAS \cite{ATLAS_RS} and CMS \cite{CMS_RS}
collaborations and is roughly 3.5 TeV with \fb{100}
of integrated luminosity.

Once we extend the RS setup to provide a model of flavor, the above conclusions regarding
the graviton KK phenomenology at colliders change drastically.  The current Tevatron bounds
do not take these effects into account, however we generally expects the direct
bounds to become much weaker.  This is due to the fact that, in warped flavor models,
KK graviton couplings to UV-localized light fermions get highly suppressed \cite{Davoudiasl:2000wi}
roughly like their Yukawa couplings \cite{KKgraviton2}, and  couplings to gauge
fields (in units of $1/\mP$) \cite{Davoudiasl:2000wi}
\begin{eqnarray}
C^{ A A G }_{ 0 0 n } & = &
e^{ k r_c \pi}
\frac{ 2 \left[ 1 - J_0 \left( x_n^G \right) \right] }{ k r_c \pi
\left( x_n^G \right)^2 | J_2 \left( x_n^G \right) | },
\end{eqnarray}
where $J_{ 0, 2 }$ denote Bessel functions,
get volume ($k r_c \pi$) suppressed.
Hence, as alluded to before, important production and decay channels become either inaccessible
or else suppressed.  Given that a TeV-scale graviton KK tower is the most generic
prediction of the RS model, largely independent of
various model building assumptions, it is worth reconsidering its phenomenology within
warped flavor models.

The reach of the LHC experiments for
the lightest graviton KK mode was accordingly
reexamined in Refs.~\cite{KKgraviton1} and \cite{KKgraviton2}.
With the light fermions basically decoupled, the dominant production
channel at the LHC is through gluon fusion, which is suppressed by $(kr_c \pi)^{-1}$.  As for the
dominant decay channels, generically we expect the right-handed top $t_R$
and the Higgs sector, including the
longitudinal gauge fields $Z_L$ and $W^\pm_L$,
to be the only important final states, due to their large overlap with the graviton KK
wavefunction \footnote{Quite often
(however not as a rule \cite{Carena:2006bn}) ,
it is the right-handed top quark that is the most IR-localized SM fermion, given that
the doublet $(t, b)_L$ is subject to precision bounds on $b$-quark couplings, as noted in the
preceding discussion of precision constraints.}.  With these assumptions, the partial widths of the
first KK graviton into pairs of $t_R$, the Higgs boson $h$, $Z_L$, and $W^\pm_L$ are given,
respectively by \cite{KKgraviton1,KKgraviton2}
\beq
\frac{\Gamma_{t_R}}{3 N_c} = \Gamma_h = \Gamma_{Z_L} = \frac{\Gamma_{W_L}}{2}
= \frac{ ({\bar c}\, x^G_1)^2 \, m^G_1}{ 960 \pi },
\label{grav-widths}
\eeq
where $N_c=3$ is the number of colors in QCD , ${\bar c}\equiv k/\mP$ (our barred notation 
for this parameter differs from that of Ref.~\cite{KKgraviton2}, to avoid confusion with 
the bulk fermion mass parameter $c$, discussed before), and final states
have been treated as massless, which is a good approximation
due to the much larger expected mass of the 
KK mode.  Also, the
decay widths to longitudinal
polarizations have been estimated using the equivalence theorem
[which is valid up to $M_{ W, Z }^2 / (m^G_1)^2$ corrections] to relate
these widths to that of the physical Higgs.  Here, we note that 
two-body decay into final states that 
include a KK mode and a heavy SM mode 
are either (i) volume suppressed, when the relevant zero mode has an  
approximately flat profile, or (ii) kinematically forbidden 
(due to a large KK fermion mass), 
or else (iii) suppressed by a power of $\vev{H}$.  If we assume that $t_R$ is highly  
IR-brane-localized (ii) applies to an otherwise potentially important $G_1 \to {\bar t_R} t_{R1}$.

\begin{figure}
\vspace{0.8cm}
\begin{center}
\includegraphics[angle=270,scale=0.32]{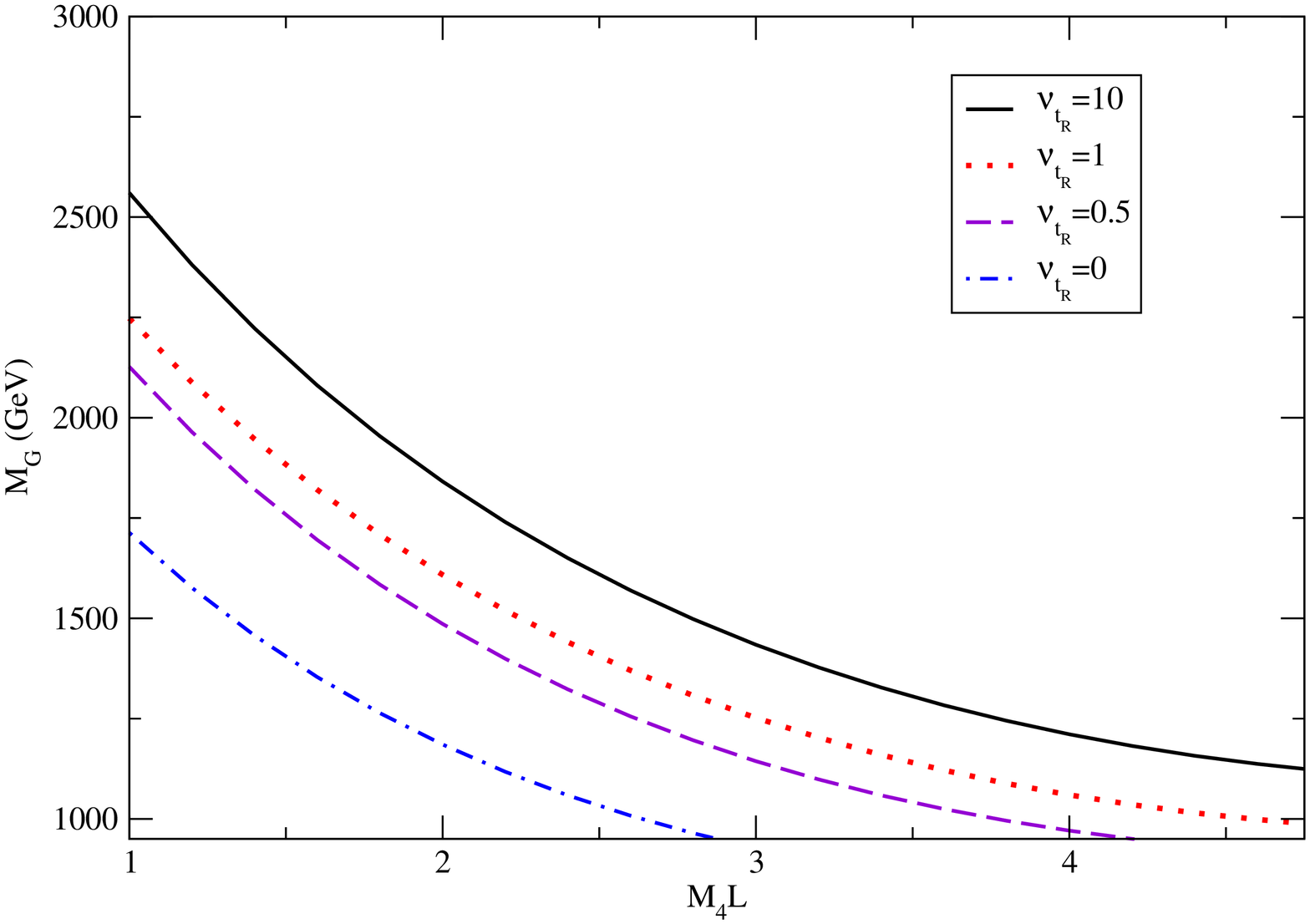}\hspace{0.2cm}
\includegraphics[angle=270,scale=0.32]{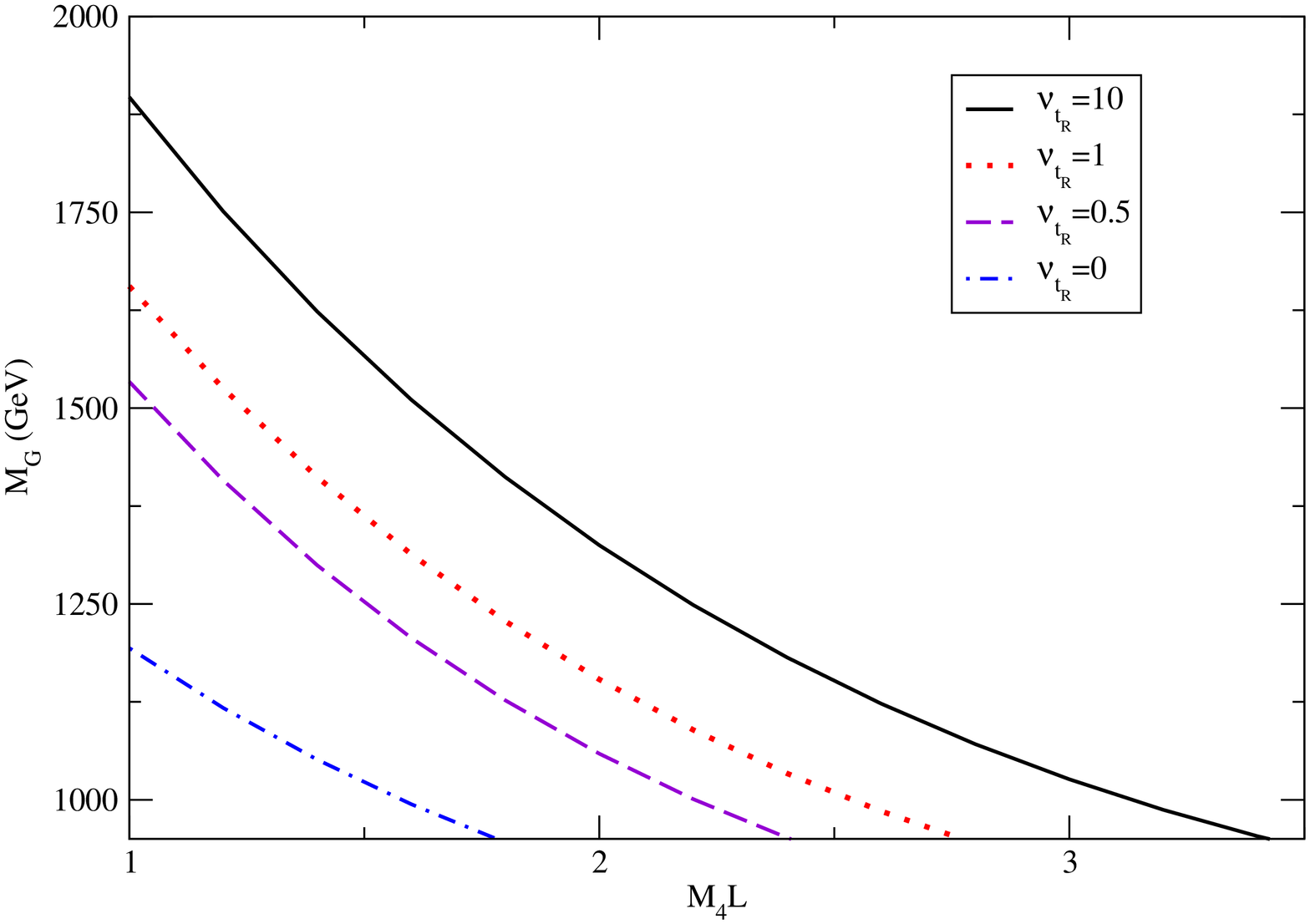}
\end{center}
\vspace{0.6cm}
\begin{center}
\includegraphics[angle=270,scale=0.32]{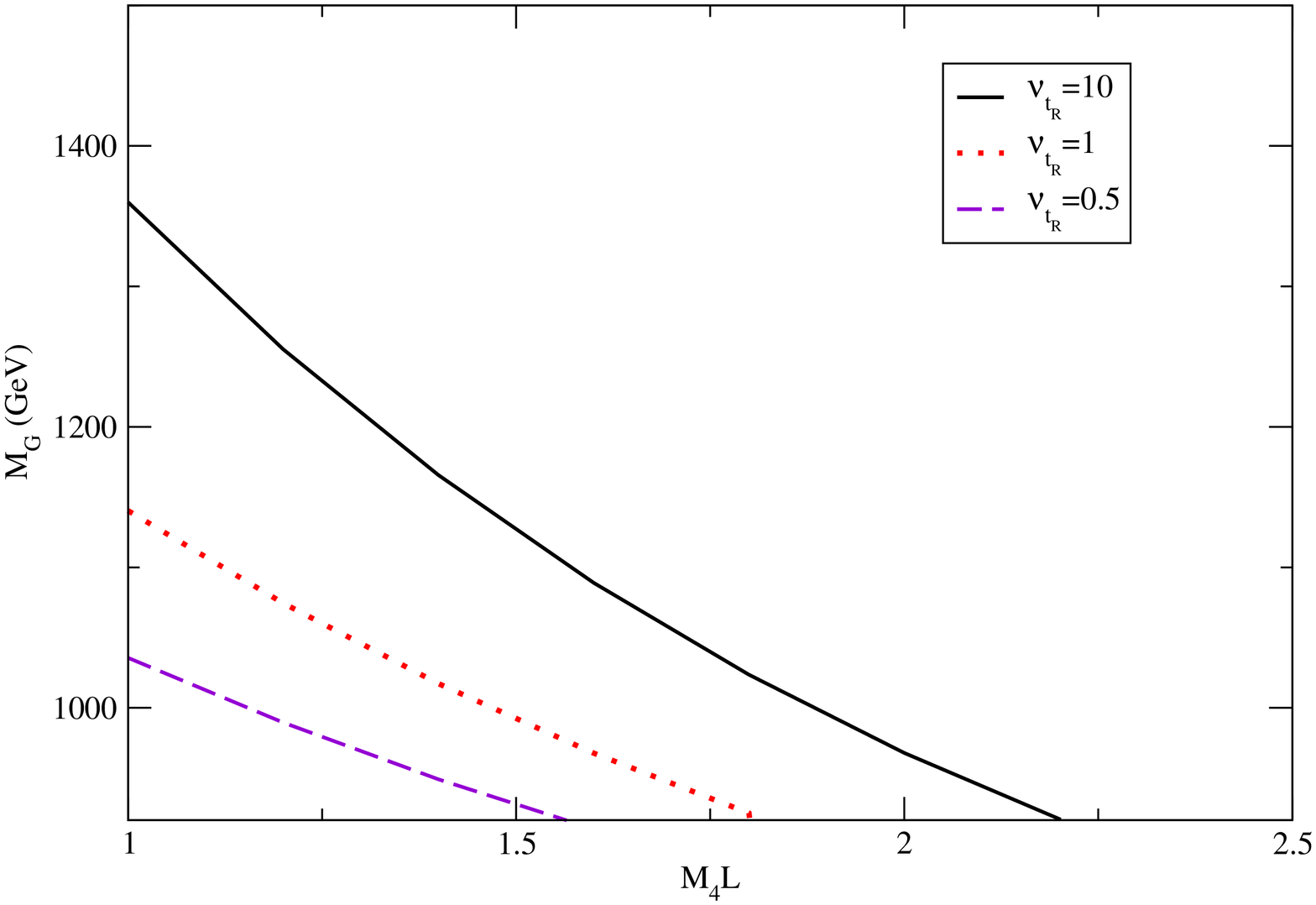}
\end{center}
\caption{The reach ($S/\sqrt{B}$ = 5) for the lightest KK graviton $G_1$,
from Ref.~\cite{KKgraviton1}, with
\fb{100} of integrated luminosity and top identification
efficiency $100\%$, $10\%$, and $1\%$, clockwise from the top.   With increasing
$\nu_{t_R}$ (corresponding to $-c_{t_R}$ in the notation of Part I) 
the profile of $t_R$ is more IR-localized;
$M_4 L$ corresponds to $1/{\bar c}$ in our notation.}
\label{kkgravtt}
\end{figure}
Given the above considerations, Ref.~\cite{KKgraviton1} focused on the $t_R {\bar t}_R$
final state to estimate the LHC reach for the lightest KK graviton.
One of the issues that affects the utility of this channel is the efficiency for top
identification,  made more challenging because of the collimated
decay products of highly boosted tops.  
We will discuss this issue in greater detail in later sections. 
The narrow width of the KK graviton,
under the above assumptions, can be helpful in limiting the background in this channel.  However,
one has to account for smearing of the measured resonance mass.

In Ref.~\cite{KKgraviton1}, the background
was taken to be all $t {\bar t}$ with invariant mass within $3\%$ of $m^G_1$, corresponding to
a typical smearing $E \times 3\%$ for the ATLAS experiment.  The top identification
efficiency was treated as uncertain in this work and the reach, given by $S/\sqrt{B}=5$, assuming
\fb{100} of integrated luminosity,
was considered for three values of efficiency, $1\%$, $10\%$, and $100\%$,
as presented in  Fig.~\ref{kkgravtt}.  These results suggest that the reach for $G_1$
may be expected to be 1.5-2~TeV, depending on top identification efficiency.

Given that the longitudinal gauge bosons $Z_L$ and $W^\pm_L$ are manifestations
of the IR-localized Higgs sector Goldstone modes, their couplings to the graviton KK modes
are substantial, as deduced from Eq.~(\ref{grav-widths}).  In particular, the process
$pp \to G_1 \to Z_L Z_L$ with both $Z$'s decaying into $\ell^\pm=e^\pm,\mu^\pm$, provides
a clean signal, unencumbered by complicated event reconstructions.  In the absence
of prompt di-lepton and di-photon channels, this process provides a ``golden mode" for
KK graviton discovery.  Based on these considerations,
Ref.~\cite{KKgraviton2} examined the
possibility of searching for warped gravitons in the $Z_L Z_L$ decay channel.

In Ref.~\cite{KKgraviton2},
it was determined that the vector boson fusion cross section for KK graviton production
is roughly an order of magnitude smaller than the gluon fusion channel considered above.
The dominant SM background to $ZZ$ production though comes from $q {\bar q}$
initial states.  Hence, one expects forward-backward cuts on pseudo-rapidity $\eta$ to
be efficient in reducing the background, mediated via $t/u$ channels, while not
affecting the signal a great deal; this was shown to be case in Ref.~\cite{KKgraviton2}, typically
yielding $S/B$ significantly larger than unity.  Note that due to negligible initial state $q {\bar q}$
coupling to $G_1$, the signal and background do not interfere to a good approximation.

The reducible SM background to the above 
KK graviton signal depends on the decay modes of the $Z$'s.  For hadronic decays
of both $Z$'s, the 4 jet QCD background was deemed too large to allow this channel to be of use.
Even the case with one $Z$ decaying hadronically and the other decaying into $\ell^\pm$
poses a serious challenge.  This is due to the large boost of the $Z$'s (similar to $t$'s in the
previous discussion) which makes the opening angle of signal di-jets of order
$M_Z/{\rm TeV} \sim 0.1$ \cite{KKgraviton2} while typical cone size for
jet reconstruction was taken to be $\ord{0.4}$ (as quoted from
Ref.~\cite{conejetalgorithm}).  This makes the relevant SM background to come form $Z+j$,
which was calculated to be an order of magnitude larger over the resonance width.  This
problem of jet-merging in the decays of highly boosted particles
is a generic challenge in searches for TeV scale resonances that
preferentially decay into heavy SM final states, and will
be encountered in our later discussions of other KK modes.

\begin{figure}
\includegraphics[width=0.49\textwidth]{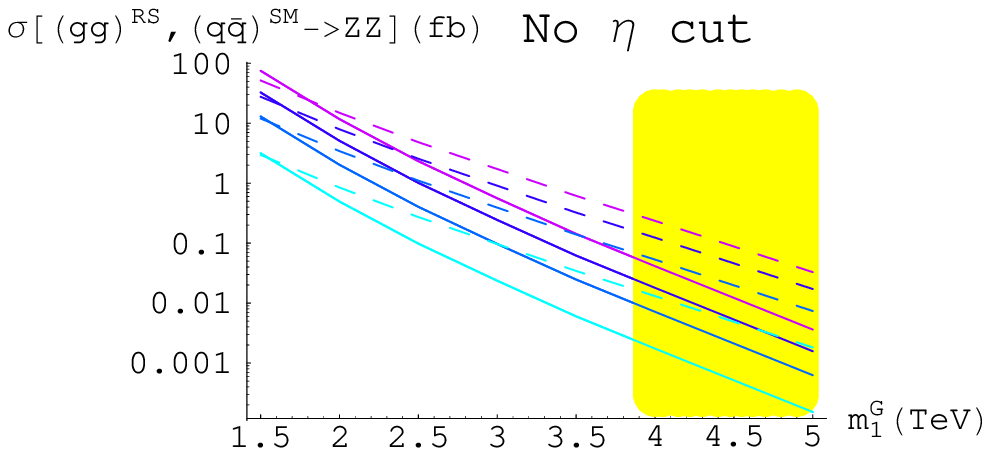}
\includegraphics[width=0.49\textwidth]{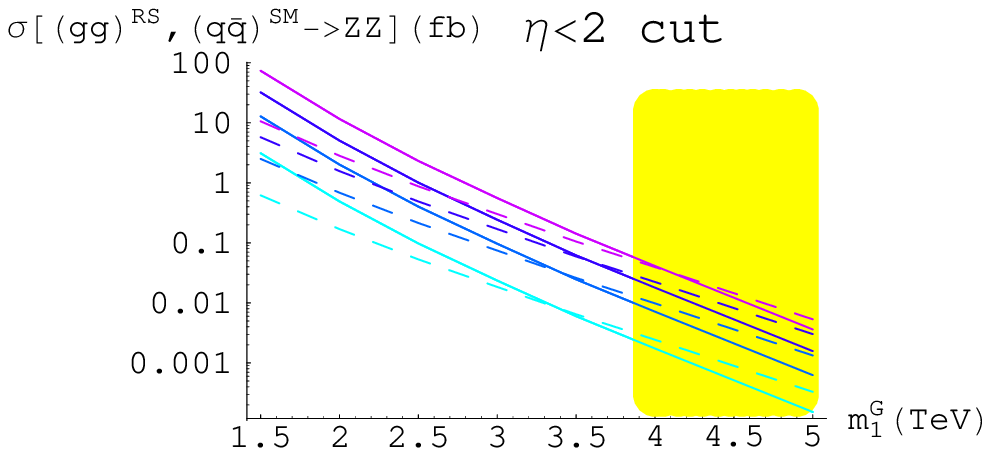}
\caption{The cross-sections (integrated over one width) for
$gg \rightarrow ZZ$ via
KK graviton (solid lines) and the corresponding SM background (dashed lines),
for ${\bar c} =0.5, 1, 1.5, 2$ (from bottom to top), with no cut on $\eta$ (left panel) and
$\eta<2$ (right panel).  The yellow region shows the expected KK graviton mass
in the simplest models based on gauge KK mass from precision tests \cite{KKgraviton2}.}
\label{gravZZ}
\end{figure}

Based on the clean $4 \ell$ final state, the signal and background were calculated
over the KK graviton width $\Gamma_G$, as shown in Fig.~\ref{gravZZ}.  We can see from
the plots that the $\eta$-cut significantly improves $S/B$; the dependence of background
on ${\bar c}$ comes from the dependence 
$\Gamma_G \propto {\bar c}^2$ that sets the integration interval.
Naive dimensional analysis suggests that higher curvature terms are suppressed by
$\Lambda \sim 24^{1/3} \, \pi \, M_5$ \cite{KKgraviton2},
which is why values of ${\bar c}>1$ were considered.  Ref.~\cite{KKgraviton2} concluded
that for ${\bar c}\lsim 2$ (for reliable calculations), the LHC reach ($S/\sqrt{B} > 4$) for $G_1$ is about
2 (3)~TeV with \fb{300 (3000)}.

Another potentially important channel for KK graviton discovery is via the $W_L^+ W_L^-$
final state, discussed in Ref.~\cite{Antipin:2007pi}.  These authors note that the KK graviton has
a branching fraction into the $W_L^\pm$ twice that for the $Z_L Z_L$ final state and that the
branching fraction of the $W^\pm$ into a leptonic final state is about 11\% compared to 3\% for
the $Z$.  However, when the highly boosted $W$'s decay into leptons,
the neutrinos will be mostly back-to-back and the missing energy
information will be lost \cite{KKgraviton2,Antipin:2007pi}.  In addition,
the hadronic decay of one of the $W$'s will again be collimated and be subject to a large
$W + j$ background \cite{KKgraviton2,Antipin:2007pi}.
The results of Ref.~\cite{Antipin:2007pi} suggest that the reach
for the KK graviton in the $W_L^\pm$ channel with \fb{300}
will not be significantly above 2~TeV, unless analysis techniques are
developed to suppress the $W+j$ background.

In case a graviton KK mode is detected at the LHC, it would be important to establish
its spin.  Ref.~\cite{Antipin:2008hj} considered whether this is feasible, in models 
that yield realistic flavor using bulk fermions, and
concluded that, with \fb{1000}, the KK graviton spin for masses up to $\sim 2$~TeV may be
confirmed.  Refs.~\cite{Cousins:2005pq,CMS_RS,Osland:2008sy,Murayama:2009jz} 
also considered this question,
but in the original RS model with the entire SM on the IR brane,
where $e^\pm$ and $\mu^\pm$ decay modes are accessible.  
Using the ``center-edge asymmetry" in the angular distribution of the final state leptons, 
Ref.~\cite{Osland:2008sy} found it feasible to identify the graviton spin at the 
2$\sigma$ level, for ${\bar c}=0.1$, up to a mass of 3.2~TeV, with \fb{100}.  
However, we again note that the original RS model
is subject to stringent bounds on cutoff-scale operators which, in the
absence of tuning, would require $m^G_1\gg 1$~TeV.

Simple warped models predict that the
lightest graviton is $3.83/2.45\simeq 1.56$ times heavier
than the lightest KK gauge boson.  Given the preceding discussion of precision data, it is then
expected that $m^G_1 \gsim 4$~TeV, corresponding to the yellow regions in Fig.~\ref{gravZZ}.
Therefore, the above analyses
suggest that, in generic models, discovering the warped KK graviton at the LHC is
quite challenging, at design or perhaps
even upgraded (SLHC \cite{Gianotti:2002xx,Bruning:2002yh}) luminosity.

\section{KK Gluons}
\label{KKgluon.SEC}

The KK gluons $g_n$ generally offer the best reach for discovery at the LHC, as indicated by
the results of Refs.~\cite{KKgluon1} and \cite{KKgluon2} that we summarize below;
we will not present the details of the analysis in these works and refer the
interested reader to these references
for more information.   First of all, the level-1
KK gauge fields are generally expected to be the lightest such excitations,
within the known (SM$\oplus$gravity) sectors; the KK masses for a gauge
field $A$ are given by Eq.~(\ref{mGn}) with
$x^G_n\to x^A_n \simeq 2.45, 5.57,\ldots$.  Secondly,
the $SU(3)_c$ coupling constant $g_s$ of the SM gluon is larger than those of
other SM gauge fields.  This leads to a larger production cross section,
which is proportional to $g_s^2$,
for the KK gluons.  However, in realistic models of flavor,
the couplings of $g_1$ to light flavors are diminished
compared to its SM counterpart ($g_0$, the gluon), suppressing
production through light quarks at colliders;
the couplings to the gluon of the SM are zero by
orthonormality of KK modes.  On the other hand,
the dominant branching fraction is set by coupling to
IR-localized top quarks, which is enhanced
compared to the SM gauge coupling.  This leads to a large width
for KK gluons $\Gamma_g \sim  m^g_1/6$ for $m^g_1 \gsim1$~ TeV
(demanded by precision data),
and renders the extraction
of a discovery signal more difficult.

The work of Ref.~\cite{KKgluon1} focused on a setup in which
$Q^3=(t,b)_L$ was quasi IR-localized, $t_R$ was basically on the
IR-brane, and all other fermions $f$ were UV-localized. Let $\xi
\equiv \sqrt{\log ({\mP/{\rm TeV}})} \approx \sqrt{k r_c \pi}\sim 5$. Then
the couplings of the lightest KK gauge field $A^{(1)}$ are given by the
following approximate relations \cite{KKgluon1}
\beq 
\hskip-2.5cm
\frac{g^{f{\bar
f} A^{(1)}}}{g_{\rm SM}} \simeq 1/\xi \; ; \;\;
\frac{g^{Q^3{\bar Q^3} A^{(1)}}}{g_{\rm SM}} \approx 1 \; ; \;\;
\frac{g^{t_R{\bar t_R} A^{(1)}}}{g_{\rm SM}} \simeq \xi
\; ; \;\; \frac{g^{A^{(0)} A^{(0)} A^{(1)}}}{g_{\rm SM}} \approx 0
\; ; \;\;
\frac{g^{H H A^{(1)}}}{g_{\rm SM}} \simeq \xi
\label{ffA1}
\eeq
where $g_{\rm SM}$ denotes a generic SM gauge
coupling, $A^{(0)}$ is a gauge field zero mode.
Note that $H$ includes both the physical Higgs ($h$) and
{\em un}physical Higgs, i.e., {\em longitudinal}
$W/Z$ by the equivalence theorem
(the derivative involved in this coupling is
similar for RS and SM cases and hence is not shown for
simplicity).  For the case of KK gluons, the last coupling in Eq.(\ref{ffA1}) 
does not exist and the fourth equality is exact, since $SU(3)_c$ is unbroken.  
Here, the effects of EWSB, expected
to be small, are ignored.  We note that,
in alternative models that possess a custodial symmetry to
protect the coupling $Z b{\bar b}$, one can arrange for the couplings of $Q^3$ and
$t_R$ to be interchanged or for both to have intermediate values \cite{Agashe:2006at}.

\begin{figure}
\includegraphics[width=0.49\textwidth]{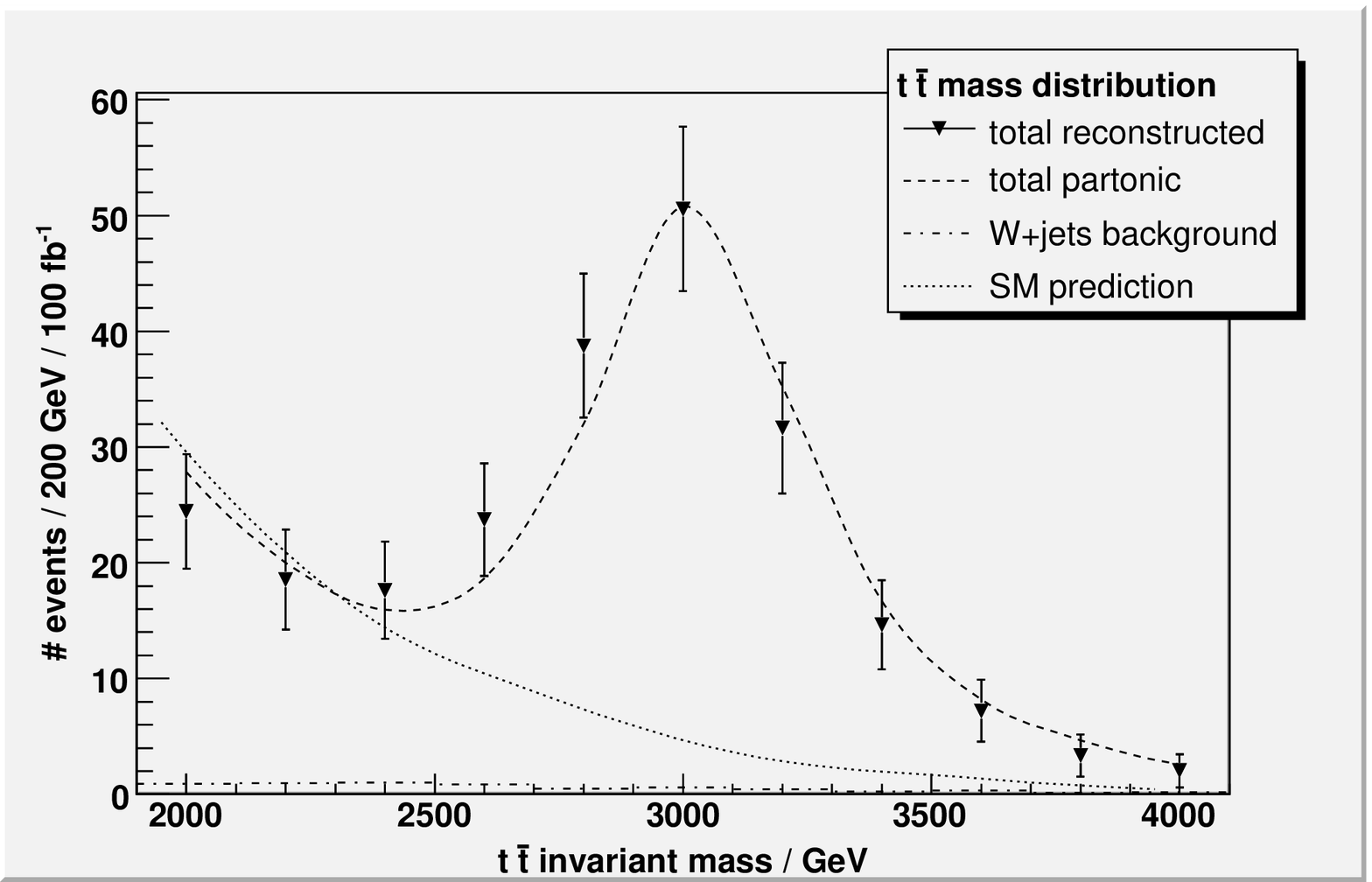}
\includegraphics[width=0.49\textwidth]{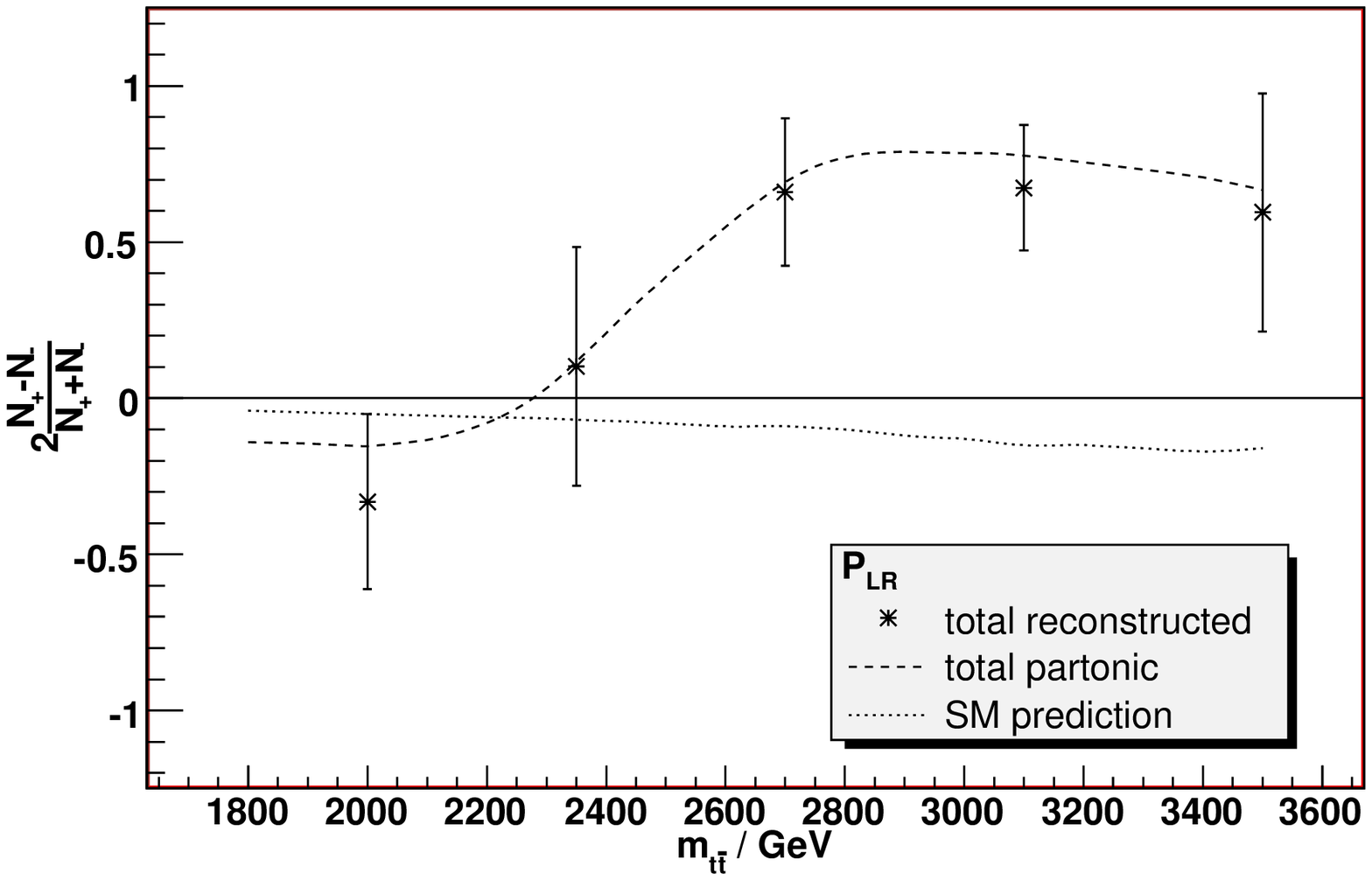}
\caption{Left panel: Invariant $t \bar t$ mass distribution
(left panel) and $P_{LR}$ (right panel).
In both panels $m^g_1 = 3$~TeV,
for parton-level signal+background (dashed) and SM prediction (dotted);
the $W jj$ background is also shown (dashed-dotted) in the left panel.  The error
bars correspond to the statistical uncertainties in the particle-level analysis.
These plots are from Ref.~\cite{KKgluon1}.}
\label{glukk1}
\end{figure}

At the LHC, the dominant $g_1$ production is through
the $u {\bar u}$ and $d \bar{d}$ initial states.  However,
the background $t \bar{t}$ is mainly from $gg$ fusion
and more forward-peaked.  Hence, a $p_T$
cut suppresses the background more than the signal.
In Ref.~\cite{KKgluon1}, the preferred reconstruction mode was
$t{\bar t} \to b {\bar b} jj \ell \nu$.  In order to minimize the impact of
parton distribution function (PDF) uncertainties, this work focused on
the differential cross section as a function of the $t {\bar t}$ invariant
mass $m_{t {\bar t}}$ and looked for a ``bump" in this distribution; due to
the large  width of the KK gluon a sharp resonance is not expected.  Particle and
parton level results from Ref.~\cite{KKgluon1} are shown
in the left panel of Fig.~\ref{glukk1}.
With a total efficiency of order 1\%, including $b$-jet-tagging (20\%)
and cut (about 20\%) efficiencies, as well as the branching fraction to the
final state (0.3),
it was concluded that $S/\sqrt{B}=11.0\,(4.2)$ and
$S/B=2.0\,(1.6)$ for $m^g_1 = 3\,(4)$~TeV
are achievable for \fb{100}, at the LHC \cite{KKgluon1}.  It is then concluded that the
LHC reach for the KK gluon $g_1$ is $m^g_1 \lsim 4$~TeV.

It is also noted in Ref.~\cite{KKgluon1}
that the enhanced coupling of $t_R$ (in this setup) to KK gluons provides an important handle
on the signal, since, whereas in the warped model the decay of the KK mode is largely
dominated by $t_R$, the SM background is dominated by QCD which
is left-right symmetric.
Here, due to the large boost of the top quark, its chirality is
preserved and can be deduced form its decay products: leptons tend to be
forward (backward)-emitted, in the
rest frame of  $t_{R (L)}$, relative to the direction of the top quark boost.  Defining
\beq
P_{LR} = 2 \times \frac{N_+ - N_-}{N_+ + N_-},
\label{PLR}
\eeq
where $N_\pm$ is the number of forward/backward positrons in the above sense,
the right panel of Fig.~\ref{glukk1} shows a
distinct asymmetry which can be correlated with
bumps in the invariant mass distribution.  Note that in the above setup, the
much smaller SM asymmetry, mostly from the left-handed weak coupling,
has the opposite sign in the signal region \cite{KKgluon1}.

\begin{figure}
\begin{center}
\includegraphics[angle=270,scale=0.3]{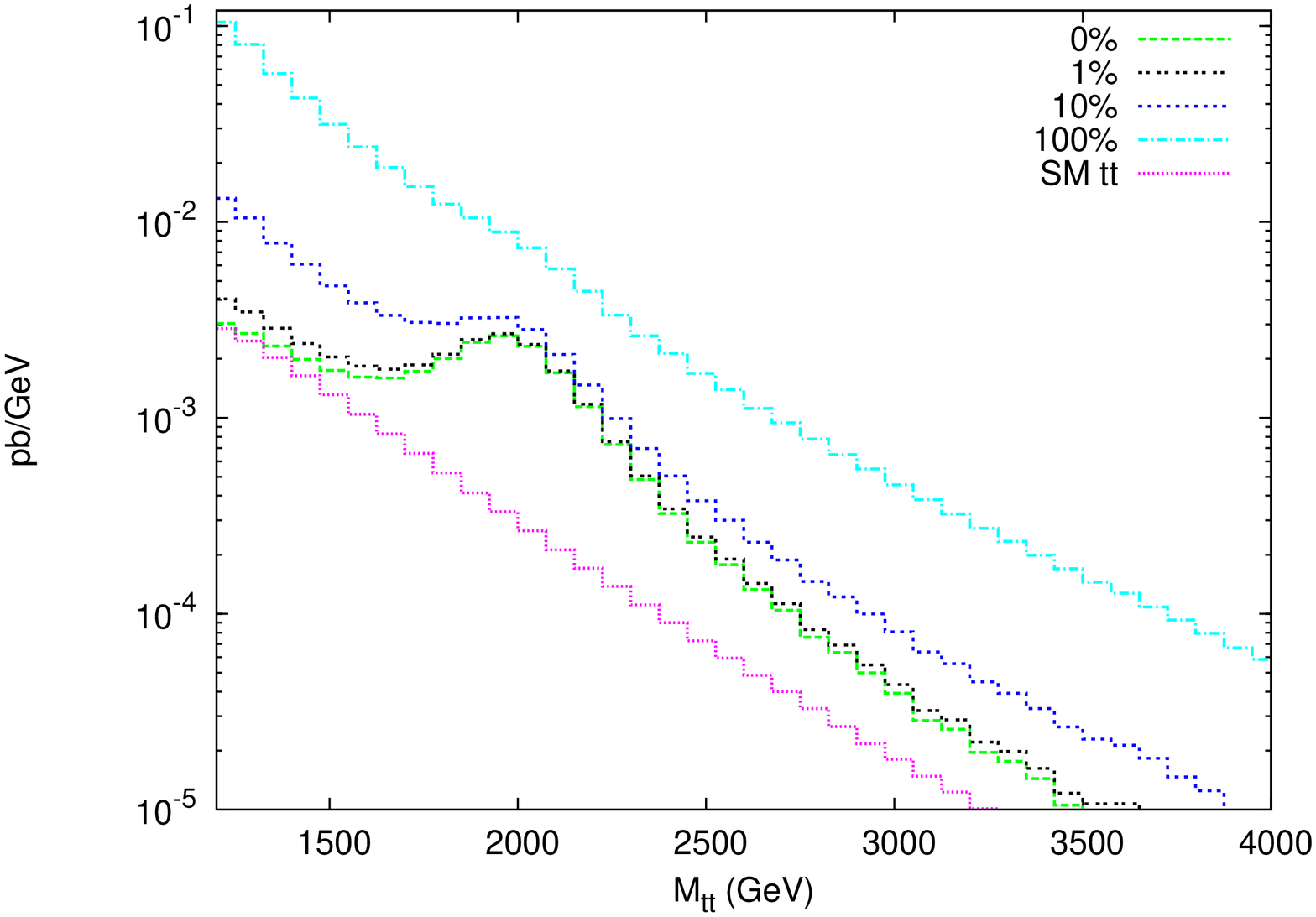}
\includegraphics[angle=270,scale=0.3]{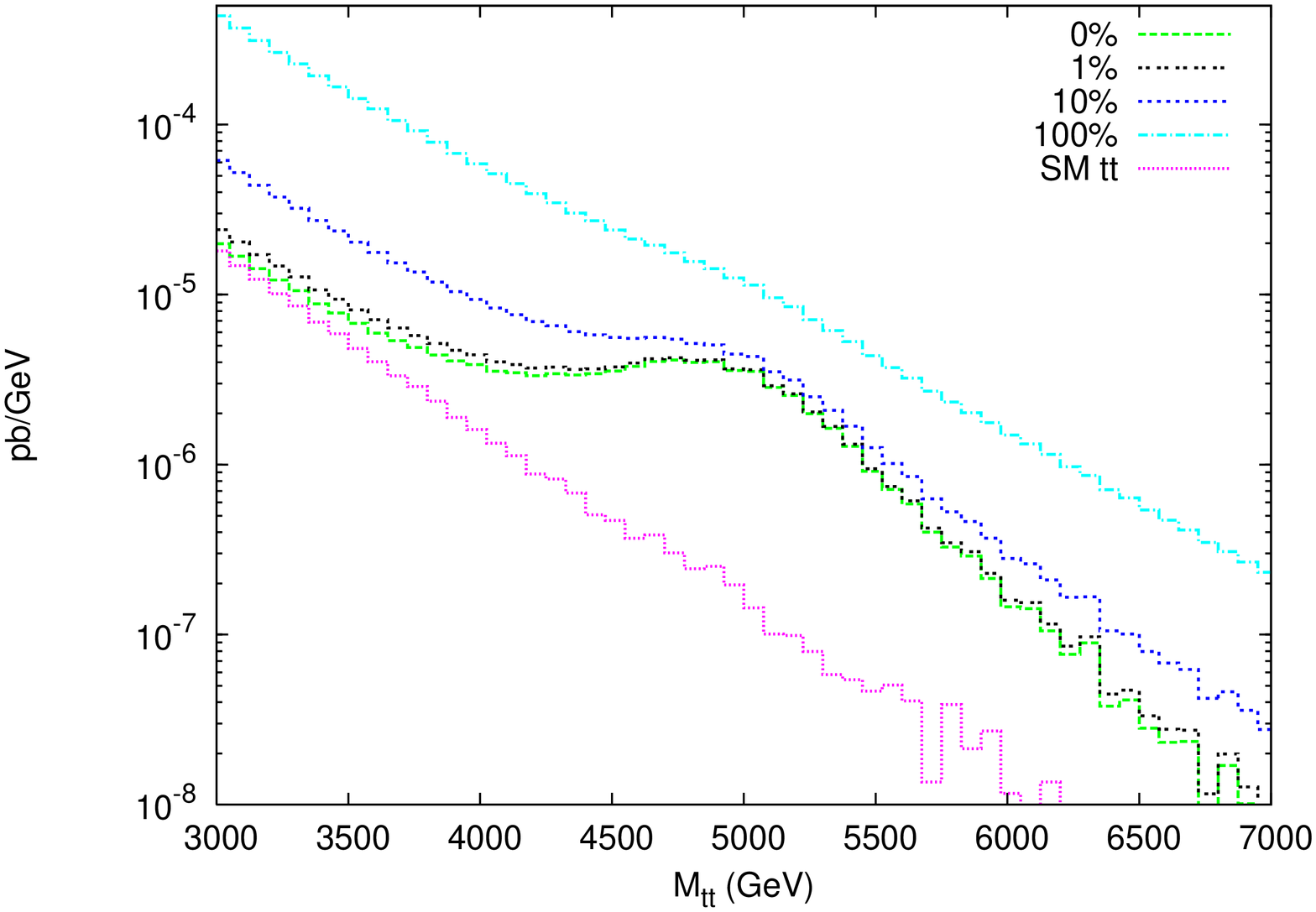}
\end{center}
\caption{$t\bar t$ invariant mass distribution for signal
plus some background, given different values of the top-tagging fake rate, for $m^g_1 =
2 (5)$~TeV in the left (right) panel.}
\label{glukk2}
\end{figure}
An independent study regarding the LHC reach for $g_1$ was performed in Ref.~\cite{KKgluon2},
using assumptions about the fermions profiles similar to that discussed above.  This resulted in
nearly the same level of suppression in the production, dominance of the $t_R$ branching
fraction, and total width.  Here, as in Ref.~\cite{KKgluon1}, it was
recognized that, for reasonable values of $m^g_1$,
the highly boosted final state tops will have collimated decay products.
Hence, the usual techniques of top identification will not be applicable
and new methods have to be developed to gain control over the background.

In Ref.~\cite{KKgluon2}, the effects of potential methods for
background suppression, generally requiring a detailed knowledge of jet morphology and
detector simulations, were parametrized by a range of signal dilution factors.  In Fig.~\ref{glukk2},
the $t \bar t$ mass distribution for $m^g_1 = 2\, (5)$~TeV is presented in the left (right) panel.  Here,
the two-object $t \bar t$ final states  are subject to QCD di-jets and $b \bar b$ backgrounds that
fake the signal.  The various curves in each panel represent different levels of fake rate and suggest
that signal extraction is possible if a background rejection factor of about 10 is achieved.  Ref.~\cite{KKgluon2}
concludes that based on their analysis, given \fb{100}, a conservative LHC reach for $g_1$ is about 5~TeV.
Note that this is a somewhat more optimistic conclusion than that reached above.  The authors of Ref.~\cite{KKgluon2}
attribute this to the different method, based on strong cuts, adopted in Ref.~\cite{KKgluon1} to identify
hadronic tops that could result in a more limited reach.  Here, we note that without a more detailed study based
on jet structure and realistic simulations of background rejection, it is difficult to determine
how conservative this extended reach is.  In any case, given that realistic values of KK gluon mass
lie above 2-3~TeV, we see that in either case, the LHC can potentially access interesting values of
$m^g_1$.
Identifying boosted tops are also studied in Refs.~\cite{otherboostedtop,toptagging}. 

The spin of the KK gluon provides another handle on the signal.  The prospects for
establishing the $1 + \cos^2\theta$ behavior of the signal top angular distribution was examined
in Ref.~\cite{KKgluon2}.  While this distribution has a characteristic forward-peaked structure,
the background is also strongly forward-peaked.  Strong $p_T$ cuts on reconstructed
tops or tagged top-jets could largely suppress the background
and enhance $S/B$.  However, we note that
a good grasp of top-tagging efficiency and background rejection
is needed before a firm conclusion can be made.  This work also considered the possibility
of establishing the chiral structure of the gluon KK coupling, in a fashion similar to that of
the preceding discussion, concluding that it is feasible to do so for $m^g_1$ up to
2-3~TeV, above which lepton isolation cannot be effectively achieved, making the
requisite measurements more challenging.

Ref.~\cite{Djouadi:2007eg} considers $t \bar t$ final state resulting from the
production of the gauge KK's. The observables are dominated by the KK gluon since
its cross section is much bigger than the EW KK's.
Similar to Ref.~\cite{KKgluon1} they show that the forward-backward asymmetry
of the top-quark can be significantly altered from the SM prediction.
Moreover, Ref.~\cite{Djouadi:2009nb} explains the $2\,\sigma$ deviation
in the $t\bar t$ forward-backward asymmetry at the Tevatron
by a 3~TeV KK gluon contribution, while the long-standing $3\,\sigma$
discripancy in the $b$-quark forward-backward asymmetry is improved
by the EW gauge KK contributions~\cite{Djouadi:2006rk}.

\underline{Associated production}: Refs.~\cite{Guchait:2007jd}~and~\cite{Djouadi:2007eg}
consider the associated production of the
KK gluon with either $t \bar t$ or $b \bar b$, with the KK gluon decaying to $t \bar t$.
Taking into account irreducible backgrounds, they find that the reach at the LHC
with $100~{\rm fb^{-1}}$ is about 3~TeV in this channel. They point out
that optimizing the cuts might lead to a better reach, although
dealing with combinatorics for the 4-top final-state
is no trivial matter.

\underline{New decay channels}: As mentioned in Part I, in models with custodial and $Z{\bar b}b$ coupling protection,  precision constraints seem to favor regions of parameter space where some of the KK excitations of the top quark are relatively light. The KK gluon can then decay into a pair of KK top ($t^1$) states with a significant branching fraction, and become a relatively broad resonance due to its large coupling to $t^1$~\cite{Carena:2007tn}.

Before closing this section, we note that, during initial phases, the LHC  
may operate at or below $\sqrt{s}=10$~TeV.  Given that the KK gluon is a promising 
mode of discovery for warped scenarios, here we would like to provide an estimate of the LHC reach for this 
state at the lower center of mass energy.  As a check of our estimate, first we consider the 
reach at  $\sqrt{s}=14$~TeV.  Consistent with the aforementioned 
results of Ref.~\cite{KKgluon1}, we find  
for $m^g_1 = 3$~TeV that the luminosity required at the LHC for a $5\,\sigma$ significance 
at $\sqrt{s}=14$~TeV is about $25~{\rm fb^{-1}}$, for cuts and efficiencies given therein. 
For $\sqrt{s}=10$~TeV, using a Monte Carlo simulation, we find that the
luminosity required for $5\,\sigma$ discovery increases to about $115~{\rm fb^{-1}}$, 
for similar cuts and efficiencies.


\section{Electroweak KK states (KK {\boldmath $W$ and $Z$})}
We restrict our discussion here to models where the electroweak gauge group in the bulk
is taken to be $SU(2)_L \otimes SU(2)_R \otimes U(1)_X$, with hypercharge being a
linear combination of $U(1)_R$ and $U(1)_X$.
The extra $SU(2)_R$ (relative to the SM) ensures suppression of the contributions to the
EWPT (especifically the $T$ parameter)~\cite{Agashe:2003zs}.
The detailed theory and LHC phenomenology of the neutral EW states are presented
in Ref.~\cite{Agashe:2007ki}, and that of the charged EW states in
Ref.~\cite{Agashe:2008jb} from which we will summarize the main aspects of their
results.
For just the SM gauge group in the bulk, see Ref.~\cite{Piai:2007ys} for a discussion
of the AdS/CFT correspondence and LHC signatures.

In these theories, there are three neutral electroweak gauge bosons
denoted as $W^3_L$, $W^3_R$ and $X$, and two charged gauge bosons denoted as
$W_L^\pm$ and $W_R^\pm$.
The $SU(2)_R \otimes U(1)_X \rightarrow U(1)_Y$
symmetry breaking by b.c.'s leaves one
combination of zero-modes in $(W^3_R,X)$ massless (the hypercharge gauge boson $B$)
while rendering the orthogonal combination ($Z_X$) massive.
In the charged sector this breaking leaves the $W_R^\pm$ without a zero-mode.
The SM Higgs doublet is promoted to a bi-doublet of
$SU(2)_L \otimes SU(2)_R$ with zero $U(1)_X$ charge and like in the SM, is responsible for
$SU(2)_L \otimes U(1)_Y \rightarrow U(1)_{EM}$ symmetry breaking by the Higgs VEV.  
This leaves one combination of zero-modes in $(W^3_L,B)$ massless (the photon $A$),
while making massive the orthogonal combination ($Z$), and in the charged sector
making the $W_L^\pm$ massive.

The bulk gauge fields can be expanded as a tower of KK states.
In each of these neutral and charged tower of states, we will restrict to the
zero and 1st KK modes only. The zero-mode is the SM, and we will denote
the 1st level KK neutral states by $A_1$, $Z_1$ and ${Z_X}_1$,
and the charged ones by $W_{ L_1 }^\pm$ and $W_{ R_1 }^\pm$.

EWSB mixes these states and the resulting mass eigenstates
are denoted by the
neutral $\tilde{A}_1$, $\tilde{Z}_1$ and $\tilde{Z}_{X_1}$,
and the charged $\tilde{W}_{L_1}$, $\tilde{W}_{R_1}$.
We will also collectively refer to the heavy neutral mass eigenstates as $Z'$,
and the charged ones as $W'$.
Note that the (EW preserving) KK masses for the first KK states
(for both the neutral and charged KK states)
are quite degenerate such that
the EWSB mixing (mass)$^2$
term is larger than KK (mass)$^2$
splitting for $m_{ KK } \stackrel{<}{\sim} 3.5$ TeV.
Hence, for the interesting range of KK masses, we expect large mixing
between $Z_1$ and ${Z_X}_1$ in the neutral sector (the $A_1$ does not mix),
with the heavy mass eigenstates roughly $50-50$ admixtures of $Z_1$ and ${Z_X}_1$
and a small admixture of $Z^{(0)}$. The lightest mass eigenstate is of course
identified as the SM $Z$ boson and is mostly the $Z^{(0)}$.
Similarly in the charged sector, the two heavy mass eigenstates are a large mixture
of $W_{R}^{(1)+}$ and $W_{L}^{(1)+}$ with a small admixture of $W^{(0)+}$.
Although the mixings between the first and zero levels are small,
it is important to keep these effects 
in Z' and W' decays to SM gauge bosons,
since they lead to $O(1)$ effects when the
small mixings are overcome by the enhancement due to
the longitudinal polarization of the energetic $Z$ and $W$ in the final state
(as expected from the equivalence theorem).

\underline{Couplings}:
As already mentioned, warped models can naturally explain the SM fermion mass hierarchy.
In such models, as shown schematically in Eq.~(\ref{ffA1}) ignoring small EWSB effects,
the light SM fermions have a small couplings to all KK's (including graviton)
based simply on the overlaps of the corresponding profiles,
while the top quark and Higgs have a large coupling to the KK's.
The exact couplings including EWSB effects are presented in 
Refs.~\cite{Agashe:2007ki,Agashe:2008jb}. 

\underline{Phenomenology}: Here we summarize the main results of
Refs.~\cite{Agashe:2007ki} and \cite{Agashe:2008jb}
that studied the KK $Z$ and $W$ comprehensively.
The results presented below are for the following choices of fermion bulk mass parameters:
$c_{ Q_L^3 } = 0.4$, $c_{ t_R } = 0$, and all the other $c$'s $> 0.5$,
and for $\xi = \sqrt{k\pi r_c} = 5.83$.
The $Zb\bar b$ coupling can be protected~\cite{Agashe:2006at} against
excessive corrections by making
the third generation quarks bidoublets under $SU(2)_L \otimes SU(2)_R$
by including extra non-SM fermions to complete the representation.
The $\wpri$ can decay to some of these extra non-SM fermions which is taken into account
in computing the BRs in Ref.~\cite{Agashe:2008jb}, but as a first-step
the analysis is based on SM final states only.

The total width of the KK $Z$ and $W$ are typically about 5 to 10\,\% of their mass.
For reasons already mentioned, they predominantly decay into heavier fermions (top),
and to the Higgs (including $W_L$ and $Z_L$, the longitudinal modes).
The main decay channels are:
$Z' \to $ $t \bar t$, $W_L W_L$, $Z_L h$, and $W' \to $ $t \bar b$, $W_L Z_L$, $W_L h$.
Although the $t\bar t$ BR can be large, the KK gluon is degenerate with the
electroweak states and have a much larger cross section into the
$t \bar t$ channel rendering this channel not very useful as a probe of the EW KK's.

\begin{figure}
\begin{center}
\includegraphics[angle=0,width=0.4\textwidth]{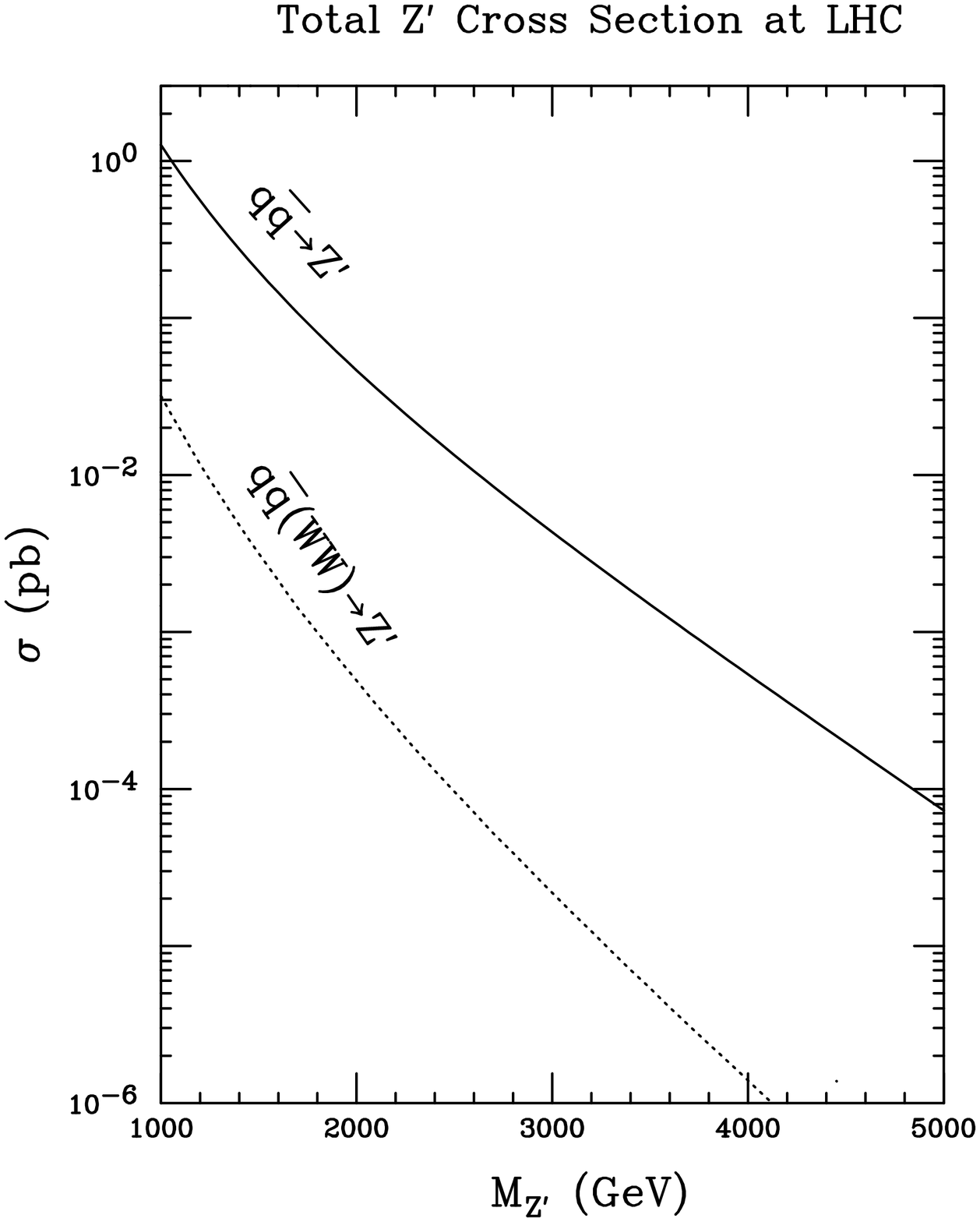}
\includegraphics[angle=0,width=0.4\textwidth,height=0.36\textheight]{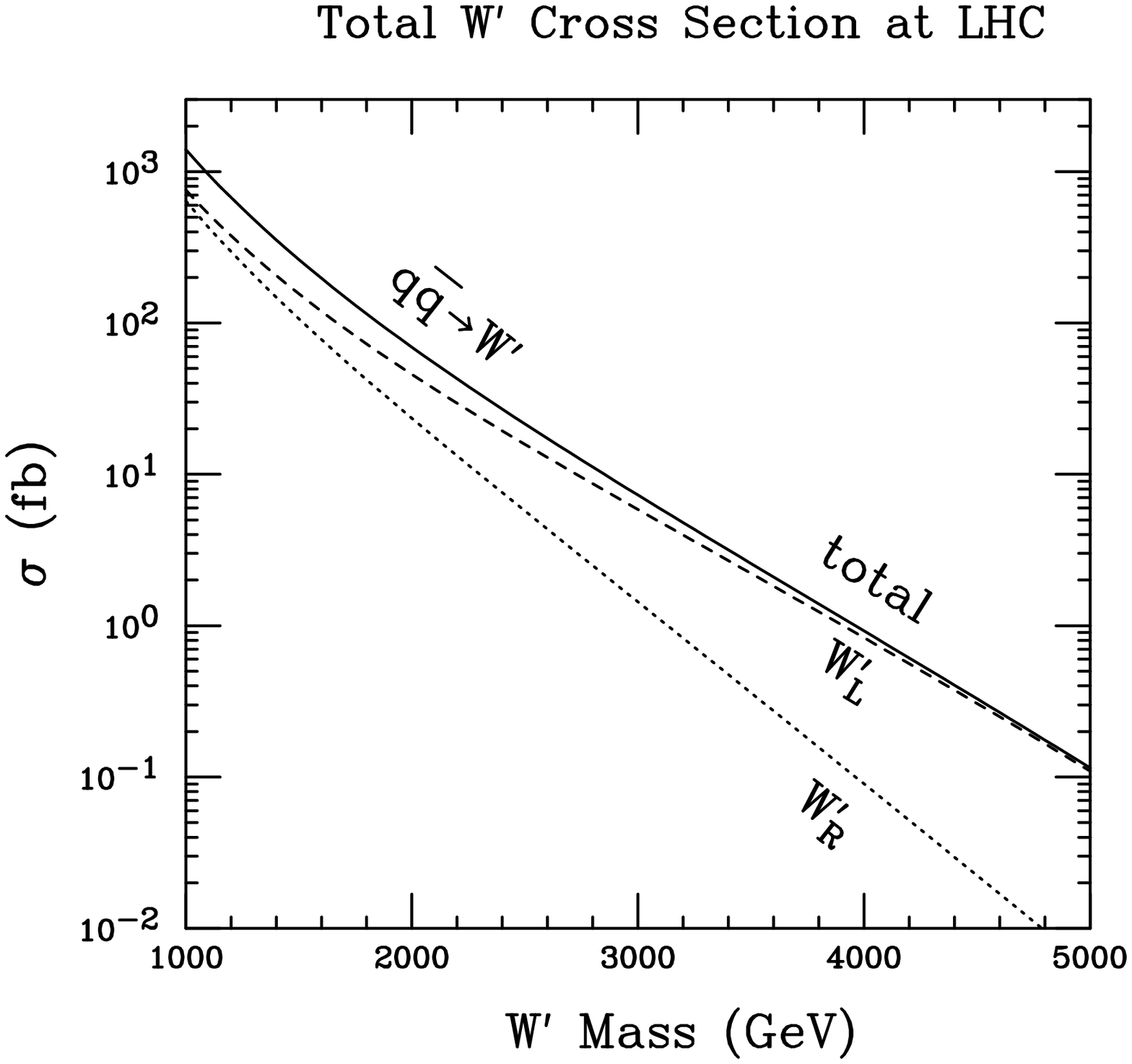}
\caption{Total cross section for $\zpri$ (left) and $\wpri$ (right) production versus
its mass (from Refs.~\cite{Agashe:2007ki}~and~\cite{Agashe:2008jb} respectively).
\label{xsmassz.FIG}}
\end{center}
\end{figure}

\underline{LHC signatures}:
We summarize here the LHC signatures analyzed in Refs.~\cite{Agashe:2007ki,Agashe:2008jb}.
The total cross section for
$pp\to Z'$ and $W'$ at the LHC with $\sqrt{s}=14~$TeV
are shown in Fig.~\ref{xsmassz.FIG} as a function of its mass.
Drell-Yan type production is dominant with vector boson fusion channel about an order of magnitude
smaller. We consider next the dominant decay modes, their signatures at the LHC and the reach
for their discovery.

\begin{figure}
\begin{center}
\includegraphics[angle=270,width=0.45\textwidth]{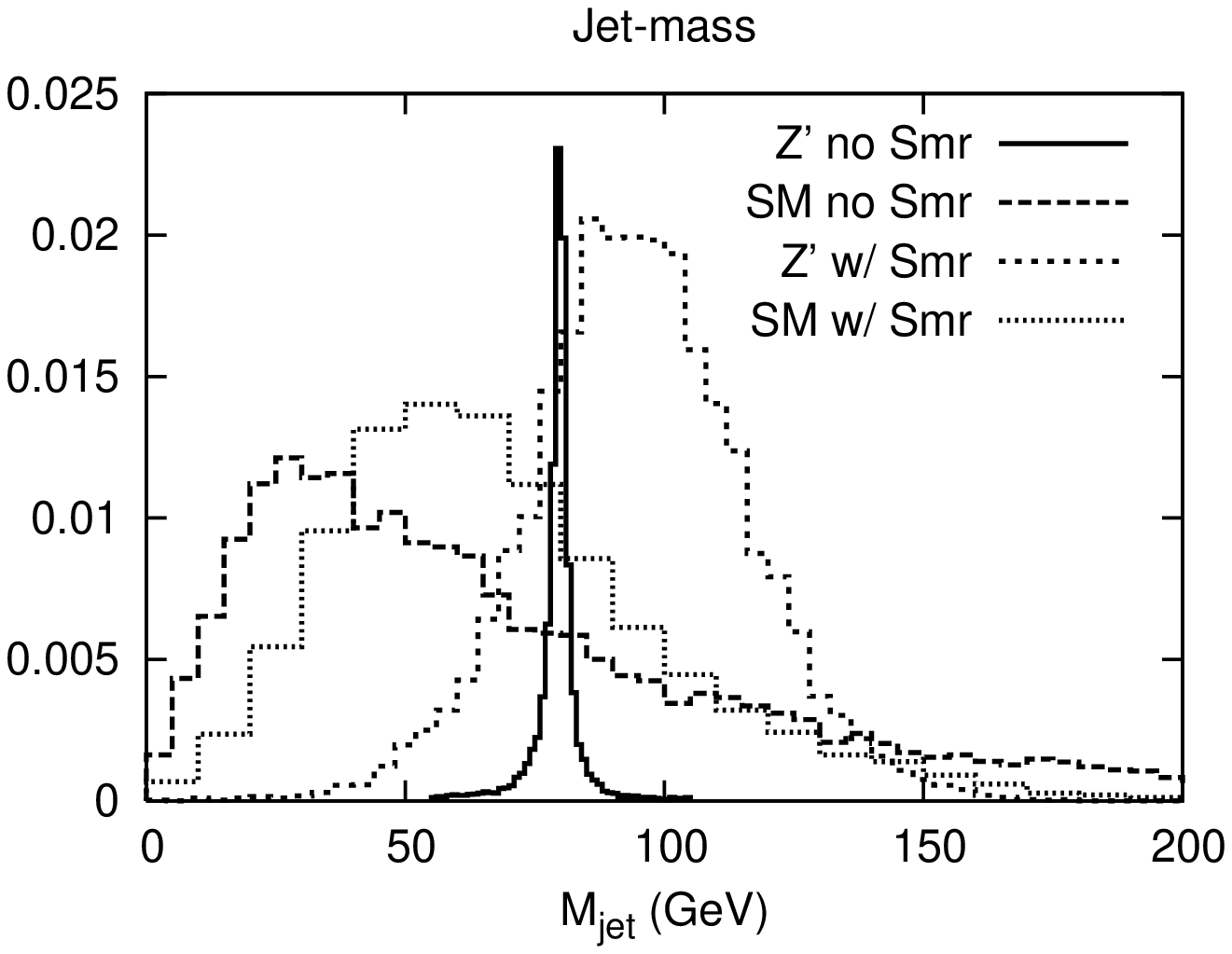}
\includegraphics[angle=270,width=0.45\textwidth]{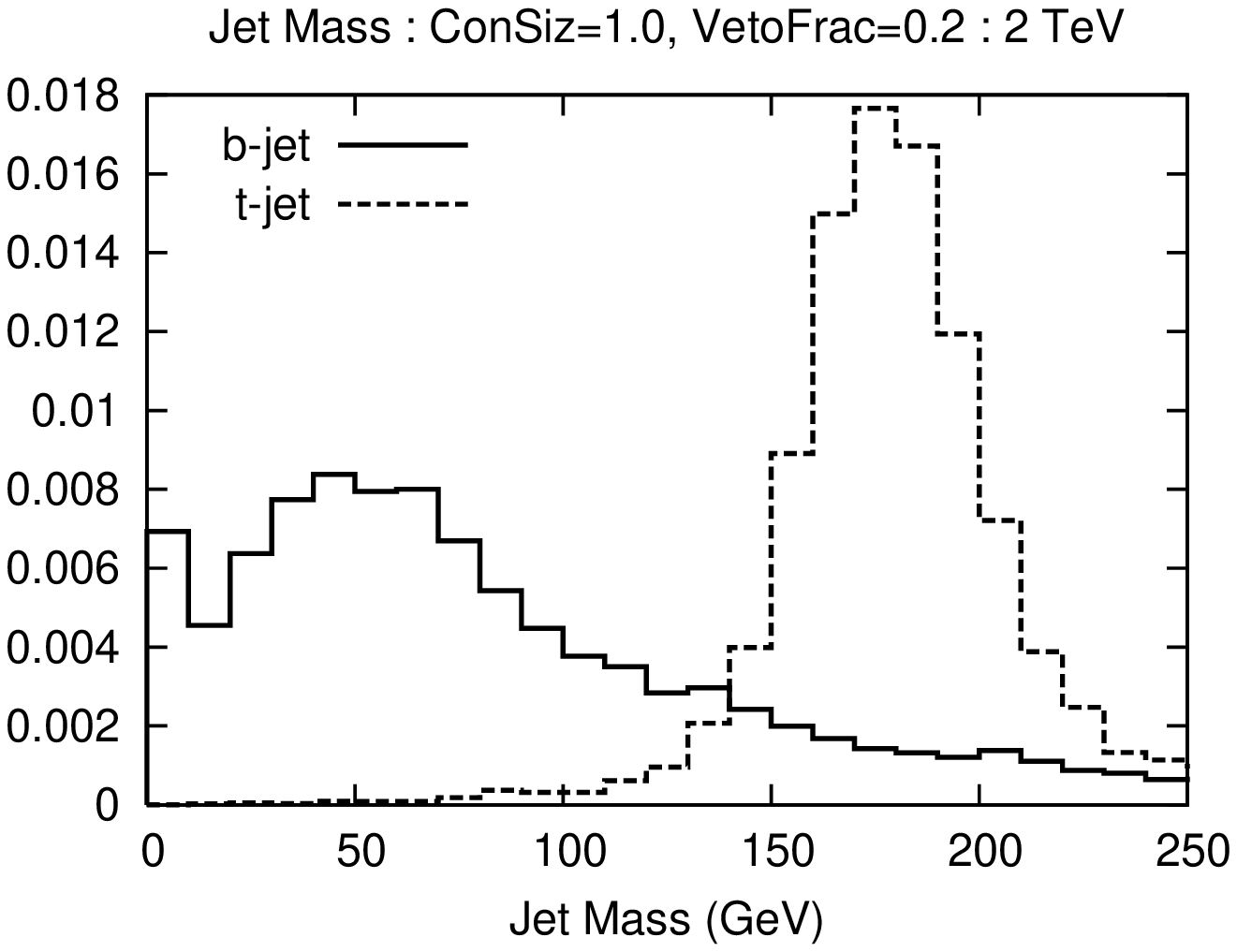}
\caption{
Jet-mass distributions for the $W\to jj$ vs. QCD jet for a cone-size of 0.4
with and without $E$, $\eta$ and $\phi$ smearing (left),
and for $t$ vs. $b$-jet for a cone-size of 1.0 (right).
Both are for a 2~TeV KK gauge boson.
\label{jetM.FIG}}
\end{center}
\end{figure}

Owing to the large mass of the KK $Z$ and $W$, the SM final states they decay into are
significantly boosted resulting in their decay products highly collimated in the lab-frame.
For example, for the semi-leptonic $Z'\to WW$, the presently typical jet reconstruction
cone size of $\Delta R=0.4$ will cause the two jets from the $W$ decay
to be reconstructed likely as a single jet (albeit a fat jet), impeding our ability
to reconstruct a $W$ mass peak.
This means that we would pick up a SM QCD jet background which typically is large,
and would require special consideration to keep it from overwhelming the signal.
In order to discriminate the merged jets of the $W$ from a QCD jet,
one can use the jet-mass, which is
the combined invariant mass of the vector sum of 4-momenta of all hadrons
making up the jet, as shown in Fig.~\ref{jetM.FIG} (left).
Techniques to discriminate against QCD jets are also studied in
Refs.~\cite{jmassRefs,Almeida:2008yp}. 
Although the two jets from $W\rightarrow jj$ are severely overlapping in the hadronic
calorimeter, it may be possible to utilize the better granularity of the 
electromagnetic calorimeter and the tracker to resolve "sub-jets"~\cite{SubJets}
and obtain a reasonable discriminating power against QCD.
Similarly, in the $W'\to t \bar b$ channel, $t\bar t$ production can become a source of
background since a highly collimated top can fake a $b$-jet.
A top and a $b$-jet can again be discriminated against by using
the jet-mass variable as shown in Fig.~\ref{jetM.FIG} (right).

Kinematic cuts can be applied to maximize the signal and suppress the background.
We refer the reader to Refs.~\cite{Agashe:2007ki} and \cite{Agashe:2008jb} for distributions
of various kinematic variables for many of the dominant final-states and cuts based on them.
\begin{table}
\begin{center}
\caption{Summary of the best channels (from Refs.~\cite{Agashe:2007ki,Agashe:2008jb})
for the $\zpri$ (top-table), and,
$\wpri$ (bottom-table), giving the luminosity and significance for the mass shown.
For the $\wpri \to t\, b$ channel the numbers without (and with) the reducible $t\bar t$
background are shown.
\label{sumry.TAB}}
\begin{tabular}{|c|c|c|c|c|}
\hline
Z' Channel&
$\mzpri$ (TeV)&
${\cal L}$ ($fb^{-1}$)&
$\frac{S}{B}$&
Significance ($\sigma$) \tabularnewline
\hline
\hline
$W\, W\to\ell\nu j\, j$&
$3$&
$1000$&
$0.2$&
$4.6$ \tabularnewline
\hline
$m_{h}=120$:  $Z\, h\to\ell\ell b\bar{b}$ &
$3$&
$1000$&
$2$&
$5.7$\tabularnewline
\hline
$m_{h}=150$: $Z\, h\to(jj)\,(jj)\,\ell\nu$ &
$3$&
$300$&
$1.2$&
$4.7$\tabularnewline
\hline
\end{tabular}
%
%
%
%
\begin{tabular}{|c|c|c|c|c|}
\hline
W' Channel&
$\mwpri$ (TeV)&
${\cal L}$ ($fb^{-1}$)&
${S}/{B}$&
Significance ($\sigma$) \tabularnewline
\hline
\hline
$t\, b\to\ell\nu b\bar{b}$&
$2$&
$1000$&
$0.4\,(0.2)$&
$3.4\,(2.5)$ \tabularnewline
\hline
$Z\, W\to\ell\ell\ell\nu$&
$3$&
$1000$&
$10$&
$6$ \tabularnewline
\hline
$m_{h}=120$:  $W\, h\to\ell\nu b\bar{b}$ &
$3$&
$300$&
$2.4$&
$6.2$ \tabularnewline
\hline
$m_{h}=150$: $W\, h\to(jj)\,\ell\nu\,(jj)$ &
$3$&
$300$&
$4$&
$8$ \tabularnewline
\hline
\end{tabular}
\end{center}
\end{table}
We summarize, in Table~\ref{sumry.TAB}, the LHC reach after suitable cuts
for the $Z'$ and $W'$ found in those studies.
Where the number of events is small, Poisson statistics is used
to find the significance and the equivalent Gaussian significance is quoted.
We see that upwards of $300~{\rm fb^{-1}}$ is needed to probe a 3~TeV KK $Z$ or $W$ state.

In what we have discussed so far, owing to the flavor connection,
the BR into experimentally clean
leptonic channels are quite small (about $10^{-3}$) rendering them useless,
and one had to work with more complicated final states.
However, if warped models are taken as a theory of flavor alone 
(and not generating the gauge-hierarchy),
the UV scale can be lowered to as low as $\ord{10^3}$~TeV ~\cite{Davoudiasl:2008hx} 
(a more detailed study of flavor constraints 
in Ref.~\cite{Bauer:2008xb} finds the minimum UV cutoff to be several $10^3$~TeV) 
instead of the usual Planck-scale.  Such ``truncated" models result in $Z'$ leptonic BR's that are 
big enough to lead to a very good significance
at the LHC, in the di-lepton channel~\cite{Davoudiasl:2008hx,Davoudiasl:2009jk}.
A big leptonic cross section due to $Z'$ exchange is also found in
Ref.~\cite{Ledroit:2007ik} where
a model is presented in which the left-handed light quarks and leptons are peaked
toward the IR brane giving much larger light fermion couplings to the $Z'$.
However, this comes at the price of requiring the bulk mass $c_L$ parameters
for the three generations to be highly degenerate in order to keep FCNC's under control.

\section{KK fermions}

The presence of KK excitations of SM fermions is a generic prediction of warped models with bulk fermions
and their discovery would be one of the ``smoking gun" signatures.
For example, see Ref.~\cite{Davoudiasl:2007zx} for how the KK fermion spectrum correlates with the
SM fermion masses.
However, Ref.~\cite{Davoudiasl:2007wf} points out that discovering these modes would likely be possible
only in a future collider and not at the (even upgraded) LHC.
This is due to the small single-production cross section and their large mass, as they
are heavier than gauge KK modes which are constrained by precision EW measurements.
For a model independent analysis at the Tevatron of heavy vector-like fermions with
significant mixings with SM fermions, see Ref.~\cite{Atre:2008iu}.

\begin{figure}
\begin{center}
\includegraphics[angle=0,width=0.6\textwidth]{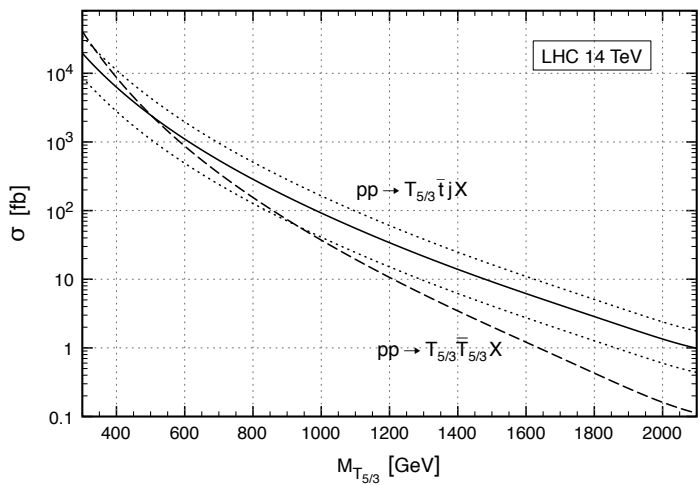}
\caption{
Production cross sections at the LHC for $T_{5/3}$ as a function of its mass.
The dashed line refers to pair-production; the solid and the two dotted curves refer to
single production for the three values of the coupling.
From Ref.~\cite{Contino:2008hi}.
\label{csT53.FIG}}
\end{center}
\end{figure}

As explained earlier, in order to obtain custodial protection of $Z b \bar b$ coupling,
the third generation quarks can be extended to be bi-doublets under $SU(2)_L \otimes SU(2)_R$,
which then necessitates the introduction of extra non-SM fermions to complete the representation 
\cite{Agashe:2006at}.  The presence of these non-SM fermions is model-dependent, but
some of them could be fairly light (100's of GeV)~\cite{Choi:2002ps,Agashe:2003zs} 
(see Ref.~\cite{Agashe:2004ci} for sample numerical values).  
Several works have considered the discovery of these light non-SM fermions at the LHC.
In models with custodial protection of $Z b \bar b$, the $t_R$ could be in either a singlet of
$SU(2)_L \otimes SU(2)_R$ or be in $(1,3) \oplus (3,1)$. In the latter case, one of the
custodial partners of the $t_R$, denoted $\tilde{b}_R$ can be fairly light,
and Ref.~\cite{Dennis:2007tv} considers its LHC signatures by looking at 4-$W$ events.
With $10~{\rm fb^{-1}}$, they show that the peak in the di-jet mass distribution stands out
above the background.
Ref.~\cite{Contino:2008hi} considers the single and pair-production of
the custodial partner $T_{5/3}$ (of the SM left-handed quark doublet) at the LHC.
In Fig.~\ref{csT53.FIG} we show from Ref.~\cite{Contino:2008hi} the cross-sections
as a function of its mass.  
They show that in the {\em same-sign} di-lepton channel, discovery at the LHC could come with less than
100~${\rm pb^{-1}}$ (20~${\rm fb^{-1}}$) of integrated luminosity for a mass of 500~GeV (1~TeV).  
See also
Ref.~\cite{AguilarSaavedra:2009es} for a study of LHC signatures of
vector-like quarks with these quantum numbers in single and multi-lepton
channels. In models with custodial protection a light
top KK mode $t^1$ plays an important role in relaxing the constraints from
EWPT~\cite{Carena:2006bn,Carena:2007ua,Medina:2007hz}.
Taking into account the $s$-channel KK gluon exchange, which interferes
constructively with the gluon diagram, this mode may be detected via pair
production up to a mass of 1.5~TeV, at the LHC with about
\fb{300}~\cite{Carena:2007tn}.

Ref.~\cite{Bouchart:2008vp} includes the effects of mixings between
SM fermions with KK excited fermions and mixings in the gauge sector, and shows that this
leads to a better fit to EW observables compared to the SM, including explaining the
discrepancy in $A_{FB}^b$. The effects from fermion mixing in $A_{FB}^b$ is in addition to
the contribution from the KK gluon already discussed in Sec.~\ref{KKgluon.SEC}.
They find, in this class of models, that
quite large values of the Higgs mass (about 500~GeV) still give acceptable EW fits.

\section{The Radion}
The radion is a scalar field associated with fluctuations in the size of the extra dimension.
The mass of the radion is dependent on the mechanism that stabilizes  
the size of the extra dimension.  This was first achieved in a simple 
model with a bulk scalar (with its own dynamics and constraints) in 
Ref.~\cite{Goldberger:1999uk}, where it can be shown that, generically, the radion 
may be expected to be the lightest new state in an RS-type setup
\cite{Goldberger:1999un}.   
This stabilization mechanism 
was further elaborated in Refs.~\cite{DeWolfe:1999cp,Csaki:2000zn}; for an alternative 
mechanism based on Casimir energy associated with a compact dimension 
see, for example, Ref.~\cite{Garriga:2002vf} 
(an earlier attempt can be found in Ref.~\cite{Goldberger:2000dv}).     
The radion interactions with SM fields, being of 5D gravitational nature, 
arise through operators of dimension-5 and higher characterized by a scale 
$\Lambda\sim$~TeV~\cite{Goldberger:1999un}.  
For the most part, the radion couplings are similar to Higgs couplings.        
The radion mass is expected to be a few tens to hundreds of GeV to have escaped
detection at LEP and be consistent with precision electroweak 
data~\cite{Csaki:2000zn,Gunion:2003px}. 
This also implies that no observable deviations from Newton's law in
torsion balance experiments are expected~\cite{Tanaka:2000er}.
However, the radion field can mix with the Higgs boson after EWSB
which involves another 
parameter, the coefficient of the curvature-scalar term~\cite{Giudice:2000av}.  

LHC signatures of the radion ($r$) have been analyzed both in
the case when SM fields are IR localized (original RS model), 
as well as when the SM is in the bulk, 
and the search methodology usually parallels Higgs searches. 
Similar to the Higgs, the main production channel is $g g \to r$, induced at 
loop level in the original RS model, given for example in Ref.~\cite{Giudice:2000av}. 
They find that the ratio of the 
radion significance (in $g g \to r$ followed by  $r \to \gamma \gamma$ and $r \to Z Z \to 4\ell$)
to the corresponding SM Higgs significance ($R_S$) at the LHC
can vary from about ten to a hundredth 
as $\Lambda$ varies from 500~GeV to 20~TeV 
(for the coefficient of the curvature-scalar term set to zero). 

For the case of the SM in the bulk there is a tree-level coupling of the radion to
massless gauge bosons (including the gluon) as pointed out in Ref.~\cite{Rizzo:2002pq}, 
and the radion couplings to all SM fields including loop induced couplings can be found 
in Ref.~\cite{Csaki:2007ns}.
\begin{figure}
\begin{center}
\includegraphics[angle=0,width=\textwidth]{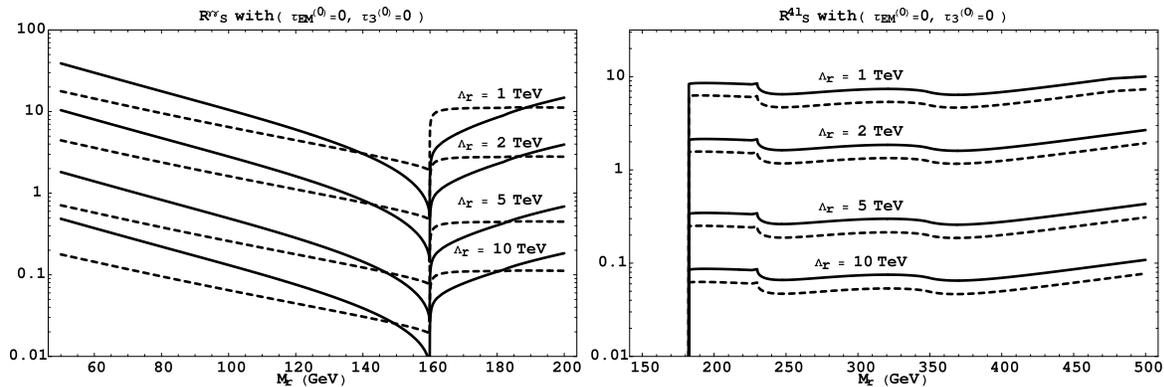}
\caption{
The ratio of the radion significance to the SM Higgs significance in the 
$\gamma \gamma$ (left) and $ZZ\to 4\ell$ (right) channels at the LHC. 
The solid lines are for the case of the SM fields in the bulk, while the dashed are 
for IR localized SM fields. 
From Ref.~\cite{Csaki:2007ns}.
\label{radion_RS.FIG}}
\end{center}
\end{figure}
$R_S$ from Ref.~\cite{Csaki:2007ns} is shown in Fig.~\ref{radion_RS.FIG} and its variation is
similar to that of the original RS case. 
Here, the curvature-scalar term is assumed to be zero.
The $\gamma \gamma$ channel (left) and $ZZ\to 4\ell$ (right)
are shown both for the SM in the bulk (solid lines) and SM fields IR localized (dashed lines).
The dependence on $\Lambda$ is also shown. 
Ref.~\cite{Toharia:2008tm} shows that the radion BR into $\gamma \gamma$ can be quite
dramatically enhanced in the bulk SM case for non-zero curvature-scalar coupling. 

Ref.~\cite{Azatov:2008vm} explores an FCNC observable mediated by
the radion $r \to t c$ at the LHC and finds that an intersting region of parameter space
can be probed with $300~{\rm fb^{-1}}$. For a related process, Ref.~\cite{Agashe:2009di} has shown
that the Higgs FCNC $BR(h\to t c)$ can be about $5\times 10^{-3}$ in warped models.  
They also find that $BR(t\to h c)$ can be about $10^{-4}$.  These effects can be looked for at the LHC.


\section{Conclusions}

We began this review with a detailed presentation of various
techniques that are useful in analyzing the physics of theories with
extra dimensions, in particular warped extra dimensions.  These are
the KK description (useful for discussing the collider phenomenology)
and techniques that allow to resum the low-energy effects of the new
physics (propagator and holographic methods), as well as their
relations.  An important application is to the determination of
indirect bounds on the scale of new physics from EW precision
constraints.  Models with bulk gauge and fermion fields (that allow
addressing the flavour puzzle) together with a custodial symmetry, can
be consistent with precision measurements with gauge KK excitation
around $3~{\rm TeV}$.  Additional flavour constraints (not reviewed
here) can result in stronger bounds, although these can be consistent
with the above KK scale with a moderate amount of fine-tuning, or with
additional flavour structure.

In surveying the collider phenomenology of warped 5D models, we
largely focused on their KK signals, both in the original RS model
(with the SM content on the IR-brane) and in models that can explain
the flavor puzzle within a 5D version of the SM. In these latter
models, suppressed couplings to the SM zero modes make the collider
discovery of the warped resonances significantly more difficult.  In
the simplest models, the KK gluons have the best prospects for
discovery, up to masses of about 4~TeV, but the KK gravitons, a
distinct feature of warped models, would likely lie outside the reach
of the LHC experiments.  Hence, the upcoming LHC experiments can probe
interesting regions of parameters in realistic warped hierarchy/flavor
models.  Improved analysis techniques, such as those relevant to the
hadronic decays of heavy boosted particles, can generally enhance the
discovery prospects for the warped KK modes.

{\it Acknowledgments: } 
J.~S. would like to acknowledge useful
discussions with M. Aybat, A. Carmona, 
R. Contino and especially M.
P\'erez-Victoria.
H.~ D. and S.~G. thank F.~Paige for helpful 
conversations.  H.~ D. and S.~G are supported by the DOE grant
DE-AC02-98CH10886.  E.P. is supported by DOE under contract
DE-FG02-92ER-40699.  J.S. is supported by the SNSF 
under contract 200021-117873.


\appendix


\section{Low-energy expansions}

We collect here some useful results for the functions involved in the
calculation of EWPT in models in AdS$_5$.  We will always consider
$L_0 \ll L_1$ and expand to leading order in $L_0/L_1$.  We also
define the dimensionless parameter $\hat{m} \equiv m L_0$, (where $m$
is the mass appearing in the IR boundary conditions).

The function $K_m(p,z)$ defined by Eqs.~(\ref{DEKS}) and (\ref{bcsK}),
which corresponds to the gauge boson holographic profile, is
explicitly given by
\beqa
K_{m}(p,z) &=& a(z)^{-1} \, \frac{J^z_{1} + B \, Y^z_{1}}{J^0_{1} + B \, Y^0_{1}}~,
\label{solK}
\eeqa
where $B = -\tilde{J}^{\rm IR}_{1}/\tilde{Y}^{\rm IR}_{1}$, and we
used the definitions of the Bessel functions, Eqs.~(\ref{Bessel}) and
(\ref{Besseltilde}), but with $m_{n} \rightarrow p$.  The vacuum
polarization for the boundary field in the holographic method is
\begin{eqnarray}
\fl
K_m^\prime(p,L_0)& =&
-\frac{L_0 \hat{m}}{L_1^2} \left[1-\frac{\hat{m}}{2}\right] 
+ p^2 L_0 \left[\log\frac{L_1}{L_0} -
  \frac{\hat{m}}{2} \right] 
+ \frac{p^4}{2} \frac{L_0 L_1^2}{2}
\left[1-\frac{5}{8} \hat{m} \right] + \cdots
\label{Pim:expansion}
\end{eqnarray}
For an IR localized Higgs, it is sufficient to consider the bulk
profile for $z \gg L_{0}$.  Expanding for small momenta and small
$\hat{m}$ one finds
\begin{eqnarray}
\fl
K_m(p,z) & =&
1- \frac{z^2}{2L_1^2} \hat{m} \left( 1 - \frac{\hat{m}}{2} \right)
+ p^2 \frac{z^2}{4}  
\left[
1+2 \log\frac{L_1}{z}
+ \hat{m} \left(\frac{z^2}{4L_1^2}-1\right) 
\right] 
\nonumber \\
\fl
&& 
+ \frac{p^4}{2}
\frac{z^2 L_1^2}{32}
\left[8-\frac{z^2}{L_1^2}
\left(5+4 \log \frac{L_1}{z} \right)
+\hat{m} \left(
-5 + 2\frac{z^2}{L_1^2} - \frac{z^4}{6 L_1^4}
\right ) \right] + \cdots
\label{Km:expansion}
\end{eqnarray} 

Similarly, the function $S(p,z)$ defined by Eqs.~(\ref{DEKS}) and
(\ref{bcsS}), which is independent of $\hat{m}$, is
explicitly given by
\beqa
S(p,z) &=& - \frac{\pi z}{2} \left( Y^0_{1} J^z_{1} - J^0_{1} Y^z_{1} \right)~,
\label{solS}
\eeqa
again with $m_{n} \rightarrow p$ in the definitions~(\ref{Bessel}). It can be
expanded for small $p$ as
\begin{equation}
\fl
S(p,z)=\frac{L_0}{2} \left( \frac{z^2}{L_0^2} -1 \right)
+ p^2 L_0^3 \left[ \frac{1}{16} \left( 1 - \frac{z^4}{L_0^2}\right)
+ \frac{z^2}{4L_0^2} \log \frac{z}{L_0} \right] + \ldots
\label{S:expansion}
\end{equation}

For completeness, we also record the fermion holographic profile [see
Eqs.~(\ref{Hol:Fermion}) and (\ref{BC:Hol:Fermion})]:
\beqa
f_{L}(p,z) &=& a(z)^{-5/2} \, \frac{J^z_{c+1/2} + 
B_{f} Y^z_{c+1/2}}{J^0_{c+1/2} + B_{f} Y^0_{c+1/2}}~,
\label{solfL}
\eeqa
where $B_{f} = -\tilde{J}^{\rm IR}_{c+1/2}/\tilde{Y}^{\rm
IR}_{c+1/2}$, and $m_{n} \rightarrow p$ in the
definitions~(\ref{Bessel}) and (\ref{Besseltilde}).  The comments
about IR localized terms made after Eq.~(\ref{Fermionf}) apply.  The
RH profile can be obtained from $p f_{R}(p,z) = O_{c} f_{L}(p,z)$,
where $O_{c}$ was defined in Eq.~(\ref{FirstOrderOp}).

Finally, we also used in the main text the auxiliary functions
\beqa
g_n(c)& \equiv &\int_{L_0}^{L_1} dz\, 
\left(\frac{z}{L_1}\right)^n 
a(z)^4 \big[f_L^0(z)\big]^2
\nonumber \\
&=& 
\left(\frac{L_0}{L_1}\right)^n \frac{2c-1}{2c-n-1}
\frac
{1-\left(\frac{L_0}{L_1}\right)^{2c-n-1}}
{1-\left(\frac{L_0}{L_1}\right)^{2c-1}}~,
\label{gn} \\
\tilde{g}_n(c) &\equiv& \int_{L_0}^{L_1} dz\, 
\log \left(\frac{z}{L_1}\right)\left(\frac{z}{L_1}\right)^n 
a(z)^4 \big[f_L^0(z)\big]^2
\nonumber \\
&=& 
\left(\frac{L_0}{L_1}\right)^n \frac{2c-1}{(2c-n-1)^2}
\frac
{1-(2c-n-1)\log\frac{L_1}{L_0}-\left(\frac{L_0}{L_1}\right)^{2c-n-1}}
{1-\left(\frac{L_0}{L_1}\right)^{2c-1}}~.
\label{gtn}
\eeqa
Note that as $c \to + \infty$, $g_n(c) \rightarrow (L_{0}/L_{1})^n$
and $\tilde{g}_n(c) \rightarrow (L_{0}/L_{1})^n \ln L_{0}/L_{1}$,
which are exponentially small for positive $n$.  Expanding the
functions of Eq.~(\ref{bulkcouplings}) for small $p^2$ using
Eq.~(\ref{Km:expansion}), one gets
\beqa
\bar{g}_i(p^2) 
&=&
\bar{g}_i(0) + p^2 \bar{g}_i^\prime(0) + \frac{1}{2} p^4
\bar{g}_i^{\prime\prime}(0) + \cdots~,
\eeqa
where 
\begin{eqnarray}
\bar{g}_i(0)&=&
1- \frac{1}{2} \hat{m} \left( 1 - \frac{1}{2} \hat{m} \right) \, g_2~, 
\\
\bar{g}_i^\prime(0)&=&
\frac{L_1^2}{4} \left[g_2-2\tilde{g}_2+\hat{m} \left( 
\frac{1}{4} g_4-g_2\right)\right]~, 
\\
\bar{g}_i^{\prime\prime}(0)&=&
\frac{L_1^4}{32}\left[
8g_2-5g_4+4\tilde{g}_4 + \hat{m} \left(
-5 g_2+2g_4-\frac{1}{6} g_6 \right)\right]~,
\end{eqnarray}
and $\hat{m}$ corresponds to $\hat{m}$, $0$ and $\hat{m}/c^2$ for
$f=W,V,A$, respectively.


\section*{References}


\begin{thebibliography}{10}


\bibitem{Randall:1999ee}
  L.~Randall and R.~Sundrum,
  Phys.\ Rev.\ Lett.\  {\bf 83}, 3370 (1999)
  [arXiv:hep-ph/9905221].

\bibitem{Goldberger:1999uk}
  W.~D.~Goldberger and M.~B.~Wise,
  Phys.\ Rev.\ Lett.\  {\bf 83}, 4922 (1999)
  [arXiv:hep-ph/9907447].

\bibitem{Davoudiasl:1999jd}
  H.~Davoudiasl, J.~L.~Hewett and T.~G.~Rizzo,
  Phys.\ Rev.\ Lett.\  {\bf 84}, 2080 (2000)
  [arXiv:hep-ph/9909255].

\bibitem{Goldberger:1999wh}
  W.~D.~Goldberger and M.~B.~Wise,
  Phys.\ Rev.\  D {\bf 60}, 107505 (1999)
  [arXiv:hep-ph/9907218].

\bibitem{Davoudiasl:1999tf}
  H.~Davoudiasl, J.~L.~Hewett and T.~G.~Rizzo,
  Phys.\ Lett.\  B {\bf 473}, 43 (2000)
  [arXiv:hep-ph/9911262].

\bibitem{Pomarol:1999ad}
  A.~Pomarol,
  Phys.\ Lett.\  B {\bf 486}, 153 (2000)
  [arXiv:hep-ph/9911294].

\bibitem{Grossman:1999ra}
  Y.~Grossman and M.~Neubert,
  Phys.\ Lett.\  B {\bf 474}, 361 (2000)
  [arXiv:hep-ph/9912408].

\bibitem{Gherghetta:2000qt}
  T.~Gherghetta and A.~Pomarol,
  Nucl.\ Phys.\  B {\bf 586}, 141 (2000)
  [arXiv:hep-ph/0003129].

\bibitem{Carena:2007ua}
  M.~S.~Carena, E.~Pont\'on, J.~Santiago and C.~E.~M.~Wagner,
  Phys.\ Rev.\  D {\bf 76}, 035006 (2007)
  [arXiv:hep-ph/0701055].



\bibitem{Huber:2000ie}
  S.~J.~Huber and Q.~Shafi,
  Phys.\ Lett.\  B {\bf 498}, 256 (2001)
  [arXiv:hep-ph/0010195];
  S.~J.~Huber,
  Nucl.\ Phys.\  B {\bf 666}, 269 (2003)
  [arXiv:hep-ph/0303183].

\bibitem{Burdman:2002gr}
  G.~Burdman,
  Phys.\ Rev.\  D {\bf 66}, 076003 (2002)
  [arXiv:hep-ph/0205329];
  Phys.\ Lett.\  B {\bf 590}, 86 (2004)
  [arXiv:hep-ph/0310144].


\bibitem{Agashe:2004ay}
  K.~Agashe, G.~Perez and A.~Soni,
  Phys.\ Rev.\ Lett.\  {\bf 93}, 201804 (2004)
  [arXiv:hep-ph/0406101].

\bibitem{Agashe:2004cp}
  K.~Agashe, G.~Perez and A.~Soni,
  Phys.\ Rev.\  D {\bf 71}, 016002 (2005)
  [arXiv:hep-ph/0408134].

\bibitem{Agashe:2006iy}
  K.~Agashe, A.~E.~Blechman and F.~Petriello,
  Phys.\ Rev.\  D {\bf 74}, 053011 (2006)
  [arXiv:hep-ph/0606021].

\bibitem{Agashe:2003zs}
  K.~Agashe, A.~Delgado, M.~J.~May and R.~Sundrum,
  JHEP {\bf 0308}, 050 (2003)
  [arXiv:hep-ph/0308036].

\bibitem{Agashe:2006at}
  K.~Agashe, R.~Contino, L.~Da Rold and A.~Pomarol,
  Phys.\ Lett.\  B {\bf 641}, 62 (2006)
  [arXiv:hep-ph/0605341].



\bibitem{Davoudiasl:2008hx}
  H.~Davoudiasl, G.~Perez and A.~Soni,
  Phys.\ Lett.\  B {\bf 665}, 67 (2008)
  [arXiv:0802.0203 [hep-ph]].

\bibitem{Randall:2001gb}
  L.~Randall and M.~D.~Schwartz,
  JHEP {\bf 0111}, 003 (2001)
  [arXiv:hep-th/0108114].

\bibitem{Csaki:2003dt}
  C.~Csaki, C.~Grojean, H.~Murayama, L.~Pilo and J.~Terning,
  Phys.\ Rev.\  D {\bf 69}, 055006 (2004)
  [arXiv:hep-ph/0305237].

\bibitem{Davoudiasl:2002ua}
  H.~Davoudiasl, J.~L.~Hewett and T.~G.~Rizzo,
  Phys.\ Rev.\  D {\bf 68}, 045002 (2003)
  [arXiv:hep-ph/0212279].

\bibitem{Carena:2002dz}
  M.~S.~Carena, E.~Pont\'on, T.~M.~P.~Tait and C.~E.~M.~Wagner,
  Phys.\ Rev.\  D {\bf 67} (2003) 096006
  [arXiv:hep-ph/0212307];
  M.~S.~Carena, A.~Delgado, E.~Pont\'on, T.~M.~P.~Tait and C.~E.~M.~Wagner,
  Phys.\ Rev.\  D {\bf 68}, 035010 (2003)
  [arXiv:hep-ph/0305188].


\bibitem{Manton:1979kb}
  N.~S.~Manton,
  Nucl.\ Phys.\  B {\bf 158}, 141 (1979).

\bibitem{Hosotani:1983xw}
  Y.~Hosotani,
  Phys.\ Lett.\  B {\bf 126}, 309 (1983).

\bibitem{Krasnikov:1991dt}
  N.~V.~Krasnikov,
  Phys.\ Lett.\  B {\bf 273}, 246 (1991).

\bibitem{Hatanaka:1998yp}
  H.~Hatanaka, T.~Inami and C.~S.~Lim,
  Mod.\ Phys.\ Lett.\  A {\bf 13}, 2601 (1998)
  [arXiv:hep-th/9805067].

\bibitem{Contino:2003ve}
  R.~Contino, Y.~Nomura and A.~Pomarol,
  Nucl.\ Phys.\  B {\bf 671}, 148 (2003)
  [arXiv:hep-ph/0306259].

\bibitem{Agashe:2004rs}
  K.~Agashe, R.~Contino and A.~Pomarol,
  Nucl.\ Phys.\  B {\bf 719}, 165 (2005)
  [arXiv:hep-ph/0412089].




\bibitem{Davoudiasl:2003zt}
  H.~Davoudiasl, J.~L.~Hewett and T.~G.~Rizzo,
  JHEP {\bf 0308} (2003) 034
  [arXiv:hep-ph/0305086].

\bibitem{ArkaniHamed:2001mi}
  N.~Arkani-Hamed, L.~J.~Hall, Y.~Nomura, D.~Tucker-Smith and N.~Weiner,
  Nucl.\ Phys.\  B {\bf 605}, 81 (2001)
  [arXiv:hep-ph/0102090].

\bibitem{Falkowski:2008yr}
 A.~Falkowski and M.~P\'erez-Victoria,
 arXiv:0810.4940 [hep-ph].

\bibitem{delAguila:2003bh}
  F.~del Aguila, M.~P\'erez-Victoria and J.~Santiago,
  JHEP {\bf 0302}, 051 (2003)
  [arXiv:hep-th/0302023].

\bibitem{Carena:2004zn}
  M.~S.~Carena, A.~Delgado, E.~Pont\'on, T.~M.~P.~Tait and C.~E.~M.~Wagner,
  Phys.\ Rev.\  D {\bf 71}, 015010 (2005)
  [arXiv:hep-ph/0410344].

\bibitem{Contino:2004vy}
  R.~Contino and A.~Pomarol,
  JHEP {\bf 0411}, 058 (2004)
  [arXiv:hep-th/0406257].

\bibitem{Buchmuller:1985jz}
  W.~Buchmuller and D.~Wyler,
  Nucl.\ Phys.\  B {\bf 268}, 621 (1986).

\bibitem{delAguila:2000aa}
  F.~del Aguila, M.~P\'erez-Victoria and J.~Santiago,
  Phys.\ Lett.\  B {\bf 492} (2000) 98
  [arXiv:hep-ph/0007160];
  JHEP {\bf 0009} (2000) 011
  [arXiv:hep-ph/0007316].

\bibitem{delAguila:2000kb}
  F.~del Aguila and J.~Santiago,
  Phys.\ Lett.\  B {\bf 493} (2000) 175
  [arXiv:hep-ph/0008143];
  JHEP {\bf 0203} (2002) 010
  [arXiv:hep-ph/0111047].

\bibitem{Barbieri:2004qk}
  R.~Barbieri, A.~Pomarol, R.~Rattazzi and A.~Strumia,
  Nucl.\ Phys.\  B {\bf 703}, 127 (2004)
  [arXiv:hep-ph/0405040].

\bibitem{Gherghetta:2006ha}
  T.~Gherghetta,
  arXiv:hep-ph/0601213.

\bibitem{Amsler:2008zzb}
  C.~Amsler {\it et al.}  [Particle Data Group],
  Phys.\ Lett.\  B {\bf 667}, 1 (2008).

\bibitem{Han:2004az}
  Z.~Han and W.~Skiba,
  Phys.\ Rev.\  D {\bf 71} (2005) 075009
  [arXiv:hep-ph/0412166];
  Z.~Han,
  Phys.\ Rev.\  D {\bf 73}, 015005 (2006)
  [arXiv:hep-ph/0510125].

\bibitem{Cacciapaglia:2006pk}
  G.~Cacciapaglia, C.~Csaki, G.~Marandella and A.~Strumia,
  Phys.\ Rev.\  D {\bf 74}, 033011 (2006)
  [arXiv:hep-ph/0604111].

\bibitem{Csaki:2002gy}
  C.~Csaki, J.~Erlich and J.~Terning,
  Phys.\ Rev.\  D {\bf 66}, 064021 (2002)
  [arXiv:hep-ph/0203034].

\bibitem{Peskin:1990zt}
  M.~E.~Peskin and T.~Takeuchi,
  Phys.\ Rev.\ Lett.\  {\bf 65}, 964 (1990);
  Phys.\ Rev.\  D {\bf 46}, 381 (1992).

\bibitem{Chacko:1999hg}
  Z.~Chacko, M.~A.~Luty and E.~Pont\'on,
  JHEP {\bf 0007}, 036 (2000)
  [arXiv:hep-ph/9909248].

\bibitem{Carena:2006bn}
  M.~S.~Carena, E.~Pont\'on, J.~Santiago and C.~E.~M.~Wagner,
  Nucl.\ Phys.\  B {\bf 759}, 202 (2006)
  [arXiv:hep-ph/0607106].

\bibitem{Agashe:2005dk}
  K.~Agashe and R.~Contino,
  Nucl.\ Phys.\  B {\bf 742} (2006) 59
  [arXiv:hep-ph/0510164].

\bibitem{Carena:2007tn}
  M.~Carena, A.~D.~Medina, B.~Panes, N.~R.~Shah and C.~E.~M.~Wagner,
  Phys.\ Rev.\  D {\bf 77}, 076003 (2008)
  [arXiv:0712.0095 [hep-ph]].

\bibitem{:2009ec}
    [Tevatron Electroweak Working Group and CDF Collaboration and D0 Collab],
  arXiv:0903.2503 [hep-ex].

\bibitem{:2008ut}
    [CDF Collaboration and D0 Collaboration],
  arXiv:0808.0147 [hep-ex].

\bibitem{Medina:2007hz}
  A.~D.~Medina, N.~R.~Shah and C.~E.~M.~Wagner,
  Phys.\ Rev.\  D {\bf 76}, 095010 (2007)
  [arXiv:0706.1281 [hep-ph]].

\bibitem{Panico:2008bx}
  G.~Panico, E.~Pont\'on, J.~Santiago and M.~Serone,
  Phys.\ Rev.\  D {\bf 77}, 115012 (2008)
  [arXiv:0801.1645 [hep-ph]].



\bibitem{Falkowski:2008fz}
  A.~Falkowski and M.~P\'erez-Victoria,
  JHEP {\bf 0812} (2008) 107
  [0806.1737 [hep-ph]];

\bibitem{Batell:2008me}
  B.~Batell, T.~Gherghetta and D.~Sword,
  Phys.\ Rev.\  D {\bf 78} (2008) 116011
  [0808.3977 [hep-ph]];

\bibitem{Delgado:2009xb}
  A.~Delgado and D.~Diego,
  0905.1095 [hep-ph].

\bibitem{Aybat:2009mk}
  S.~M.~Aybat and J.~Santiago,
  arXiv:0905.3032 [hep-ph].

\bibitem{Gherghetta:2009qs}
  T.~Gherghetta and D.~Sword,
  arXiv:0907.3523 [hep-ph].


\bibitem{Cacciapaglia:2004rb}
  G.~Cacciapaglia, C.~Csaki, C.~Grojean and J.~Terning,
  Phys.\ Rev.\  D {\bf 71}, 035015 (2005)
  [arXiv:hep-ph/0409126].

\bibitem{Cacciapaglia:2006gp}
  G.~Cacciapaglia, C.~Csaki, G.~Marandella and J.~Terning,
  Phys.\ Rev.\  D {\bf 75}, 015003 (2007)
  [arXiv:hep-ph/0607146].

\bibitem{Grojean:2006wq}
  C.~Grojean,
  Comptes Rendus Physique {\bf 8}, 1068 (2007).

\bibitem{Csaki:2005vy}
  C.~Csaki, J.~Hubisz and P.~Meade,
  arXiv:hep-ph/0510275.

\bibitem{Csaki:2008zd}
  C.~Csaki, A.~Falkowski and A.~Weiler,
  JHEP {\bf 0809}, 008 (2008)
  [arXiv:0804.1954 [hep-ph]].

\bibitem{Casagrande:2008hr}
  S.~Casagrande, F.~Goertz, U.~Haisch, M.~Neubert and T.~Pfoh,
  JHEP {\bf 0810}, 094 (2008)
  [arXiv:0807.4937 [hep-ph]].

\bibitem{Blanke:2008zb}
  M.~Blanke, A.~J.~Buras, B.~Duling, S.~Gori and A.~Weiler,
  JHEP {\bf 0903} (2009) 001
  [arXiv:0809.1073 [hep-ph]].

\bibitem{Agashe:2008uz}
  K.~Agashe, A.~Azatov and L.~Zhu,
  Phys.\ Rev.\  D {\bf 79} (2009) 056006
  [arXiv:0810.1016 [hep-ph]].

\bibitem{Blanke:2008yr}
  M.~Blanke, A.~J.~Buras, B.~Duling, K.~Gemmler and S.~Gori,
  JHEP {\bf 0903} (2009) 108
  [arXiv:0812.3803 [hep-ph]].

\bibitem{Albrecht:2009xr}
  M.~E.~Albrecht, M.~Blanke, A.~J.~Buras, B.~Duling and K.~Gemmler,
  arXiv:0903.2415 [hep-ph].

\bibitem{Gedalia:2009ws}
  O.~Gedalia, G.~Isidori and G.~Perez,
  arXiv:0905.3264 [hep-ph].

\bibitem{Agashe:2009di}
  K.~Agashe and R.~Contino,
  arXiv:0906.1542 [hep-ph].

\bibitem{Azatov:2009na}
  A.~Azatov, M.~Toharia and L.~Zhu,
  arXiv:0906.1990 [hep-ph].








\bibitem{Cacciapaglia:2007fw}
  G.~Cacciapaglia, C.~Csaki, J.~Galloway, G.~Marandella, J.~Terning
  and A.~Weiler, 
  JHEP {\bf 0804}, 006 (2008)
  [arXiv:0709.1714 [hep-ph]].


\bibitem{Fitzpatrick:2007sa}
  A.~L.~Fitzpatrick, G.~Perez and L.~Randall,
  arXiv:0710.1869 [hep-ph].

\bibitem{Cheung:2007bu}
  C.~Cheung, A.~L.~Fitzpatrick and L.~Randall,
  JHEP {\bf 0801}, 069 (2008)
  [arXiv:0711.4421 [hep-th]].

\bibitem{Fitzpatrick:2008zza}
  A.~L.~Fitzpatrick, L.~Randall and G.~Perez,
  Phys.\ Rev.\ Lett.\  {\bf 100}, 171604 (2008).

\bibitem{Santiago:2008vq}
  J.~Santiago,
  JHEP {\bf 0812}, 046 (2008)
  [arXiv:0806.1230 [hep-ph]].

\bibitem{Csaki:2008eh}
  C.~Csaki, A.~Falkowski and A.~Weiler,
  Phys.\ Rev.\  D {\bf 80}, 016001 (2009)
  [arXiv:0806.3757 [hep-ph]].

\bibitem{Agashe:2009tu}
  K.~Agashe,
  arXiv:0902.2400 [hep-ph].

\bibitem{Csaki:2009wc}
  C.~Csaki, G.~Perez, Z.~Surujon and A.~Weiler,
  arXiv:0907.0474 [hep-ph].

\bibitem{Chen:2009gy}
  M.~C.~Chen, K.~T.~Mahanthappa and F.~Yu,
  arXiv:0907.3963 [hep-ph].



\bibitem{Hassanain:2009at}
  B.~Hassanain, J.~March-Russell and J.~G.~Rosa,
  JHEP {\bf 0907}, 077 (2009)
  [arXiv:0904.4108 [hep-ph]].

\bibitem{Perelstein:2009qi}
  M.~Perelstein and A.~Spray,
  arXiv:0907.3496 [hep-ph].


\bibitem{CDF_RS}
  T.~Aaltonen {\it et al.}  [CDF Collaboration],
  Phys.\ Rev.\ Lett.\  {\bf 102}, 091805 (2009)
  [arXiv:0811.0053 [hep-ex]].

\bibitem{D0_RS}
  V.~M.~Abazov {\it et al.}  [D0 Collaboration],
  Phys.\ Rev.\ Lett.\  {\bf 100}, 091802 (2008)
  [arXiv:0710.3338 [hep-ex]].

\bibitem{KKgraviton2}
  K.~Agashe, H.~Davoudiasl, G.~Perez and A.~Soni,
  Phys.\ Rev.\  D {\bf 76}, 036006 (2007)
  [arXiv:hep-ph/0701186].

\bibitem{ATLAS_RS}
  B.~C.~Allanach, K.~Odagiri, M.~J.~Palmer, M.~A.~Parker, A.~Sabetfakhri and B.~R.~Webber,
  JHEP {\bf 0212}, 039 (2002)
  [arXiv:hep-ph/0211205].

\bibitem{CMS_RS}
  I.~Belotelov {\it et al.}, CERN-CMS-NOTE-2006-104.

\bibitem{Davoudiasl:2000wi}
  H.~Davoudiasl, J.~L.~Hewett and T.~G.~Rizzo,
  Phys.\ Rev.\  D {\bf 63}, 075004 (2001)
  [arXiv:hep-ph/0006041].

\bibitem{KKgraviton1}
  A.~L.~Fitzpatrick, J.~Kaplan, L.~Randall and L.~T.~Wang,
  arXiv:hep-ph/0701150.

\bibitem{conejetalgorithm}
J.E. Huth {\em et al.},
{\it Proceedings of Research Directions For The Decade: Snowmass Accord 1990}.

\bibitem{Antipin:2007pi}
  O.~Antipin, D.~Atwood and A.~Soni,
  Phys.\ Lett.\  B {\bf 666}, 155 (2008)
  [arXiv:0711.3175 [hep-ph]].

\bibitem{Antipin:2008hj}
  O.~Antipin and A.~Soni,
  JHEP {\bf 0810}, 018 (2008)
  [arXiv:0806.3427 [hep-ph]].

\bibitem{Cousins:2005pq}
  R.~Cousins, J.~Mumford, J.~Tucker and V.~Valuev,
  JHEP {\bf 0511}, 046 (2005).

\bibitem{Osland:2008sy}
  P.~Osland, A.~A.~Pankov, N.~Paver and A.~V.~Tsytrinov,
  Phys.\ Rev.\  D {\bf 78}, 035008 (2008)
  [arXiv:0805.2734 [hep-ph]].

\bibitem{Murayama:2009jz}
  H.~Murayama and V.~Rentala,
  arXiv:0904.4561 [hep-ph].

\bibitem{Gianotti:2002xx}
  F.~Gianotti {\it et al.},
  Eur.\ Phys.\ J.\ C {\bf 39}, 293 (2005)
  [arXiv:hep-ph/0204087].

\bibitem{Bruning:2002yh}
  O.~Bruning {\it et al.},
CERN-LHC-PROJECT-REPORT-626, Dec 2002. 98pp.

\bibitem{KKgluon1}
K.~Agashe, A.~Belyaev, T.~Krupovnickas, G.~Perez and J.~Virzi,
  Phys.\ Rev.\  D {\bf 77}, 015003 (2008)
  [arXiv:hep-ph/0612015].

\bibitem{KKgluon2}
B.~Lillie, L.~Randall and L.~T.~Wang,
  JHEP {\bf 0709}, 074 (2007)
  [arXiv:hep-ph/0701166].

\bibitem{otherboostedtop}
  V.~Barger, T.~Han and D.~G.~E.~Walker,
  Phys.\ Rev.\ Lett.\  {\bf 100}, 031801 (2008)
  [arXiv:hep-ph/0612016];
%
  B.~Lillie, J.~Shu and T.~M.~P.~Tait,
  Phys.\ Rev.\  D {\bf 76}, 115016 (2007)
  [arXiv:0706.3960 [hep-ph]];
%
  U.~Baur and L.~H.~Orr,
Phys.\ Rev.\  D {\bf 76}, 094012 (2007)
  [arXiv:0707.2066 [hep-ph]]
and
%
  Phys.\ Rev.\  D {\bf 77}, 114001 (2008)
  [arXiv:0803.1160 [hep-ph]];
  Y.~Bai and Z.~Han,
  arXiv:0809.4487 [hep-ph].

\bibitem{toptagging}
  J.~Thaler and L.~T.~Wang,
  JHEP {\bf 0807}, 092 (2008)
  [arXiv:0806.0023 [hep-ph]];
  D.~E.~Kaplan, K.~Rehermann, M.~D.~Schwartz and B.~Tweedie,
  arXiv:0806.0848 [hep-ph];
%
 L.~G.~Almeida, S.~J.~Lee, G.~Perez, I.~Sung and J.~Virzi,
 arXiv:0810.0934 [hep-ph].




\bibitem{Djouadi:2007eg}
  A.~Djouadi, G.~Moreau and R.~K.~Singh,
  Nucl.\ Phys.\  B {\bf 797}, 1 (2008)
  [arXiv:0706.4191 [hep-ph]].

\bibitem{Djouadi:2009nb}
  A.~Djouadi, G.~Moreau, F.~Richard and R.~K.~Singh,
  arXiv:0906.0604 [hep-ph].

\bibitem{Djouadi:2006rk}
  A.~Djouadi, G.~Moreau and F.~Richard,
  Nucl.\ Phys.\  B {\bf 773}, 43 (2007)
  [arXiv:hep-ph/0610173].

\bibitem{Guchait:2007jd}
  M.~Guchait, F.~Mahmoudi and K.~Sridhar,
  Phys.\ Lett.\  B {\bf 666}, 347 (2008)
  [arXiv:0710.2234 [hep-ph]].


\bibitem{Agashe:2007ki}
  K.~Agashe {\it et al.},
  Phys.\ Rev.\  D {\bf 76}, 115015 (2007)
  [arXiv:0709.0007 [hep-ph]].

\bibitem{Agashe:2008jb}
  K.~Agashe, S.~Gopalakrishna, T.~Han, G.~Y.~Huang and A.~Soni,
  arXiv:0810.1497 [hep-ph].

\bibitem{Piai:2007ys}
  M.~Piai,
  arXiv:0704.2205 [hep-ph].

\bibitem{jmassRefs}
D.~Benchekroun, C.~Driouichi, A.~Hoummada, SN-ATLAS-2001-001, ATL-COM-PHYS-2000-020,
EPJ Direct 3, 1 (2001);
  W.~Skiba and D.~Tucker-Smith,
  Phys.\ Rev.\  D {\bf 75}, 115010 (2007)
  [arXiv:hep-ph/0701247];
See also talk by Gustaaf Brooijmans at the
"Workshop on Possible Parity Restoration at High Energy", Beijing (China) June 11-12, 2007.

\bibitem{Almeida:2008yp}
  L.~G.~Almeida, S.~J.~Lee, G.~Perez, G.~Sterman, I.~Sung and J.~Virzi,
  arXiv:0807.0234 [hep-ph].

\bibitem{SubJets}
M.~Strassler, ``Unusual Physics Signatures at the LHC,'' 
talk presented at the 2007 Phenomenology Symposium - Pheno 07, 
University of Wisconsin, Madison, May 7-9, 2007; 
Frank Paige, 2007, private communication; 
The ATLAS Collaboration, ATL-PHYS-PUB-2009-081, ATL-COM-PHYS-2009-255; 
The CMS Collaboration CMS PAS JME-09-001, CMS PAS EXO-09-002. 

\bibitem{Bauer:2008xb}
  M.~Bauer, S.~Casagrande, L.~Grunder, U.~Haisch and M.~Neubert,
  arXiv:0811.3678 [hep-ph].

\bibitem{Davoudiasl:2009jk}
  H.~Davoudiasl, S.~Gopalakrishna and A.~Soni,
  arXiv:0908.1131 [hep-ph].

\bibitem{Ledroit:2007ik}
  F.~Ledroit, G.~Moreau and J.~Morel,
  JHEP {\bf 0709}, 071 (2007)
  [arXiv:hep-ph/0703262].

\bibitem{Davoudiasl:2007zx}
  H.~Davoudiasl and A.~Soni,
  Phys.\ Rev.\  D {\bf 76}, 095015 (2007)
  [arXiv:0705.0151 [hep-ph]].

\bibitem{Davoudiasl:2007wf}
  H.~Davoudiasl, T.~G.~Rizzo and A.~Soni,
  Phys.\ Rev.\  D {\bf 77}, 036001 (2008)
  [arXiv:0710.2078 [hep-ph]].

\bibitem{Atre:2008iu}
  A.~Atre, M.~Carena, T.~Han and J.~Santiago,
  arXiv:0806.3966 [hep-ph].

\bibitem{Choi:2002ps}
  K.~w.~Choi and I.~W.~Kim,
  Phys.\ Rev.\  D {\bf 67}, 045005 (2003)
  [arXiv:hep-th/0208071].


\bibitem{Agashe:2004ci}
  K.~Agashe and G.~Servant,
  Phys.\ Rev.\ Lett.\  {\bf 93}, 231805 (2004)
  [arXiv:hep-ph/0403143];
  JCAP {\bf 0502}, 002 (2005)
  [arXiv:hep-ph/0411254].

\bibitem{Dennis:2007tv}
  C.~Dennis, M.~Karagoz Unel, G.~Servant and J.~Tseng,
  arXiv:hep-ph/0701158.

\bibitem{Contino:2008hi}
  R.~Contino and G.~Servant,
  JHEP {\bf 0806}, 026 (2008)
  [arXiv:0801.1679 [hep-ph]].

\bibitem{AguilarSaavedra:2009es}
 J.~A.~Aguilar-Saavedra,
 arXiv:0907.3155 [hep-ph].

\bibitem{Bouchart:2008vp}
  C.~Bouchart and G.~Moreau,
  Nucl.\ Phys.\  B {\bf 810}, 66 (2009)
  [arXiv:0807.4461 [hep-ph]].




\bibitem{Goldberger:1999un}
  W.~D.~Goldberger and M.~B.~Wise,
  Phys.\ Lett.\  B {\bf 475}, 275 (2000)
  [arXiv:hep-ph/9911457].

\bibitem{DeWolfe:1999cp}
  O.~DeWolfe, D.~Z.~Freedman, S.~S.~Gubser and A.~Karch,
  Phys.\ Rev.\  D {\bf 62}, 046008 (2000)
  [arXiv:hep-th/9909134].

\bibitem{Csaki:2000zn}
  C.~Csaki, M.~L.~Graesser and G.~D.~Kribs,
  Phys.\ Rev.\  D {\bf 63}, 065002 (2001)
  [arXiv:hep-th/0008151].

\bibitem{Garriga:2002vf}
  J.~Garriga and A.~Pomarol,
  Phys.\ Lett.\  B {\bf 560}, 91 (2003)
  [arXiv:hep-th/0212227].

\bibitem{Goldberger:2000dv}
  W.~D.~Goldberger and I.~Z.~Rothstein,
  Phys.\ Lett.\  B {\bf 491}, 339 (2000)
  [arXiv:hep-th/0007065].

\bibitem{Gunion:2003px}
  J.~F.~Gunion, M.~Toharia and J.~D.~Wells,
  Phys.\ Lett.\  B {\bf 585}, 295 (2004)
  [arXiv:hep-ph/0311219].

\bibitem{Tanaka:2000er}
  T.~Tanaka and X.~Montes,
  Nucl.\ Phys.\  B {\bf 582}, 259 (2000)
  [arXiv:hep-th/0001092].


\bibitem{Giudice:2000av}
  G.~F.~Giudice, R.~Rattazzi and J.~D.~Wells,
  Nucl.\ Phys.\  B {\bf 595}, 250 (2001)
  [arXiv:hep-ph/0002178].

\bibitem{Rizzo:2002pq}
  T.~G.~Rizzo,
  JHEP {\bf 0206}, 056 (2002)
  [arXiv:hep-ph/0205242].

\bibitem{Csaki:2007ns}
  C.~Csaki, J.~Hubisz and S.~J.~Lee,
  Phys.\ Rev.\  D {\bf 76}, 125015 (2007)
  [arXiv:0705.3844 [hep-ph]].

\bibitem{Toharia:2008tm}
  M.~Toharia,
  Phys.\ Rev.\  D {\bf 79}, 015009 (2009)
  [arXiv:0809.5245 [hep-ph]].

\bibitem{Azatov:2008vm}
  A.~Azatov, M.~Toharia and L.~Zhu,
  arXiv:0812.2489 [hep-ph].




\end{thebibliography}
\end{document}